\documentclass[twocolumn,showpacs,showkeys,superscriptaddress,prc,nofootinbib]{revtex4-2}

\usepackage{hyperref}
\usepackage[hyphenbreaks]{breakurl}
\hypersetup{breaklinks=true,colorlinks=true, citecolor=blue, urlcolor=blue, linkcolor=blue}
\PassOptionsToPackage{hyphens}{url}
\usepackage{CJK}
\usepackage{mathrsfs}
\usepackage{amssymb}
\usepackage{amsmath}
\usepackage{latexsym}
\usepackage{graphicx}
\usepackage{dcolumn}
\usepackage{bm}
\usepackage{graphicx}
\usepackage{float}
\usepackage[normalem]{ulem}
\usepackage{color,xcolor}
\usepackage{multirow}
\usepackage{enumerate}
\usepackage{ulem}
\usepackage{textcomp}

\newcolumntype{d}[1]{Dc{.}{.}{#1}}
\begin{document}
	\begin{CJK*}{UTF8}{}
		
		\title{ 
			Robust linear correlations related to neutron skin thickness
		}
		\author{Y. Lei ({\CJKfamily{gbsn}雷杨})}
		\email[]{leiyang19850228@gmail.com}
		\affiliation{School of Nuclear Science and Technology, Southwest University of Science and Technology, Mianyang 621010, China}
		\author{X. Lian ({\CJKfamily{gbsn}连新})}
		\affiliation{College of Physics, Sichuan University, Chendu 610065, China}
		\author{C. L. Bai ({\CJKfamily{gbsn}白春林})}
		\email[]{bclphy@scu.edu.cn}
		\affiliation{College of Physics, Sichuan University, Chendu 610065, China}
		\date{\today}
		
		\begin{abstract}
			We observe various robust linear correlations related to neutron skin thickness ($\Delta R_{\rm np}$) within different interaction ensembles, including newly proposed random Skyrme ensemble. The robust linear correlation between $\Delta R_{\rm np}$, or charge radius difference of mirror nuclei ($\Delta R_{\rm mirr}$), and the isospin asymmetry ($I=\frac{N-Z}{A}$) becomes apparent as the model space is enlarged. Shape coexistence, or shape effect on charge radius, is considered to explain the experimental deviation of ${}^{18}$O/Ne and some odd-$A$ $\Delta R_{\rm mirr}$s from the $\Delta R_{\rm mirr}-I$ linearity. The slopes of the linear $\Delta R_{\rm mirr}-I$ and $\Delta R_{\rm np}-I$ correlations ($C_{\rm np}$ and $C_{\rm mirr}$, respectively) are also robustly and linearly correlated to the slope of the symmetry energy ($L$). These linear correlations are further understood with the similar formulation between between $L$ and the symmetry energy coefficient ($J$). The linear correlations between $C_{\rm np}-L$ and $C_{\rm mirr}-L$ are also adopted to constrain $L$ to $20\sim36$ MeV with 1$\sigma$ confidence. Considering the deviation of ${}^{18}$O/Ne $\Delta R_{mirr}$ due to shape coexistence, the 1$\sigma$ range for $L$ is further narrowed to $28\sim36$ MeV, suggesting a relatively soft equation of state for nuclear matter.
		\end{abstract}
		\maketitle
	\end{CJK*}
	\section{Introduction}\label{sec-int}	
	The nuclear equation of state (EOS) is crucial for understanding the structure of neutron stars \cite{ns} and the dynamical processes in binary compact-star mergers and core-collapse supernovae. These events provide conditions of high densities and temperatures, essential for nucleosynthesis beyond the iron group \cite{eos-super}. Additionally, the density dependence of the nuclear symmetry energy within the EOS plays a significant role in studying drip lines, nuclear masses, and collectivities of neutron-rich nuclei \cite{sym-stru-1,sym-stru-2,sym-stru-3,sym-stru-4,sym-stru-5,sym-stru-6,sym-stru-7,sym-stru-8,sym-stru-9}. 
	
	The neutron skin thickness, $\Delta R_{\rm np}$, defined as the difference between the root-mean-squared radii of the neutron and proton density distributions. It represents a delicate equilibrium between the inward pressure arising from surface tension and the outward pressure due to degeneracy. This equilibrium is reminiscent of the one found in neutron stars, albeit with the inward pressure stemming from gravity in that context. Consequently, the bulk properties of neutron stars are likely related to the neutron skin of nuclei \cite{brown2000,PhysRevLett.86.5647,PhysRevC.86.015803,Hagen2016,brown-rch,PhysRevLett.120.172702,PhysRevC.100.015802}.	
	
	Specifically, $\Delta R_{\rm np}$ is believed to be positively and linearly correlated with the parameter $L$ \cite{rnp-1,rnp-2,explain_phys_rep}, which represents the slope of the symmetry energy against nucleon density at saturation density $\rho_0$. $L$ plays a pivotal role in extending our understanding to high-density scenarios, where unique astrophysical objects and events may potentially exist and manifest themselves \cite{to-high-density}. For neutron stars, $L$---indicating the softness or stiffness of the EOS---determines their radius \cite{ns}. In symmetric nuclear matter, $L$ is proportional to the pressure of pure neutron matter at $\rho_0$ \cite{L-pressure}. However, $L$ cannot be directly measured experimentally, so data from terrestrial nuclear laboratory experiments and astrophysical observations are used to constrain it. Various studies have reported different constraints for $L$, such as $88\pm 25$ MeV from isospin diffusion data in heavy-ion collisions \cite{CLW}, $58.9\pm 16$ MeV from 29 previous analyses \cite{Li-L}, $58.7\pm 28.1$ MeV from averaging over more than 53 results \cite{rmp-con-L}, $59.8\pm 4.1$ MeV using a Bayesian approach \cite{Bay-L}, $57.7\pm 19$ MeV from neutron-star observations following the binary neutron star merger GW170817 \cite{NS-L}, $50.3\sim 89.4$ MeV and $29.0\sim 82.0$ MeV from the pygmy dipole resonance of $^{68}$Ni and $^{132}$Sn, respectively \cite{PhysRevC.81.041301}, and $37\pm 18$ MeV from isovector giant quadrupole resonance energies \cite{PhysRevC.87.034301}. Admittedly, not all experimental data could be included here. Nevertheless, given these varying constraints from different data sources, it remains an open question whether $L$ is smaller or larger, implying a softer or stiffer EOS, and correspondingly, smaller or larger neutron star radii.
	
	
	Obviously, the correlation between $\Delta R_{\rm np}$ and $L$ is also valuable for constraining $L$ through nuclear radius measurements. The basic approach involves constructing an ensemble within a specific many-body theory framework (typically the Skyrme-Hartree-Fock model or covariant energy density functional theory), where different interaction or energy density functional parametrizations are applied to a nucleus with an experimentally determined $\Delta R_{\rm np}$. Within this ensemble, a quantitative correlation between $\Delta R_{\rm np}$ and $L$ is anticipated. Consequently, experimental values of $\Delta R_{\rm np}$ can offer constraints on $L$. Many studies have been conducted along this line of thinking \cite{CLW-np,PhysRevC.93.064303,WL-np,Fan2015,RZZ-np,LBA-np}.
	
	
	However, measuring the neutron radius is more difficult and less accurate than measuring the proton radius because experimentalists are more adept at manipulating electromagnetic interactions, suitable for proton exploration, than strong or electroweak probes sensitive to neutron density. Therefore, an alternative to $\Delta R_{\rm np}$ for constraining $L$ is desirable. Wang, Li, and Brown proposed that the charge radii difference between mirror nuclei ($\Delta R_{\rm mirr}$) should also be correlated with $L$, considering that under nuclear isospin conservation, $\Delta R_{\rm mirr} \simeq \Delta R_{\rm np}$ \cite{WL-rch,brown-rch}. Since $\Delta R_{\rm mirr}$ only involves proton density probes, it may provide more precise data for constraining $L$. Studies using mirror nuclei have deduced $L$ values within various ranges \cite{brown-rch-2,PhysRevLett.127.182503,An2023}, and a recent comprehensive survey further identified some mirror pairs that may be less suitable for $L$ calibration, and emphasized the pairing effect in weakening the correlation between $\Delta R_{\rm mirr}$ and $L$ \cite{LYF-rch}.
	
\begin{figure}[!htb]
	\includegraphics[angle=0,width=0.45\textwidth]{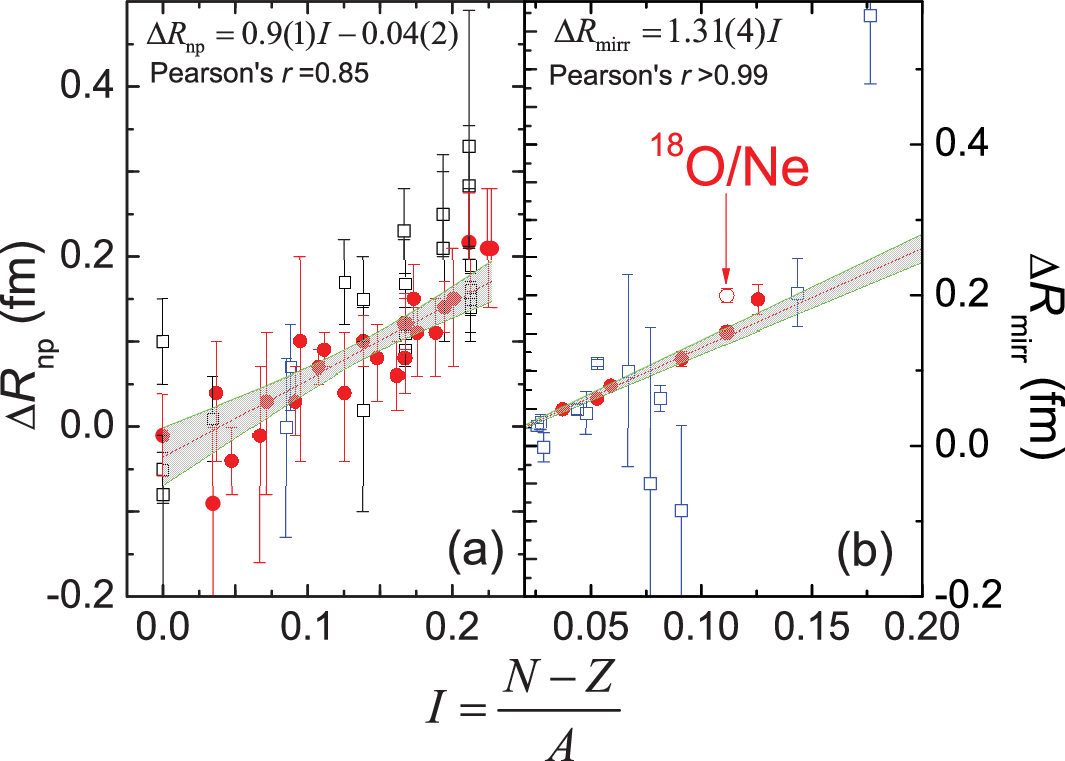}
	\caption{(Color online)
		Experimental $\Delta R_{\rm np}$ and $\Delta R_{\rm mirr}$ against $I=\frac{N-Z}{A}$, in Panel (a) and (b), respectively. For each nucleus, the most recent data available to us are shown as red filled circles ($\color{red} \bullet$) for even-even nuclei and blue open boxes ($\color{blue}\Box$) for odd-mass nuclei. In Panel (a), recent data are taken from Refs. \cite{zenihiro2018directdeterminationneutronskin,doi:10.1142/S0218301304002168,PhysRevC.46.1825,PhysRevLett.129.042501,PhysRevLett.131.202302}, and other $\Delta R_{\rm np}$ data from Refs. \cite{PhysRevLett.126.172502,zenihiro2018directdeterminationneutronskin,doi:10.1142/S0218301304002168,npa567.521,PhysRevC.19.1855,PhysRevC.21.1488,PhysRevC.46.1825,PhysRevLett.108.112502,PhysRevLett.126.172502} are represented by black open boxes ($\Box$). It can be seen that the recent $\Delta R_{\rm np}$ data suggests a linear trend with $I$, characterized by a Pearson correlation coefficient of $r\simeq 0.85$. For the recent $\Delta R_{\rm np}$ data, a linear fit was performed, yielding $\Delta R_{\rm np}=0.9(1)I-0.04(2)$, indicated by the gray shaded 1$\sigma$-confidence band. In Panel (b), $\Delta R_{\rm mirr}$ data are from Ref. \cite{rc13,rc21,PhysRevLett.127.182503,PhysRevLett.132.162502,ZHAO2024139082}, and a strong linear correlation is observed, with a Pearson correlation coefficient $r>0.99$ for even-even nuclei, except for the $^{18}$O/Ne mirror pair. Therefore, the data for $^{18}$O/Ne is highlighted by red open circles ($\color{red} \circ$). For this general linear correlation, a proportional fit was performed, constrained by $\Delta R_{\rm mirr} \equiv 0$ at $I=0$. This fitting yields $\Delta R_{\rm mirr}=1.31(4)I$, indicated by the gray shaded 1$\sigma$-confidence band. We also note that the systematics of odd-mass $\Delta R_{\rm mirr}$ data shows more scatter, which will be discussed in Sec. \ref{sec:shape}, along with the deviation of the $^{18}$O/Ne mirror pair.
	}\label{fig:exp}
	\end{figure}

	Furthermore, a linear correlation between $\Delta R_{\rm np}$/$\Delta R_{\rm mirr}$ and the isospin asymmetry $I = \frac{N-Z}{A}$ has been noted \cite{rc21,PhysRevLett.87.082501,rch-I}. As shown in Figure \ref{fig:exp}, except for $\Delta R_{\rm mirr}$ of the $^{18}$O/Ne pair, all recently measured data for even-even nuclei follow this general linear correlation. The liquid drop model  is often used to explain this linear relationship \cite{explain_phys_rep}. It has also been explained and reproduced by mean field, $ab~initio$ coupled cluster, and auxiliary field diffusion Monte Carlo models \cite{rch-I,MYERS1969395,MYERS1974186,MYERS1980267,PETHICK1996173}.
	
	Consequently, a many-body theory, when based on appropriate parameters related to $L$ and capable of yielding experimentally consistent $\Delta R_{\rm np}$ or $\Delta R_{\rm mirr}$ values for nuclei with available data, might be expected to exhibit this linear correlation between $\Delta R_{\rm np}$ (or $\Delta R_{\rm mirr}$) and the isospin asymmetry $I=\frac{N-Z}{A}$, with a slope and intercept that align with experimental findings. Therefore, the existence, slope, and intercept of this linear correlation within a many-body framework could potentially offer constraints on $L$, by considering the available experimental $\Delta R_{\rm np}$ or $\Delta R_{\rm mirr}$ data.
	
	A key question is whether the linear correlation between $\Delta R_{\rm np}$ (or $\Delta R_{\rm mirr}$) and the isospin asymmetry $I=\frac{N-Z}{A}$  persists in many-body theory ensembles, even without finely tuned Hamiltonians or Lagrangians related to reasonable $L$s.  It is worth noting that many-body calculations \cite{rch-I,MYERS1969395,MYERS1974186,MYERS1980267,PETHICK1996173} that reproduce this linear trend employ diverse interactions, model spaces, and even theoretical frameworks. This observation might suggest that this correlation could be largely independent of interaction and modeling details, indicating a degree of robustness. However, such linearity is not necessarily guaranteed, at least from a microscopic perspective. Therefore, it is pertinent to investigate if this correlation is indeed a robust, simple feature emerging from the complex many-body system, or if it relies on specific or sophisticated nucleon-nucleon interactions.
	
	Nuclear robust properties can be effectively explored within a random interaction ensemble of nuclear structure models \cite{KOTA2001223,ZELEVINSKY2004311,ZHAO20041,RevModPhys.81.539}, where all two-body (and even one-body) interaction matrix elements are assigned random values. Using random interactions, one can perform many-body calculations repeatedly. Statistical analysis of these calculations reveals how {\it simple} regularities can emerge from {\it complex} nuclei, even when employing interactions significantly different from realistic interactions \cite{Sherrill01042005}. Examples of such regularities are the predominance of $I^\pi=0^+$ ground states \cite{PhysRevLett.80.2749,PhysRevC.70.054322}, the collective-like motions \cite{PhysRevLett.84.420,PhysRevC.104.054319}, odd-even staggering in proton-neutron interactions \cite{PhysRevC.91.054319}, and robust correlations among nuclear observables \cite{PhysRevC.93.024319,Qin2018}.
	
	This work employs the idea of random interactions to explore potential robust correlations between $\Delta R_{\rm np}$, $\Delta R_{\rm mirr}$, $I$, $L$, and related quantities. Section \ref{sec:r-I} discusses the robustness of the linear relationships between $\Delta R_{\rm np}$ and $I$, as well as $\Delta R_{\rm mirr}$ and $I$, using the shell model and the Skyrme-Hartree-Fock (SHF) model. Section \ref{sec:c-L} shows that the slopes of the $\Delta R_{\rm np}-I$ and $\Delta R_{\rm mirr}-I$ linear relationships also exhibit a robust linear correlation with $L$.  An analytical explanation is presented. Section \ref{sec:L-con} utilizes the slopes of the $\Delta R_{\rm np}-I$ and $\Delta R_{\rm mirr}-I$ linear relationships to constrain $L$, considering the shape coexistence in $^{18}$O/Ne.  The influence of nuclear shape on the $\Delta R_{\rm mirr}-I$ relationship is emphasized, and this is used to explain the off-systematics behaviors observed in odd-$A$ $\Delta R_{\rm mirr}$. Section \ref{sec:sum} summarizes this work.

\section{$\Delta R_{\rm np}-I$ and $\Delta R_{\rm mirr}-I$ linearity}\label{sec:r-I}
\subsection{random quasi-particle ensemble (RQE)}	
	To examine the robustness of the linear relationship between $\Delta R_{\rm np}$ (or $\Delta R_{\rm mirr}$) and $I$, several considerations were taken into account before performing the random interaction calculations.
	
	First, we need to decide which random-interaction ensemble to adopt. We chose to use the random quasi-particle ensemble (RQE) \cite{PhysRevLett.80.2749} without isospin symmetry. In the RQE, the statistics of two-body interaction matrix elements is invariant under particle-hole transformation. This ensures consistent interaction statistics govern shell-model calculations within the model space, facilitating a wider range of accessible $I$ in the RQE.
	
	Second, we need to consider the model space for RQE calculations. Many-body calculations within a single major shell of the harmonic oscillator always yield a constant expectation value for the $\hat r^2$ operator, resulting in zero $\Delta R_{\rm np}$. With two adjacent major shells, $\langle\hat r^2\rangle$ becomes a linear combination of particle occupations across the two shells, which might lead to a trivial linear correlation between $\Delta R_{\rm np}$ (or $\Delta R_{\rm mirr}$) and $I$. From a microscopic perspective, the nonlinear contribution to $\Delta R_{\rm np}$ is expected to arise from one-body correlations across two major shells, specifically, the non-zero off-diagonal matrix elements of $\langle nl|r^2|n^\prime l\rangle$. For nuclei heavier than those of the $sd$ shell, this nonlinear contribution may be negligible. To investigate the influence of nonlinear contributions from off-diagonal matrix elements on the linearity of $\Delta R_{\rm np}-I$ and $\Delta R_{\rm mirr}-I$ correlations, our shell-model calculations are performed within a single-particle model space including single-particle orbits from two major shells but restricted to the same orbital angular momentum $l$ and total angular momentum $j$. Specifically, we perform our RQE calculations in spaces of $\{1d_{5/2},2d_{5/2}\}$, $\{1f_{7/2},2f_{7/2}\}$, $\{1h_{9/2},2h_{9/2}\}$, $\{1h_{11/2},2h_{11/2}\}$, $\{1j_{13/2},2j_{13/2}\}$, and $\{1j_{15/2},2j_{15/2}\}$, denoted by $d_{5/2}$, $f_{7/2}$, $h_{9/2}$, $h_{11/2}$, $j_{13/2}$ and $j_{15/2}$ model spaces in this paper.
	
\begin{figure}[!htb]
		\includegraphics[angle=0,width=0.45\textwidth]{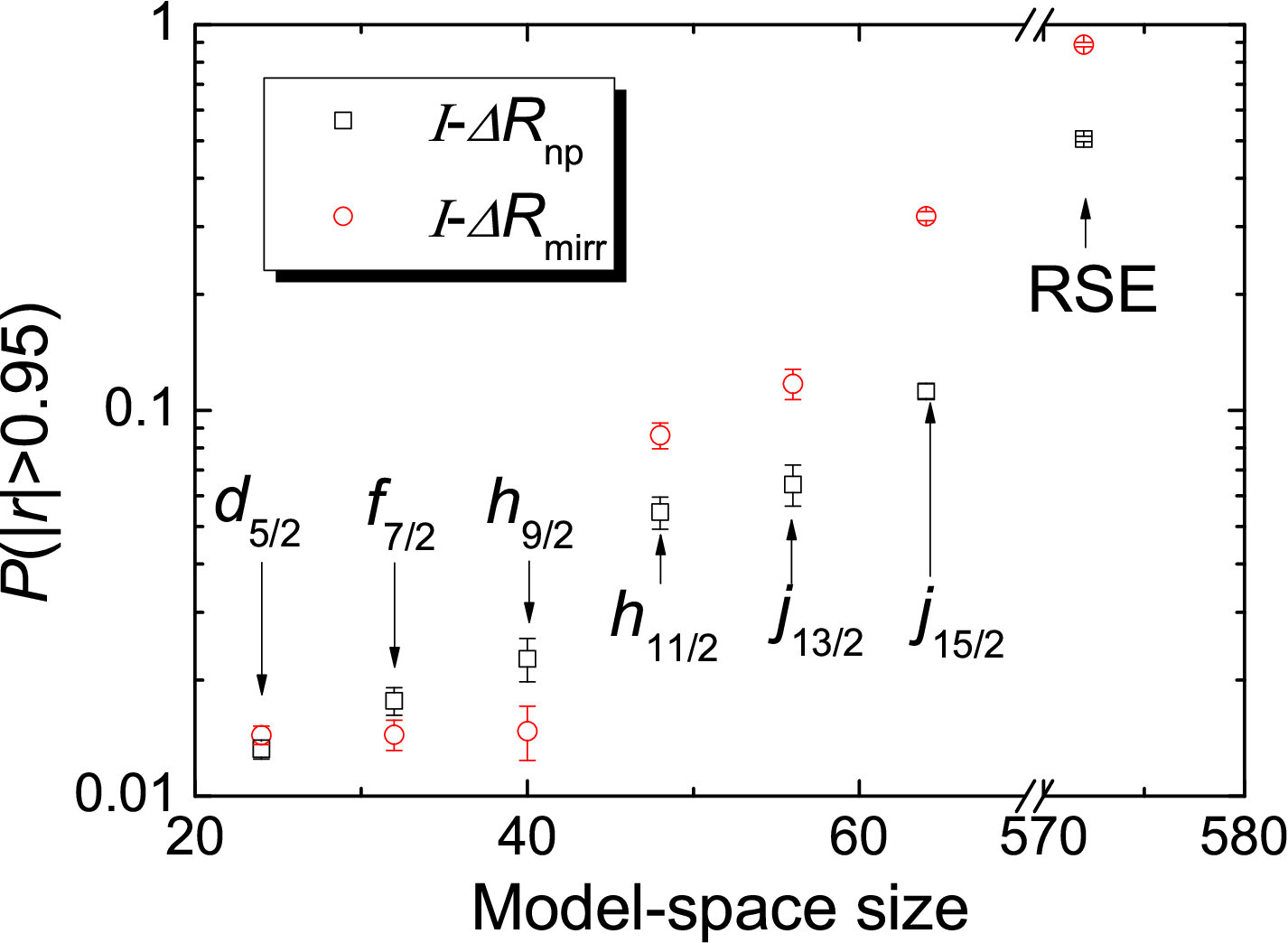}
		\caption{(Color online)
			Probabilities of Pearson's $|r|>0.95$, $P(|r|>0.95)$, for $\Delta R_{\rm np}-I$ and $\Delta R_{\rm mirr}-I$ correlations, against model-space sizes within the RQE (random quasi-particle ensemble) and RSE (random Skyrme ensemble) calculations. For RQE, corresponding model-spaces are labeled. The error bar corresponds to the statistic error of $|r|$ counting. According to our numerical test, if the our many-body calculations provided independent random number as $\Delta R_{\rm np}$ and $\Delta R_{\rm mirr}$ output, there would have been normally less than $10^{-5}$ order of possibility to produce $|r|>0.95$, i.e., we would not have observed a single linear correlation within 10000 random-interaction runs.
		}\label{fig:rgt9}
	\end{figure}

	Third, using the RQE, we repeatedly perform over 1000 shell-model calculations for each calculable\footnote{Note: ``Calculable" means the shell-model calculation for a specific pseudo nuclei takes no more than 64GB memory with a 36-thread openmp run of BIGSTICK \cite{johnson2018bigstickflexibleconfigurationinteractionshellmodel}.} pseudo nucleus in these model spaces. For a pseudo nucleus with $Z$ protons and $N$ neutrons, the expectation values of $\hat r^2$ for the resulting ground states are calculated for both proton and neutron distributions, denoted as $\langle \hat r^2_\pi\rangle(Z,N)$ and $\langle \hat r^2_\nu\rangle(Z,N)$, respectively. The squared charge radius $r^2_{\rm ch}(Z,N)$ is calculated based on the method described in Ref. \cite{ZHAO2024139082}. Thus, we define $\Delta R_{\rm np}(Z,N)=\sqrt{\langle \hat r^2_\nu\rangle}(Z,N)-\sqrt{\langle \hat r^2_\pi\rangle}(Z,N)$, and $\Delta R_{\rm mirr}(Z,N)= \sqrt{r^2_{\rm ch}}(N,Z)-\sqrt{r^2_{\rm ch}}(Z,N)$. Furthermore, we calculate the Pearson's $r$ coefficient \cite{pearson1895} between $\Delta R_{\rm np} (Z,N)$ (or $\Delta R_{\rm mirr} (Z,N)$) and $I=\frac{N-Z}{A}$ for each random interaction. In this context, $|r|=1$ indicates perfect linearity, while $|r|=0$ implies no linear correlation between the radius differences and isospin asymmetry $I$. Subsequently, we count the occurrences of $|r|$ in each model space. The probabilities of $|r|>0.95$, $P(|r|>0.95)$, are then shown in Fig. \ref{fig:rgt9} against the model-space sizes, which correspond to the maximum nucleon number in each model space.	
	
	As shown in Fig. \ref{fig:rgt9}, our RQE calculations indicate that linear $\Delta R_{\rm np}-I$ and $\Delta R_{\rm mirr}-I$ correlations with $|r|>0.95$ occur with a probability greater than 1\%. This probability suggests a notable predominance of these linear correlations within the RQE framework. This is particularly significant considering that the probability of achieving $|r|>0.95$ would typically be on the order of $10^{-5}$ if our shell-model calculations were to produce independent random numbers as outputs for $\Delta R_{\rm np}$ and $\Delta R_{\rm mirr}$, as demonstrated by our numerical tests.	
	
	We also observed that as the model space size increases, the probability $P(|r|>0.95)$ increases significantly for both $\Delta R_{\rm np}-I$ and $\Delta R_{\rm mirr}-I$ correlations. This suggests that the linearity of these correlations becomes more pronounced with increasing model space size. This trend appears to be more evident for the $\Delta R_{\rm mirr}-I$ correlation.

	\subsection{random Skyrme ensemble (RSE)}
		
	Due to the computational demands of shell-model calculations, the model space in RQE calculations is limited. To extend our investigation, we employ Skyrme-Hartree-Fock (SHF) calculations using the lowest 10 major shells of an isotropic harmonic oscillator. This significantly expands our model space to 572 single-particle levels, considerably larger than those employed in our RQE calculations. These SHF calculations were performed using the HFBTHO code \cite{hfbtho}, with zero-range pairing interactions disabled. This is because pairing interactions can generally weaken the linear correlation between between $\Delta R_{\rm mirr}$ and $L$\cite{LYF-rch}. We also used a prolately deformed initial basis with $\beta=0.2$, as most of deformed nuclei tend to exhibit prolate shapes.
	
	For these calculations, we collected 160 previously proposed Skyrme parametrizations, primarily from Ref. \cite{PhysRevC.85.035201}.  We then performed SHF calculations for all even-even nuclei shown in Fig. \ref{fig:exp}, as well as $^{22}$Si, $^{26}$S, $^{30}$Ar, $^{34}$Ca, $^{38}$Ti, $^{28}$S, and $^{50}$Ni. The latter nuclei were included to broaden the range of $I$ in our investigation of the linear correlation of $\Delta R_{\rm mirr}-I$, as in Ref. \cite{rch-I}. In this paper, we focus on even-even nuclei because odd-$A$ $\Delta R_{\rm mirr}$ values can deviate substantially from the general $\Delta R_{\rm mirr} -I$ linear trend, as seen in Fig. \ref{fig:exp}. Furthermore, many-body calculations for odd-$A$ nuclei are considerably more complex due to the unpaired nucleon. This set of calculations constitutes what we term the ``Skyrme ensemble".
	
	Within this Skyrme ensemble, we observed a clear linear correlation between $\Delta R_{\rm np}-I$ and $\Delta R_{\rm mirr}-I$ with Pearson's $r>0.85$ for {\it all} considered parametrizations. This is consistent with experimental observations depicted in Fig. \ref{fig:exp}. We also noticed that the probability of achieving $r>0.995$ for $\Delta R_{\rm np}-I$ linearity is 5\%, which is notably smaller than the 69\% observed for $\Delta R_{\rm mirr}-I$ linearity.  This suggests that, within the Skyrme ensemble, Pearson's $r$ values for the $\Delta R_{\rm np}-I$ correlation are generally lower than those for the $\Delta R_{\rm mirr}-I$ correlation. This is in line with our RQE findings in Fig. \ref{fig:rgt9}, suggesting that the linearity of the $\Delta R_{\rm mirr}-I$ correlation may become more prominent with larger model spaces.
	
	Such a high probability of obtaining $r$ values close to 1 might indicate a possible interaction-independent property of quantum many-body systems described by Skyrme forces. Thus, we proceeded to further randomize the Skyrme parametrizations. A typical Skyrme force is characterized by 10 parameters ($t_{0\sim 3}$, $x_{0\sim 3}$, $W_0$ and $1/\sigma$), as detailed in Eq. (2) of Ref. \cite{An2023}. For the 160 Skyrme parametrizations in our Skyrme ensemble, the mean values and covariance matrix of these 10 parameters are presented in Table \ref{tab:cov}. Following the method described in Ref. \cite{mul-gau-rand}, we employed linear transformations of independent Gaussian random numbers to generate random Skyrme parametrizations that statistically resemble the original set of 160 Skyrme parameters, maintaining similar means, fluctuations, and correlations, as listed in Table \ref{tab:cov}.

	\begin{table*}
		\caption{
		Mean values and covariance matrix of the 160 Skyrme parametrizations in the Skyrme ensemble.
		The notation of Skyrme parameters follow Eq. (2) of Ref. \cite{An2023}.
		}\label{tab:cov}
		\begin{tabular}{c|ccccccccccccccccccccccccc}
			\hline\hline
			  & $\langle t_0\rangle $ & $\langle t_1 \rangle$ & $\langle t_2\rangle$ & $\langle t_3\rangle$ & $\langle x_0\rangle$ & $\langle x_1\rangle$ & $\langle x_2\rangle$ & $\langle x_3\rangle$ & $\langle W_0\rangle$ & $\langle 1/\sigma\rangle$ \\
			 \hline
			 mean & $-$2.03$\times 10^{3}$ & 394 & $-$156 & 1.3$\times 10^{4}$ & 0.443 & $-$0.461 & 0.471 & 0.594 & 134 & 4.11 \\
			 \hline
			 covariance & $t_0$ & $t_1$ & $t_2$ & $t_3$ & $x_0$ & $x_1$ & $x_2$ & $x_3$ & $W_0$ & $1/\sigma$ \\
			 \hline
			 $t_0$ & 2.39$\times 10^{5}$ & $-$1.21$\times 10^{4}$ & 4.26$\times 10^{4}$ & $-$5.39$\times 10^{5}$ & $-$29.5 & 18.6 & 247 & $-$86.3 & $-$2.95$\times 10^{3}$ & $-$821 \\
			 $t_1$ & $-$1.21$\times 10^{4}$ & 1.32$\times 10^{4}$ & 276 & $-$1.52$\times 10^{5}$ & 0.6 & $-$12.3 & $-$33.1 & 14 & 1.21$\times 10^{3}$ & 65.2 \\
			 $t_2$ & 4.26$\times 10^{4}$ & 276 & 1.41$\times 10^{5}$ & $-$2.56$\times 10^{5}$ & $-$77.8 & $-$229 & 119 & $-$124 & 4.97$\times 10^{3}$ & $-$140 \\
			 $t_3$ & $-$5.39$\times 10^{5}$ & $-$1.52$\times 10^{5}$ & $-$2.56$\times 10^{5}$ & 8.35$\times 10^{6}$ & 157 & 373 & $-$1.22$\times 10^{3}$ & 49.8 & $-$1.54$\times 10^{4}$ & 1.49$\times 10^{3}$ \\
			 $x_0$ & $-$29.5 & 0.6 & $-$77.8 & 157 & 0.146 & 0.0574 & $-$0.114 & 0.241 & $-$0.399 & 0.134 \\
			 $x_1$ & 18.6 & $-$12.3 & $-$229 & 373 & 0.0574 & 0.688 & $-$0.046 & 5.04$\times 10^{-3}$ & $-$17.9 & $-$0.12 \\
			 $x_2$ & 247 & $-$33.1 & 119 & $-$1.22$\times 10^{3}$ & $-$0.114 & $-$0.046 & 46 & $-$0.176 & $-$7.23 & $-$1.01 \\
			 $x_3$ & $-$86.3 & 14 & $-$124 & 49.8 & 0.241 & 5.04$\times 10^{-3}$ & $-$0.176 & 0.484 & 4.04 & 0.375 \\
			 $W_0$ & $-$2.95$\times 10^{3}$ & 1.21$\times 10^{3}$ & 4.97$\times 10^{3}$ & $-$1.54$\times 10^{4}$ & $-$0.399 & $-$17.9 & $-$7.23 & 4.04 & 1.82$\times 10^{3}$ & 13.2 \\
			 $1/\sigma$ & $-$821 & 65.2 & $-$140 & 1.49$\times 10^{3}$ & 0.134 & $-$0.12 & $-$1.01 & 0.375 & 13.2 & 2.99 \\
			\hline\hline
		\end{tabular}
	\end{table*}
	
	Beside the consideration of statistic nature of random Skyrme parameters, we also need to make sure the existence of the nuclear matter governed by such randomized Skyrme force, which may be demonstrated by corresponding saturation density ($\rho_0$). Unlike 160 Skyrme parametrizations in the Skyrme ensemble 
	, which were well adjusted targeting realistic nuclear properties, the random Skyrme parametrization dose not necessarily provide $\rho_0$ close to 0.16 fm$^{-3}$, i.e., the realistic value. The density should be the root of saturation condition equation of
	
	Beyond the statistical nature of random Skyrme parameters, we also ensured the existence of nuclear matter for each randomized Skyrme force by checking for a corresponding saturation density ($\rho_0$). $\rho_0$ should be a root of the saturation condition equation:
	\begin{equation}\label{eq:sat}
		\begin{aligned}
			\frac{\hbar^2}{5 m} \left(\frac{3\pi^2}{2}\right)^{2/3} \rho_0^{-1 / 3}+\frac{3}{8} t_0+\frac{1}{16} t_3(\alpha+1) \rho_0^\alpha&\\
			+\frac{1}{16}\left[3 t_1+t_2\left(5+4 x_2\right)\right]  \left(\frac{3\pi^2}{2}\right)^{2/3}  \rho_0^{2 / 3}&=0.
		\end{aligned}
	\end{equation}
	Unlike the 160 Skyrme parametrizations in the Skyrme ensemble which were carefully fitted to reproduce realistic nuclear properties, the random Skyrme parametrizations might not necessarily yield a $\rho_0$ root close to the realistic value of 0.16 fm$^{-3}$.  In some extreme cases, there might not even be a real positive root, suggesting that nuclear matter might not exist under such a Skyrme force. We considered such parametrizations to be unphysical and therefore excluded them from our ensemble investigation.
	
	We emphasize that this work is aimed at constraining $L$, which is related to Skyrme parametrization by
	\begin{equation}\label{eq:L-def}
		\begin{aligned}
		L=&\frac{\hbar^2}{3 m} c \rho_0^{2 / 3}-\frac{3}{8} t_0\left(1+2 x_0\right) \rho_0\\
		&-\frac{1}{16} t_3\left(1+2 x_3\right)(\alpha+1) \rho_0^{\alpha+1}\\
		&-\frac{5}{24}\left[3 t_1 x_1-t_2\left(4+5 x_2\right)\right] c \rho_0^{5 / 3}.
		\end{aligned}
	\end{equation}
	To fairly assess the predominance of linear $\Delta R_{\rm np}-I$ and $\Delta R_{\rm mirr}-I$ correlations for different $L$ values, and to mitigate potential biases from intrinsic $L$ distribution within the random Skyrme calculations, we sampled approximately 4000 Skyrme parametrizations for each $\Delta L=5$ MeV interval across the range $L= 0\sim 200$ MeV. These parametrizations were then used in SHF calculations for the same set of even-even nuclei as in the RQE. Considering that prolate deformation is common in nuclei, we employed an initial single-particle basis with a prolate deformation of $\beta=0.2$, which likely favors a prolate HF solution. This set of random interaction calculations forms what we denote as the random Skyrme ensemble (RSE) in this paper. For each parametrization within the RSE, we calculated the Pearson's $r$ for both $\Delta R_{\rm np}-I$ and $\Delta R_{\rm mirr}-I$ correlations. We found that there is a 50.6(9)\% probability of achieving strong linearity ($r>0.95$) for $\Delta R_{\rm np}-I$ and a higher probability of 89(1)\% for $\Delta R_{\rm mirr}-I$. These notably high probabilities are even more significant than those from our RQE calculations. For comparison, we have also included these percentages in Fig. \ref{fig:rgt9}. The larger model space in the SHF calculations does indeed appear to strengthen the predominance of linear $\Delta R_{\rm np}-I$ and $\Delta R_{\rm mirr}-I$ correlations.
	
	For realistic nuclear systems, nucleons move within an effectively infinite Hilbert space. Based on our observations in Fig. \ref{fig:rgt9}, robust linear correlations of $\Delta R_{\rm np}-I$ and $\Delta R_{\rm mirr}-I$ are reasonably expected, consistent with experimental findings as shown in Fig. \ref{fig:exp}.
	
\section{$C_{\rm np}-L$ and $C_{\rm mirr}-L$ linearity}	\label{sec:c-L}	
	Having presented evidence for the robust linear correlations of $\Delta R_{\rm np}-I$ and $\Delta R_{\rm mirr}-I$, we also remind the reported linear relationship between $\Delta R_{\rm np}$ (or $\Delta R_{\rm mirr}$) and the symmetry energy slope $L$ \cite{CLW-np,PhysRevC.93.064303,WL-np,Fan2015,RZZ-np,LBA-np,brown-rch-2,PhysRevLett.127.182503,An2023}.  Given the transitive nature of linearity, one might expect $L$ to also exhibit a linear correlation with $I$ or with the slopes of the $\Delta R_{\rm np}-I$ and $\Delta R_{\rm mirr}-I$ correlations (denoted as $C_{\rm np}$ and $C_{\rm mirr}$, respectively).  Since $L$ is understood to be an observable reflecting a bulk property of nuclear matter, and thus  independent of specific nucleon numbers like $N$ and $Z$ (i.e., $I$), we anticipate observing linearity primarily in the $C_{\rm np}-L$ and $C_{\rm mirr}-L$ correlations.  Therefore, exploring these potential linear relationships seems worthwhile. If such linear correlations are indeed present, and given that $C_{\rm np}$ and $C_{\rm mirr}$ might be experimentally accessible, this could provide a potentially new avenue for constraining the value of $L$

	\subsection{verification}
	
	\begin{figure}[!htb]
		\includegraphics[angle=0,width=0.45\textwidth]{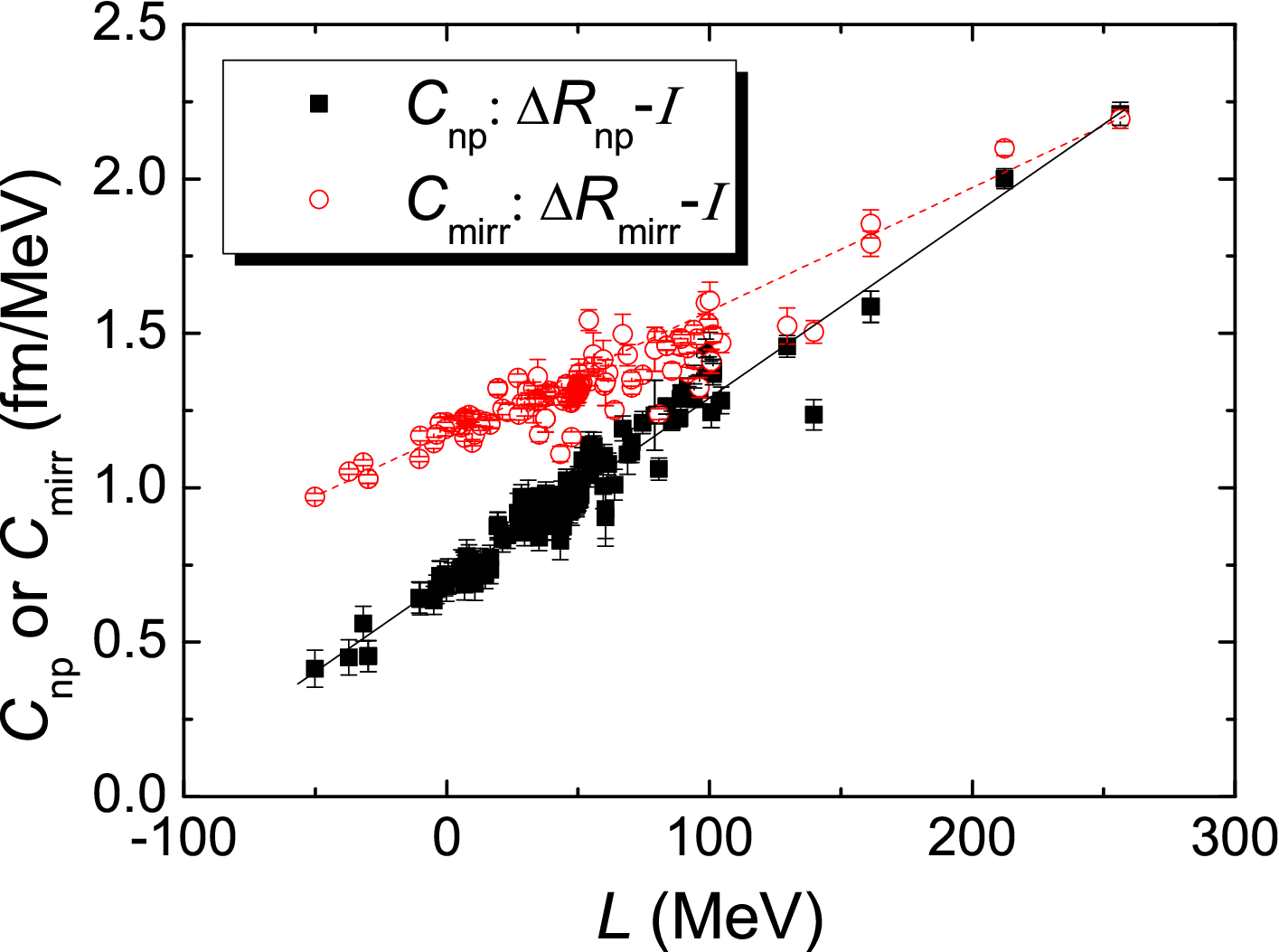}
		\caption{(Color online)
			$C_{\rm np}$ and $C_{\rm mirr}$ (slopes of the $\Delta R_{\rm np}-I$ and $\Delta R_{\rm mirr}-I$ linear correlations, respectively) as a function of $L$ in the Skyrme ensemble, which includes 160 previously proposed Skyrme parametrizations.  The approximate linear trends for both $C_{\rm np}-L$ and $C_{\rm mirr}-L$ correlations are visually apparent, highlighted by the black solid and red dashed lines as visual guides.			
		}\label{fig:L-c-161}
	\end{figure}
	
	We now investigate the linearity of the $C_{\rm np}-L$ and $C_{\rm mirr}-L$ correlations within the Skyrme ensemble, which comprises 160 existing Skyrme parametrizations. As demonstrated previously, all these Skyrme parametrizations yield Pearson's $r$ values greater than 0.85 for both $\Delta R_{\rm np}-I$ and $\Delta R_{\rm mirr}-I$ correlations, suggesting that deviations from linear fits for these correlations should be reasonably small.  Therefore, we proceed to perform linear fitting for the $\Delta R_{\rm np}-I$ correlation and proportional fitting for $\Delta R_{\rm mirr}-I$ correlation.  For the proportional fitting of $\Delta R_{\rm mirr}-I$, we omit the intercept, recognizing that $\Delta R_{\rm mirr}\equiv 0$ at $I=0$. Figure \ref{fig:L-c-161} then displays the plots of the extracted slopes, $C_{\rm np}$ and $C_{\rm mirr}$, as a function of $L$ for each parametrization in the Skyrme ensemble. These plots indicate a general linear trend for both $C_{\rm np}-L$ and $C_{\rm mirr}-L$ correlations.  It is also worth noting that for lower $L$ values ($<100$ MeV), $C_{\rm np}$ tends to be systematically smaller than $C_{\rm mirr}$, while the difference between $C_{\rm np}$ and $C_{\rm mirr}$ gradually diminishes as $L$ increases.

	\begin{figure}[!htb]
		\includegraphics[angle=0,width=0.45\textwidth]{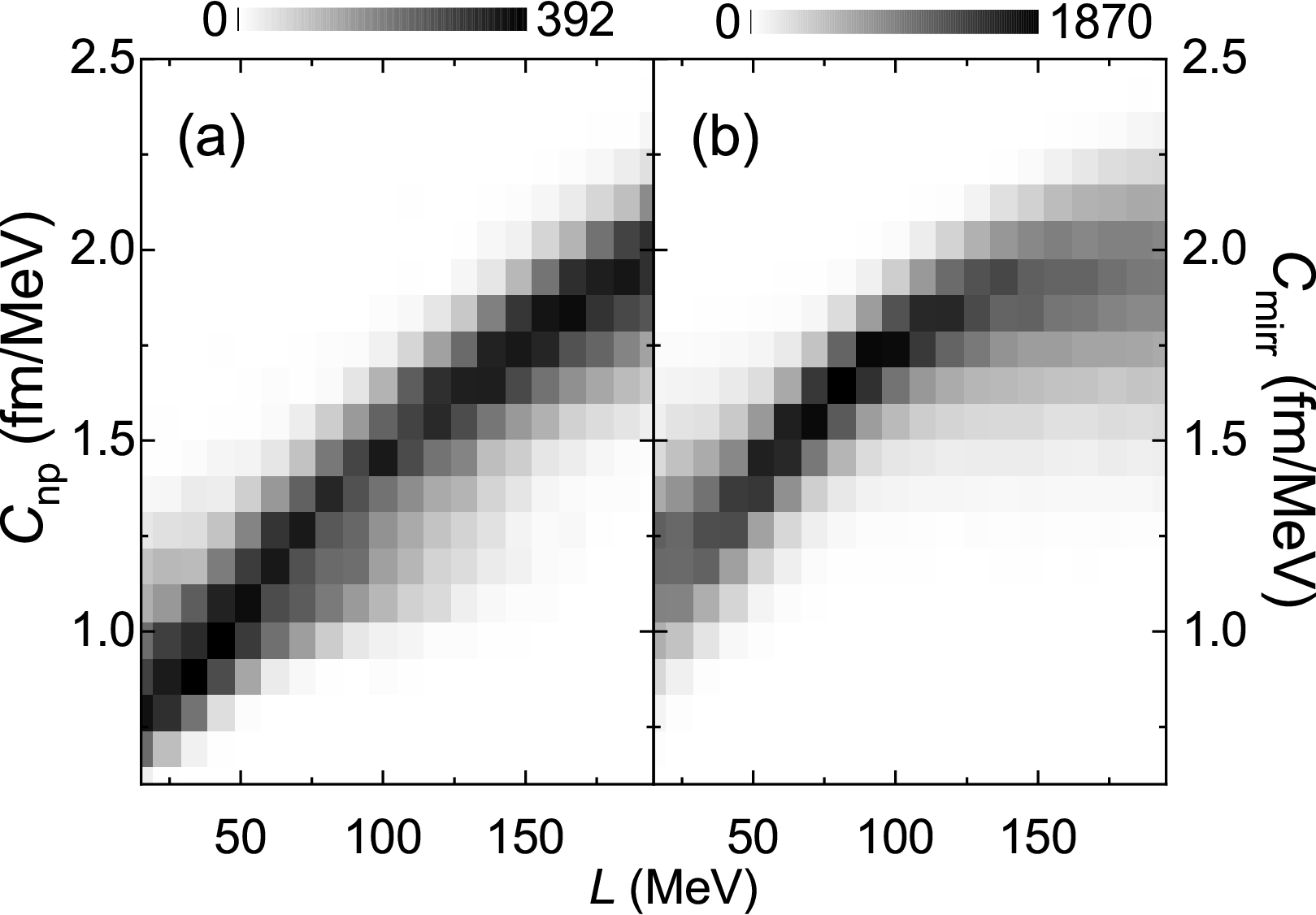}
		\caption{						
			Grey scale maps illustrating the two-dimensional counting of $\{C_{\rm np},L\}$ and $\{C_{\rm mirr},L\}$ data pairs from the RSE. Only $C_{\rm np}$ and $C_{\rm mirr}$ values with corresponding Pearson's $r$ greater than 0.85 are included. Panel (a) shows a clear linear correlation between $C_{\rm np}$ and $L$, consistent with Fig. \ref{fig:L-c-161}. Panel (b) presents the correlation between $C_{\rm mirr}$ and $L$. Unlike the $C_{\rm np}-L$ correlation, the linear trend in the $C_{\rm mirr}-L$ correlation appears to weaken at larger $L$ values.
		}\label{fig:L-c-rse}
	\end{figure}

	We also observe the linear correlation between $C_{\rm np}$ and $L$, and between $C_{\rm mirr}$ and $L$ in the RSE. For Skyrme parametrizations in the RSE that exhibit Pearson's $r$ values above 0.85, we perform two-dimensional counting of $\{C_{\rm np},L\}$ and $\{C_{\rm mirr},L\}$ data pairs. The results are shown in Fig. \ref{fig:L-c-rse} using gray scale mapping. As shown in Fig. \ref{fig:L-c-rse}(a), a linear correlation between $C_{\rm np}$ and $L$ is apparent. However, as depicted in Fig. \ref{fig:L-c-rse}(b), the linear correlation of $C_{\rm mirr}-L$ seems to be maintained primarily for $L<150$ MeV. Beyond $L=150$ MeV, the slope of this correlation appears to decrease.  Consistent with observations from the Skyrme ensemble using well-established Skyrme parametrizations, for lower $L$ values ($<100$ MeV), $C_{\rm np}$ tends to be systematically smaller than $C_{\rm mirr}$, whereas at higher $L$ values, they become quantitatively comparable.

	\subsection{explanation}
	
	\begin{figure}[!htb]
		\includegraphics[angle=0,width=0.45\textwidth]{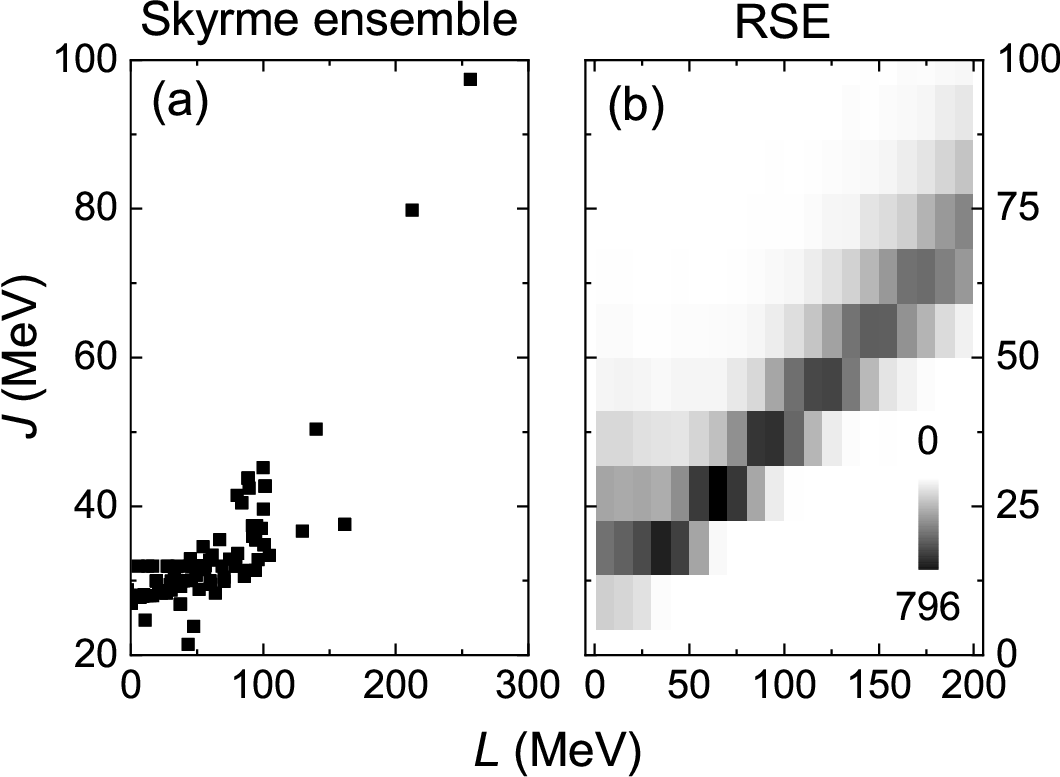}
		\caption{
			 $J$ against $L$ in the Skyrme ensemble and RSE ensemble, as shown in Panel (a) and (b) respectively. Panel (b) shows a two-dimensional counting for $\{J,L\}$ combinations.
		}\label{fig:J-L}
	\end{figure}

	To understand the possible origin of the robust linear correlations observed for $C_{\rm np}-L$ and $C_{\rm mirr}-L$, we focus on $C_{\rm np}-L$, assuming that the robust linearity of $C_{\rm mirr}-L$ reflects a similar underlying mechanism.  Following the formalism from the supplement of Ref. \cite{rch-I}, we have:
	\begin{equation}\label{eq:c-L}
		C_{\rm np}=\frac{3}{2}r_0\frac{J}{Q^*},
	\end{equation}
	where $J$ is the symmetry energy coefficient, $r_0$ is the oscillation parameter, and $Q^*$ represents the effective surface stiffness coefficient, which describes the resistance to the pulling apart of neutrons and protons at the nuclear surface.  Both $J$ and $L$ are functions of the Skyrme parametrization.  Furthermore, it's worth noting that $Q^*$ is related to specific $N$ and $Z$ values within many-body calculations (as in Ref. \cite{KOHLER1976301}). Given that the linear correlation of $\Delta R_{\rm np}-L$ is observed to be robust across the nuclear chart, one might expect $C_{\rm np}$ to be relatively stable with respect to variations in $N$ and $Z$. This would further suggest that the $N$ and $Z$ dependence of $Q^*$ may not significantly impact $C_{\rm np}$. Therefore, for simplicity and following the approach of Ref. \cite{rch-I}, we can consider $Q^*$ to be approximately constant compared to the variation of $J$. Consequently, if $C_{\rm np}$ and $L$ are linearly correlated, Eq. (\ref{eq:c-L}) implies a corresponding linear correlation between $J$ and $L$ within the $L$ range where we observe the $C_{\rm np}-L$ linearity, as shown in Figs. \ref{fig:L-c-161} and \ref{fig:L-c-rse}(a). To examine the linearity of the $J-L$ correlation, we plot $J$, calculated using
	\begin{equation}\label{eq:j}
		\begin{aligned}
			J=&\frac{\hbar^2}{6 m} c \rho^{2 / 3}-\frac{1}{8} t_0\left(1+2 x_0\right) \rho\\
			&-\frac{1}{48} t_3\left(1+2 x_3\right) \rho^{\alpha+1}\\
			&-\frac{1}{24}\left[3 t_1 x_1-t_2\left(4+5 x_2\right)\right] c \rho^{5 / 3}
		\end{aligned}
	\end{equation}
	against $L$, calculated using Eq. (\ref{eq:L-def}), for each of the 160 parametrizations in the Skyrme ensemble in Fig. \ref{fig:J-L}(a).  While some fluctuation in the $J-L$ correlation is present for $L<50$ MeV, a roughly linear trend appears to emerge in the range $L=50\sim 300$ MeV. In the RSE, we repeated similar calculations and present the $\{J,L\}$ pairs as a two-dimensional histogram in Fig. \ref{fig:J-L}(b). The linearity of $J-L$ becomes more evident, which may offer a possible explanation for the robust linear $C_{\rm np}-L$ correlation.

	We now consider why $L$ and $J$ exhibit a linear correlation. Let us rewrite $J$ from Eq. (\ref{eq:j}) as a sum of four terms:
	\begin{equation}\label{eq:j-decoupled}
		\begin{aligned}
			J=&J_1+J_2+J_3+J_4\\
			J_1=&\frac{\hbar^2}{6 m} c \rho^{2 / 3}\\
			J_2=&-\frac{1}{8} t_0\left(1+2 x_0\right) \rho\\
			J_3=&-\frac{1}{48} t_3\left(1+2 x_3\right) \rho^{\alpha+1}\\
			J_4=&-\frac{1}{24}\left[3 t_1 x_1-t_2\left(4+5 x_2\right)\right] c \rho^{5 / 3}.
		\end{aligned}
	\end{equation}
	Correspondingly, we can also express $L$ from Eq. (\ref{eq:L-def}) in terms of $J_1$, $J_2$, $J_3$, and $J_4$ as:
	\begin{equation}\label{eq:l}
		L=2J_1+3J_2+4(1+\sigma)J_3+5J_4.
	\end{equation}
	We observe that both $L$ and $J$ are linear combinations of $J_1$, $J_2$, $J_3$, and $J_4$.  In a numerical test, assuming $J_1$, $J_2$, $J_3$, $J_4$, and $\sigma$ are independent random numbers following a normal distribution, we found that the probability of obtaining a Pearson's $r>0.9$ for ten $\{J,L\}$ data pairs, calculated using Eqs. (\ref{eq:j-decoupled}) and (\ref{eq:l}), is approximately 49(2)\%. In contrast, if ten $\{J,L\}$ data pairs are independently sampled from normal distributions, this probability is only around 1.90(4)\%. The similarity in the linear combination structure for $J$ and $L$ as shown in Eqs. (\ref{eq:j-decoupled}) and (\ref{eq:l}) likely enhances the linear correlation between them, and consequently, the linearity between $L$ and $C_{\rm np}$. This formulation similarity may offer a partial explanation for the observed robust linear correlations in Figs. \ref{fig:L-c-161} and \ref{fig:L-c-rse}.

	\section{$L$ constraint}\label{sec:L-con}
	\subsection{$C_{\rm np}$ and $C_{\rm mirr}$ constraint}\label{sec:c-con}
	\begin{figure}[!htb]
		\includegraphics[angle=0,width=0.45\textwidth]{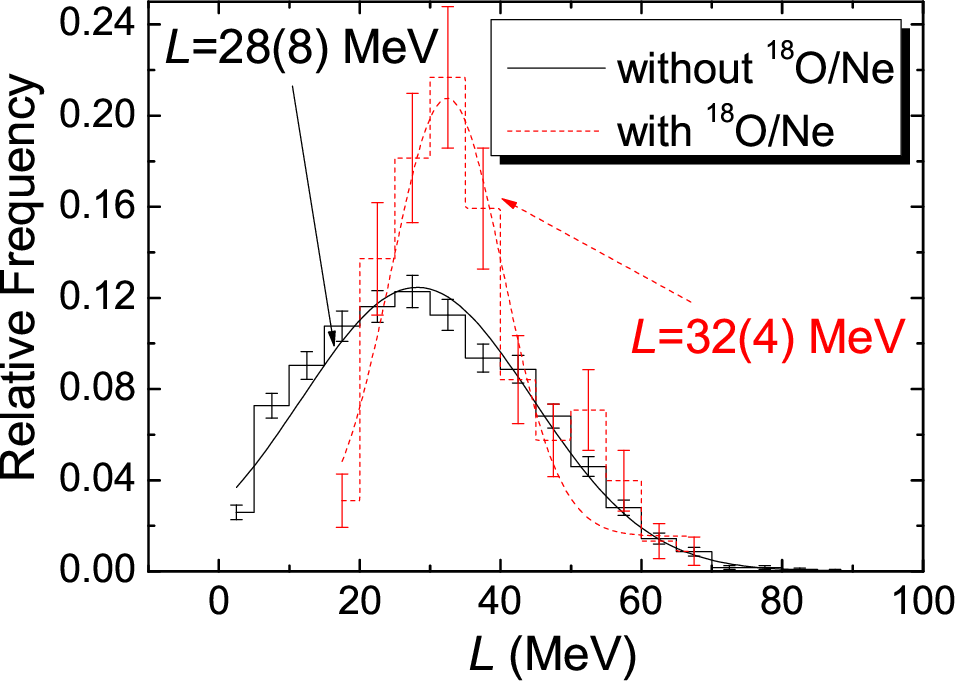}
		\caption{(Color online)			
			Distributions of the symmetry energy slope parameter $L$ for two subsets of Skyrme parametrizations within the RSE. The black solid histograms correspond to parametrizations that reproduce the linearity of $\Delta R_{\rm np}-I$ and $\Delta R_{\rm mirr}-I$ with Pearson's $r > 0.85$ and $0.99$, respectively, and yield corresponding $C_{\rm np}$ and $C_{\rm mirr}$ values within experimental uncertainties. The red dashed histograms correspond to those parametrizations that additionally reproduce $\Delta R_{\rm mirr}$ for the $^{18}$O/Ne mirror pair within experimental uncertainties, considering the potential shape coexistence of $^{18}$O. Step histograms are from sample counts, with error bars representing statistical uncertainties. The curves represent Gaussian fits, providing constraints on $L=28\pm 8$ MeV and $32\pm 4$ MeV with and without including the $\Delta R_{\rm mirr}$ data for the $^{18}$O/Ne mirror pair, respectively.
		}\label{fig:Lcon}
	\end{figure}
	
	We further emphasize that the linear correlations of $C_{\rm np}-L$ and $C_{\rm mirr}-L$ are not only novel, but also potentially valuable for constraining $L$. First, experimental data in Fig. \ref{fig:exp} indicates that $C_{\rm np}$ is smaller than $C_{\rm mirr}$. This suggests a possible upper limit for $L$ around 100 MeV, as they appear to become comparable at $L=100$ MeV, as shown in Fig. \ref{fig:L-c-161}. This observation may be crucial, especially considering that $L \approx 100$ MeV is often discussed as a boundary between soft and stiff EOS \cite{Pb-L,PhysRevLett.127.182503}. Second, the robust linearity of the $C_{\rm np}-L$ and $C_{\rm mirr}-L$ correlations, along with their considerable slopes, makes $C_{\rm np}$ and $C_{\rm mirr}$ sensitive indicators of $L$, potentially enabling constraints with smaller uncertainty. Third, $C_{\rm np}$ and $C_{\rm mirr}$ are comprehensive observables, since they incorporate information from all experimental $R_{\rm np}$ and $R_{\rm mirr}$ data. Consequently, $L$ constraints from these observables could largely mitigate the interference from density anomalies in specific nuclei with exotic structures.
	
	More precisely, we sample the Skyrme parametrizations in the RSE, which could reproduce the linearity of $\Delta R_{\rm np}-I$ and $\Delta R_{\rm mirr}-I$ with Pearson's $r>0.85$ and $0.99$ respectively, and corresponding $C_{\rm np}$ and $C_{\rm mirr}$ within experimental error. Approximately 2000 parametrizations met these sampling criteria.	The $L$ distribution of these sampled parametrization demonstrate the reasonable range of $L$. In Fig. \ref{fig:Lcon}, the black solid step line presents such a distribution. The distribution brings an peak, which can be fitted by an Gaussian function. The fitting curve is presented in Fig. \ref{fig:Lcon} with black solid line with peak center at $28.2$ and width $16.4$. Thus, according to the experimental slope of $\Delta R_{\rm np}-I$ and $\Delta R_{\rm mirr}-I$, $L=28^{+8}_{-8}$ MeV with 1$\sigma$ confidence.
	
	To perform this constraint, we sampled Skyrme parametrizations from the RSE that could reproduce the linearity of $\Delta R_{\rm np}-I$ and $\Delta R_{\rm mirr}-I$ (Pearson's $r>0.85$ and $0.99$, respectively), and yield $C_{\rm np}$ and $C_{\rm mirr}$ values consistent with experimental $C_{\rm np}=0.9(1)$ fm/MeV and $C_{\rm mirr}=1.31(4)$ fm/MeV from Fig. \ref{fig:exp}. The $L$ distribution of these sampled parametrizations demonstrates a plausible range for $L$. In Fig. \ref{fig:Lcon}, the black solid step histogram represents this distribution. The distribution exhibits a peak, which is well described by a Gaussian function. The Gaussian fit is shown as a black solid curve in Fig. \ref{fig:Lcon}, with a peak center at $28$ MeV and a width of $16$ MeV. Therefore, based on the slopes of experimental $\Delta R_{\rm np}-I$ and $\Delta R_{\rm mirr}-I$ correlation, we estimate $L=28 \pm 8$ MeV at the 1$\sigma$ confidence level.
	
	\subsection{shape effect on $\Delta R_{\rm mirr}$ deviations}\label{sec:shape}
	
	Given the robust linearity of the $\Delta R_{\rm mirr}-I$ correlation, and the clear deviation of experimental $\Delta R_{\rm mirr}$ for $^{18}$O/Ne from this linear trend, we expect that Skyrme parametrizations reproducing the experimental linear $\Delta R_{\rm mirr}-I$ behavior would likely fail to reproduce the experimental $\Delta R_{\rm np}$ of $^{18}$O/Ne. To verify this, we make an attempt to select Skyrme parametrizations that could simultaneously reproduce the linear $\Delta R_{\rm np}-I$ and $\Delta R_{\rm mirr}-I$ correlations with experimental slopes, and the experimental $\Delta R_{\rm np}$ of $^{18}$O/Ne within a 1$\sigma$ confidence interval. Among approximately 150,000 parametrizations in the RSE, we found none that met these selection criteria. This suggests a significant discrepancy between $^{18}$O/Ne and other even-even nuclei regarding the $\Delta R_{\rm mirr}-I$ linearity.  More importantly, this discrepancy appears to prevent us from finding suitable Skyrme parametrizations, and consequently reasonable constraints on $L$, capable of reproducing all experimental $\Delta R_{\rm np}$ and $\Delta R_{\rm mirr}$ data for even-even nuclei. This outcome is not satisfactory.
	
	To address this apparent discrepancy and to obtain an $L$ constraint that is consistent with the $^{18}$O/Ne $\Delta R_{\rm mirr}$ data, it seems necessary to explain the deviation of the $^{18}$O/Ne mirror pair observed in Fig. \ref{fig:exp}. We recall that $^{18}$O and $^{18}$Ne are both candidates for shape coexistence \cite{coexistence_rmp}.  Recent studies also indicate that shape coexistence might be a widespread phenomenon across the nuclide chart \cite{PhysRevC.110.054318}. Different nuclear shapes could naturally lead to different charge radii. Thus, it is conceivable that shape coexistence in the $^{18}$O/Ne pair results in multiple $\Delta R_{\rm mirr}$ values. One of these values might follow the general linear trend of $\Delta R_{\rm mirr}-I$, potentially originating from one or two local minima in the Hartree-Fock potential energy surface of $^{18}$O/Ne. Another $\Delta R_{\rm mirr}$ value might then agree with the experimental $\Delta R_{\rm mirr}$ of $^{18}$O/Ne, corresponding to Hartree-Fock ground states. However, as mentioned earlier, our initial SHF calculations started from a basis with $\beta=0.2$, which may primarily converge to a single prolate deformation and thus potentially hinder the consistent reproduction of both the linear $\Delta R_{\rm mirr}-I$ correlation and the specific $\Delta R_{\rm mirr}$ of $^{18}$O/Ne.
	
	\begin{figure}[!htb]
		\includegraphics[angle=0,width=0.45\textwidth]{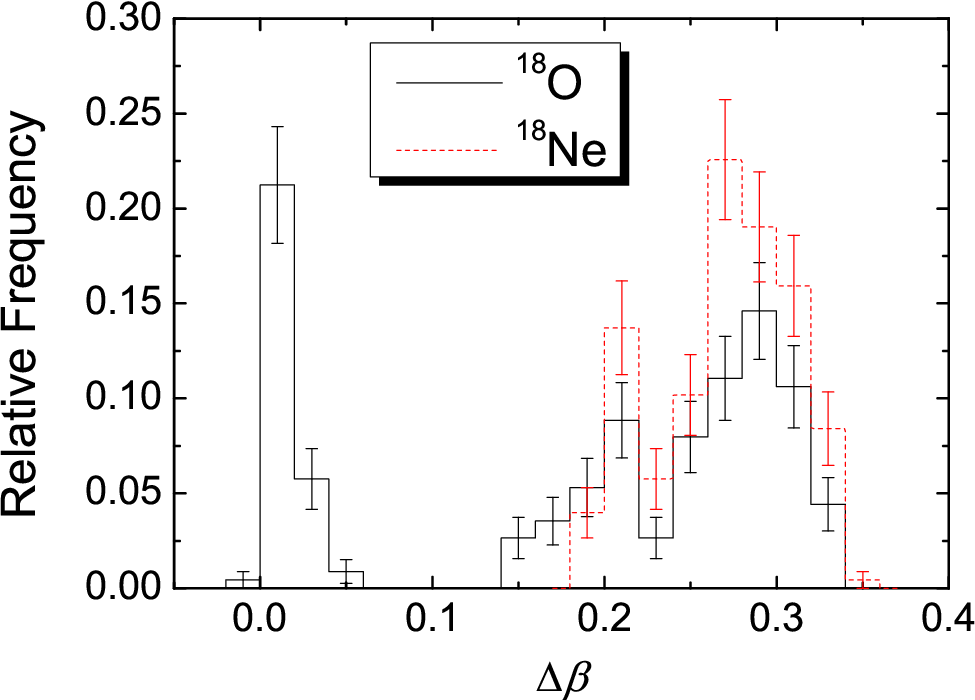}
		\caption{(Color online)
			Distributions of $\beta$ differences ($\Delta\beta$) for $^{18}$O and $^{18}$Ne between two shapes, obtained from RSE parametrizations that reproduce both the experimental $\Delta R_{\rm mirr}-I$ linearity and the experimental $\Delta R_{\rm mirr}$ of $^{18}$O/Ne (within 1$\sigma$ confidence). Non-zero $\Delta\beta$ values are indicative of shape coexistence.
		}\label{fig:dbeta}
	\end{figure}

	To further explore this shape coexistence picture, we repeated the SHF calculations for $^{18}$O/Ne with the $\sim$2000 parametrizations sampled from the RSE, which could reproduce the experimental linear correlations of $\Delta R_{\rm np}-I$ and $\Delta R_{\rm mirr}-I$, as described in Sec. \ref{sec:c-con}. To allow for the possibility of shape coexistence, these calculations started from a single-particle basis with $\beta=-0.2\sim 0.2$ with 0.01 interval, where negative $\beta$ values indicate oblate shapes. As expected, for each parametrization, the SHF calculations may yield several $\Delta R_{\rm mirr}$ values for $^{18}$O/Ne. Given the robust $\Delta R_{\rm mirr}-I$ linearity, we anticipated that one of these values would align with the experimental $\Delta R_{\rm mirr}-I$ trend. Notably, among these $\sim$2000 parametrizations, approximately 200 also produced a $\Delta R_{\rm mirr}$ value consistent with experimental data for $^{18}$O/Ne. Thus, these $\sim$200 parametrizations yielded at least two $\Delta R_{\rm mirr}$ values, presumably originating from different shapes of $^{18}$O and/or $^{18}$Ne. To illustrate the shape difference, we calculated the $\beta$ differences ($\Delta\beta$) of $^{18}$O and $^{18}$Ne between two shapes, one that produces $\Delta R_{\rm mirr}$ following the experimental $\Delta R{\rm mirr}-I$ linearity, and the other that reproduces the experimental $\Delta R_{\rm mirr}$ of $^{18}$O/Ne, within 1$\sigma$ confidence. The distribution of these $\Delta\beta$ values from these $\sim$200 parametrizations is shown in Fig. \ref{fig:dbeta}. For $^{18}$O, we observe three peaks in the distribution, around $\Delta \beta\sim 0$, 0.2, and 0.3. These peaks may correspond to scenarios of no shape coexistence, spherical-deformed coexistence (between $\beta\simeq 0$ and $\pm 0.2$), and prolate-oblate coexistence (between $\beta\simeq 0.2$ and $-0.1$, or $\beta\simeq 0.1$ and $-0.2$). For $^{18}$Ne, two peaks are located at $\Delta \beta\sim 0.2$ and 0.3, similarly hinting at spherical-deformed and prolate-oblate coexistence. Therefore, to reasonably reproduce the linear $\Delta R_{\rm np}-I$ and $\Delta R_{\rm mirr}-I$ correlations, along with the $\Delta R_{\rm mirr}$ of $^{18}$O/Ne using a single Skyrme parametrization, incorporating shape coexistence, at least for $^{18}$Ne, appears to be necessary.  Shape coexistence in $^{18}$O may also be occasionally required.
		
	Finally, using the $\sim$200 Skyrme parametrizations sampled above, considering the experimental $\Delta R_{\rm mirr}$ fit for $^{18}$O/Ne, we can also estimate another constraint on $L$.  The distribution of $L$ values for these parametrizations is plotted in Fig. \ref{fig:Lcon}. A Gaussian fit to this distribution suggests a value of $L=32(4)$ MeV within the 1$\sigma$ confidence.
	
	Given the considerable influence of nuclear shape on the $\Delta R_{\rm mirr}-I$ linearity, we propose that it may also contribute to the less systematic behavior observed for odd-$A$ $\Delta R_{\rm mirr}$ values. For doubly-even mirror pair, $\Delta R_{\rm mirr}$ corresponds to a charge-radius difference between two even-like-nucleon systems, while for odd-$A$ pair, it corresponds to difference between odd-like-nucleon system and even-like-nucleon system. The shapes of even-like-nucleon systems are typically governed by nuclear collectivity, and are generally stable within a local isotopic region. However, the deformation of odd-like-nucleon systems can be significantly influenced by the unpaired single nucleon. The specific orbit occupied by this unpaired nucleon can introduce discontinuities in nuclear structure, potentially leading to discontinuous shape evolution. This distinction between even and odd systems is also reflected in phenomena such as the shape staggering observed in charge radii of Hg and Cu isotopes \cite{Marsh2018,deGroote2020}. Based on this reasoning, $\Delta R_{\rm mirr}$ for even-even nuclei primarily reflects the difference in collectivity between two even-like-nucleon systems, which may follow a continuous local trend, and captures the general regularity of finite nuclear matter, i.e., the $\Delta R_{\rm mirr}-I$ linearity. In contrast, for odd-$A$ nuclei, the discontinuities introduced by the unpaired nucleon can lead to a more scattered, or less systematically organized, $\Delta R_{\rm mirr}$ behavior. Thus, odd-$A$ $\Delta R_{\rm mirr}$ values may exhibit larger deviations from the $\Delta R_{\rm mirr}-I$ linearity compared to even-even nuclei.

	\section{summary}\label{sec:sum}
	
	In summary, we have observed various linear correlations related to neutron skin thickness ($\Delta R_{\rm np}$) in random interaction ensembles. These include random quasi-particle ensemble (RQE) with shell model calculation, and the newly proposed random Skyrme ensemble (RSE) with Hatree-Fock calculations, as well as the Skyrme ensemble based on 160 previously proposed and physically-tuned Skyrme parametrizations. The RSE is designed to align with the statistical properties (means and covariance matrix) of the Skyrme ensemble.
	
	The linear correlation between $\Delta R_{\rm np}$, or charge radius difference of mirror nuclei ($\Delta R_{\rm mirr}$), and the isospin asymmetry ($I=\frac{N-Z}{A}$) become more obvious in the random interaction, as increasing model space. In the Skyrme ensemble, this linear correlation is reproduced by all Skyrme parametrizations. The robustness of this linear correlation suggests that it reflects a nature of finite nuclear matter. 
	
	In Sec. \ref{sec:shape}, we have further explain the obvious deviation of ${}^{18}$O/Ne $\Delta R_{mirr}$ from the $\Delta R_{mirr}-I$ linearity, considering that ${}^{18}$O and ${}^{18}$Ne are both candidates for shape coexistence. Following a similar line of reasoning, we also attribute the less systematic behavior of $\Delta R_{\rm mirr}$ in odd-$A$ mirror nuclei to the shape variations arising from the single-particle motion of the unpaired nucleon.

	Within the Skyrme ensemble, and the RSE, the slopes of the linear $\Delta R_{\rm np}-I$ and $\Delta R_{\rm mirr}-I$ correlations ($C_{\rm np}$ and $C_{\rm mirr}$, respectively) are also robustly and linearly correlated to the slope of the symmetry energy ($L$) against nucleon density at equilibrium density in the nuclear equation of state. This correlation is further explained with the similar formulation between between $L$ and the symmetry energy coefficient ($J$), as described by Eqs. (\ref{eq:j-decoupled}) and (\ref{eq:l}). 
	
	The linear $C_{\rm np}-L$ and $C_{\rm mirr}-L$ correlations have been used to constrain $L$ to $20\sim36$ MeV, based on the sampling in the RSE. Considering the deviation of ${}^{18}$O/Ne $\Delta R_{mirr}$, the 1$\sigma$ range of $L$ is further reduced to $28\sim36$ MeV, tentatively suggesting a relatively soft EOS.

	

\begin{thebibliography}{84}%
		\makeatletter
		\providecommand \@ifxundefined [1]{%
			\@ifx{#1\undefined}
		}%
		\providecommand \@ifnum [1]{%
			\ifnum #1\expandafter \@firstoftwo
			\else \expandafter \@secondoftwo
			\fi
		}%
		\providecommand \@ifx [1]{%
			\ifx #1\expandafter \@firstoftwo
			\else \expandafter \@secondoftwo
			\fi
		}%
		\providecommand \natexlab [1]{#1}%
		\providecommand \enquote  [1]{``#1''}%
		\providecommand \bibnamefont  [1]{#1}%
		\providecommand \bibfnamefont [1]{#1}%
		\providecommand \citenamefont [1]{#1}%
		\providecommand \href@noop [0]{\@secondoftwo}%
		\providecommand \href [0]{\begingroup \@sanitize@url \@href}%
		\providecommand \@href[1]{\@@startlink{#1}\@@href}%
		\providecommand \@@href[1]{\endgroup#1\@@endlink}%
		\providecommand \@sanitize@url [0]{\catcode `\\12\catcode `\$12\catcode
			`\&12\catcode `\#12\catcode `\^12\catcode `\_12\catcode `\%12\relax}%
		\providecommand \@@startlink[1]{}%
		\providecommand \@@endlink[0]{}%
		\providecommand \url  [0]{\begingroup\@sanitize@url \@url }%
		\providecommand \@url [1]{\endgroup\@href {#1}{\urlprefix }}%
		\providecommand \urlprefix  [0]{URL }%
		\providecommand \Eprint [0]{\href }%
		\providecommand \doibase [0]{https://doi.org/}%
		\providecommand \selectlanguage [0]{\@gobble}%
		\providecommand \bibinfo  [0]{\@secondoftwo}%
		\providecommand \bibfield  [0]{\@secondoftwo}%
		\providecommand \translation [1]{[#1]}%
		\providecommand \BibitemOpen [0]{}%
		\providecommand \bibitemStop [0]{}%
		\providecommand \bibitemNoStop [0]{.\EOS\space}%
		\providecommand \EOS [0]{\spacefactor3000\relax}%
		\providecommand \BibitemShut  [1]{\csname bibitem#1\endcsname}%
		\let\auto@bib@innerbib\@empty
		\bibitem [{\citenamefont {Lattimer}\ and\ \citenamefont {Prakash}(2007)}]{ns}%
		\BibitemOpen
		\bibfield  {author} {\bibinfo {author} {\bibfnamefont {J.~M.}\ \bibnamefont
				{Lattimer}}\ and\ \bibinfo {author} {\bibfnamefont {M.}~\bibnamefont
				{Prakash}},\ }\bibfield  {title} {\bibinfo {title} {Neutron star
				observations: Prognosis for equation of state constraints},\ }\href
		{https://doi.org/https://doi.org/10.1016/j.physrep.2007.02.003} {\bibfield
			{journal} {\bibinfo  {journal} {Physics Reports}\ }\textbf {\bibinfo {volume}
				{442}},\ \bibinfo {pages} {109} (\bibinfo {year} {2007})},\ \bibinfo {note}
		{the Hans Bethe Centennial Volume 1906-2006}\BibitemShut {NoStop}%
		\bibitem [{\citenamefont {Oertel}\ \emph
			{et~al.}(2017{\natexlab{a}})\citenamefont {Oertel}, \citenamefont {Hempel},
			\citenamefont {Kl\"ahn},\ and\ \citenamefont {Typel}}]{eos-super}%
		\BibitemOpen
		\bibfield  {author} {\bibinfo {author} {\bibfnamefont {M.}~\bibnamefont
				{Oertel}}, \bibinfo {author} {\bibfnamefont {M.}~\bibnamefont {Hempel}},
			\bibinfo {author} {\bibfnamefont {T.}~\bibnamefont {Kl\"ahn}},\ and\ \bibinfo
			{author} {\bibfnamefont {S.}~\bibnamefont {Typel}},\ }\bibfield  {title}
		{\bibinfo {title} {Equations of state for supernovae and compact stars},\
		}\href {https://doi.org/10.1103/RevModPhys.89.015007} {\bibfield  {journal}
			{\bibinfo  {journal} {Rev. Mod. Phys.}\ }\textbf {\bibinfo {volume} {89}},\
			\bibinfo {pages} {015007} (\bibinfo {year} {2017}{\natexlab{a}})}\BibitemShut
		{NoStop}%
		\bibitem [{\citenamefont {Alex~Brown}(2000{\natexlab{a}})}]{sym-stru-1}%
		\BibitemOpen
		\bibfield  {author} {\bibinfo {author} {\bibfnamefont {B.}~\bibnamefont
				{Alex~Brown}},\ }\bibfield  {title} {\bibinfo {title} {Neutron radii in
				nuclei and the neutron equation of state},\ }\href
		{https://doi.org/10.1103/PhysRevLett.85.5296} {\bibfield  {journal} {\bibinfo
				{journal} {Phys. Rev. Lett.}\ }\textbf {\bibinfo {volume} {85}},\ \bibinfo
			{pages} {5296} (\bibinfo {year} {2000}{\natexlab{a}})}\BibitemShut {NoStop}%
		\bibitem [{\citenamefont {Typel}\ and\ \citenamefont
			{Brown}(2001)}]{sym-stru-2}%
		\BibitemOpen
		\bibfield  {author} {\bibinfo {author} {\bibfnamefont {S.}~\bibnamefont
				{Typel}}\ and\ \bibinfo {author} {\bibfnamefont {B.~A.}\ \bibnamefont
				{Brown}},\ }\bibfield  {title} {\bibinfo {title} {Neutron radii and the
				neutron equation of state in relativistic models},\ }\href
		{https://doi.org/10.1103/PhysRevC.64.027302} {\bibfield  {journal} {\bibinfo
				{journal} {Phys. Rev. C}\ }\textbf {\bibinfo {volume} {64}},\ \bibinfo
			{pages} {027302} (\bibinfo {year} {2001})}\BibitemShut {NoStop}%
		\bibitem [{\citenamefont {Furnstahl}(2002)}]{sym-stru-3}%
		\BibitemOpen
		\bibfield  {author} {\bibinfo {author} {\bibfnamefont {R.}~\bibnamefont
				{Furnstahl}},\ }\bibfield  {title} {\bibinfo {title} {Neutron radii in
				mean-field models},\ }\href
		{https://doi.org/https://doi.org/10.1016/S0375-9474(02)00867-9} {\bibfield
			{journal} {\bibinfo  {journal} {Nuclear Physics A}\ }\textbf {\bibinfo
				{volume} {706}},\ \bibinfo {pages} {85} (\bibinfo {year} {2002})}\BibitemShut
		{NoStop}%
		\bibitem [{\citenamefont {Steiner}\ \emph
			{et~al.}(2005{\natexlab{a}})\citenamefont {Steiner}, \citenamefont {Prakash},
			\citenamefont {Lattimer},\ and\ \citenamefont {Ellis}}]{sym-stru-4}%
		\BibitemOpen
		\bibfield  {author} {\bibinfo {author} {\bibfnamefont {A.}~\bibnamefont
				{Steiner}}, \bibinfo {author} {\bibfnamefont {M.}~\bibnamefont {Prakash}},
			\bibinfo {author} {\bibfnamefont {J.}~\bibnamefont {Lattimer}},\ and\
			\bibinfo {author} {\bibfnamefont {P.}~\bibnamefont {Ellis}},\ }\bibfield
		{title} {\bibinfo {title} {Isospin asymmetry in nuclei and neutron stars},\
		}\href {https://doi.org/https://doi.org/10.1016/j.physrep.2005.02.004}
		{\bibfield  {journal} {\bibinfo  {journal} {Physics Reports}\ }\textbf
			{\bibinfo {volume} {411}},\ \bibinfo {pages} {325} (\bibinfo {year}
			{2005}{\natexlab{a}})}\BibitemShut {NoStop}%
		\bibitem [{\citenamefont {Todd-Rutel}\ and\ \citenamefont
			{Piekarewicz}(2005)}]{sym-stru-5}%
		\BibitemOpen
		\bibfield  {author} {\bibinfo {author} {\bibfnamefont {B.~G.}\ \bibnamefont
				{Todd-Rutel}}\ and\ \bibinfo {author} {\bibfnamefont {J.}~\bibnamefont
				{Piekarewicz}},\ }\bibfield  {title} {\bibinfo {title} {Neutron-rich nuclei
				and neutron stars: A new accurately calibrated interaction for the study of
				neutron-rich matter},\ }\href {https://doi.org/10.1103/PhysRevLett.95.122501}
		{\bibfield  {journal} {\bibinfo  {journal} {Phys. Rev. Lett.}\ }\textbf
			{\bibinfo {volume} {95}},\ \bibinfo {pages} {122501} (\bibinfo {year}
			{2005})}\BibitemShut {NoStop}%
		\bibitem [{\citenamefont {Centelles}\ \emph {et~al.}(2009)\citenamefont
			{Centelles}, \citenamefont {Roca-Maza}, \citenamefont {Vi\~nas},\ and\
			\citenamefont {Warda}}]{sym-stru-6}%
		\BibitemOpen
		\bibfield  {author} {\bibinfo {author} {\bibfnamefont {M.}~\bibnamefont
				{Centelles}}, \bibinfo {author} {\bibfnamefont {X.}~\bibnamefont
				{Roca-Maza}}, \bibinfo {author} {\bibfnamefont {X.}~\bibnamefont {Vi\~nas}},\
			and\ \bibinfo {author} {\bibfnamefont {M.}~\bibnamefont {Warda}},\ }\bibfield
		{title} {\bibinfo {title} {Nuclear symmetry energy probed by neutron skin
				thickness of nuclei},\ }\href
		{https://doi.org/10.1103/PhysRevLett.102.122502} {\bibfield  {journal}
			{\bibinfo  {journal} {Phys. Rev. Lett.}\ }\textbf {\bibinfo {volume} {102}},\
			\bibinfo {pages} {122502} (\bibinfo {year} {2009})}\BibitemShut {NoStop}%
		\bibitem [{\citenamefont {Warda}\ \emph {et~al.}(2009)\citenamefont {Warda},
			\citenamefont {Vi\~nas}, \citenamefont {Roca-Maza},\ and\ \citenamefont
			{Centelles}}]{sym-stru-7}%
		\BibitemOpen
		\bibfield  {author} {\bibinfo {author} {\bibfnamefont {M.}~\bibnamefont
				{Warda}}, \bibinfo {author} {\bibfnamefont {X.}~\bibnamefont {Vi\~nas}},
			\bibinfo {author} {\bibfnamefont {X.}~\bibnamefont {Roca-Maza}},\ and\
			\bibinfo {author} {\bibfnamefont {M.}~\bibnamefont {Centelles}},\ }\bibfield
		{title} {\bibinfo {title} {Neutron skin thickness in the droplet model with
				surface width dependence: Indications of softness of the nuclear symmetry
				energy},\ }\href {https://doi.org/10.1103/PhysRevC.80.024316} {\bibfield
			{journal} {\bibinfo  {journal} {Phys. Rev. C}\ }\textbf {\bibinfo {volume}
				{80}},\ \bibinfo {pages} {024316} (\bibinfo {year} {2009})}\BibitemShut
		{NoStop}%
		\bibitem [{\citenamefont {Carbone}\ \emph
			{et~al.}(2010{\natexlab{a}})\citenamefont {Carbone}, \citenamefont {Col\`o},
			\citenamefont {Bracco}, \citenamefont {Cao}, \citenamefont {Bortignon},
			\citenamefont {Camera},\ and\ \citenamefont {Wieland}}]{sym-stru-8}%
		\BibitemOpen
		\bibfield  {author} {\bibinfo {author} {\bibfnamefont {A.}~\bibnamefont
				{Carbone}}, \bibinfo {author} {\bibfnamefont {G.}~\bibnamefont {Col\`o}},
			\bibinfo {author} {\bibfnamefont {A.}~\bibnamefont {Bracco}}, \bibinfo
			{author} {\bibfnamefont {L.-G.}\ \bibnamefont {Cao}}, \bibinfo {author}
			{\bibfnamefont {P.~F.}\ \bibnamefont {Bortignon}}, \bibinfo {author}
			{\bibfnamefont {F.}~\bibnamefont {Camera}},\ and\ \bibinfo {author}
			{\bibfnamefont {O.}~\bibnamefont {Wieland}},\ }\bibfield  {title} {\bibinfo
			{title} {Constraints on the symmetry energy and neutron skins from pygmy
				resonances in $^{68}\mathrm{Ni}$ and $^{132}\mathrm{Sn}$},\ }\href
		{https://doi.org/10.1103/PhysRevC.81.041301} {\bibfield  {journal} {\bibinfo
				{journal} {Phys. Rev. C}\ }\textbf {\bibinfo {volume} {81}},\ \bibinfo
			{pages} {041301} (\bibinfo {year} {2010}{\natexlab{a}})}\BibitemShut
		{NoStop}%
		\bibitem [{\citenamefont {Chen}\ \emph {et~al.}(2010)\citenamefont {Chen},
			\citenamefont {Ko}, \citenamefont {Li},\ and\ \citenamefont
			{Xu}}]{sym-stru-9}%
		\BibitemOpen
		\bibfield  {author} {\bibinfo {author} {\bibfnamefont {L.-W.}\ \bibnamefont
				{Chen}}, \bibinfo {author} {\bibfnamefont {C.~M.}\ \bibnamefont {Ko}},
			\bibinfo {author} {\bibfnamefont {B.-A.}\ \bibnamefont {Li}},\ and\ \bibinfo
			{author} {\bibfnamefont {J.}~\bibnamefont {Xu}},\ }\bibfield  {title}
		{\bibinfo {title} {Density slope of the nuclear symmetry energy from the
				neutron skin thickness of heavy nuclei},\ }\href
		{https://doi.org/10.1103/PhysRevC.82.024321} {\bibfield  {journal} {\bibinfo
				{journal} {Phys. Rev. C}\ }\textbf {\bibinfo {volume} {82}},\ \bibinfo
			{pages} {024321} (\bibinfo {year} {2010})}\BibitemShut {NoStop}%
		\bibitem [{\citenamefont {Alex~Brown}(2000{\natexlab{b}})}]{brown2000}%
		\BibitemOpen
		\bibfield  {author} {\bibinfo {author} {\bibfnamefont {B.}~\bibnamefont
				{Alex~Brown}},\ }\bibfield  {title} {\bibinfo {title} {Neutron radii in
				nuclei and the neutron equation of state},\ }\href
		{https://doi.org/10.1103/PhysRevLett.85.5296} {\bibfield  {journal} {\bibinfo
				{journal} {Phys. Rev. Lett.}\ }\textbf {\bibinfo {volume} {85}},\ \bibinfo
			{pages} {5296} (\bibinfo {year} {2000}{\natexlab{b}})}\BibitemShut {NoStop}%
		\bibitem [{\citenamefont {Horowitz}\ and\ \citenamefont
			{Piekarewicz}(2001)}]{PhysRevLett.86.5647}%
		\BibitemOpen
		\bibfield  {author} {\bibinfo {author} {\bibfnamefont {C.~J.}\ \bibnamefont
				{Horowitz}}\ and\ \bibinfo {author} {\bibfnamefont {J.}~\bibnamefont
				{Piekarewicz}},\ }\bibfield  {title} {\bibinfo {title} {Neutron star
				structure and the neutron radius of $^{208}pb$},\ }\href
		{https://doi.org/10.1103/PhysRevLett.86.5647} {\bibfield  {journal} {\bibinfo
				{journal} {Phys. Rev. Lett.}\ }\textbf {\bibinfo {volume} {86}},\ \bibinfo
			{pages} {5647} (\bibinfo {year} {2001})}\BibitemShut {NoStop}%
		\bibitem [{\citenamefont {Tsang}\ \emph {et~al.}(2012)\citenamefont {Tsang},
			\citenamefont {Stone}, \citenamefont {Camera}, \citenamefont {Danielewicz},
			\citenamefont {Gandolfi}, \citenamefont {Hebeler}, \citenamefont {Horowitz},
			\citenamefont {Lee}, \citenamefont {Lynch}, \citenamefont {Kohley},
			\citenamefont {Lemmon}, \citenamefont {M\"oller}, \citenamefont {Murakami},
			\citenamefont {Riordan}, \citenamefont {Roca-Maza}, \citenamefont
			{Sammarruca}, \citenamefont {Steiner}, \citenamefont {Vida\~na},\ and\
			\citenamefont {Yennello}}]{PhysRevC.86.015803}%
		\BibitemOpen
		\bibfield  {author} {\bibinfo {author} {\bibfnamefont {M.~B.}\ \bibnamefont
				{Tsang}}, \bibinfo {author} {\bibfnamefont {J.~R.}\ \bibnamefont {Stone}},
			\bibinfo {author} {\bibfnamefont {F.}~\bibnamefont {Camera}}, \bibinfo
			{author} {\bibfnamefont {P.}~\bibnamefont {Danielewicz}}, \bibinfo {author}
			{\bibfnamefont {S.}~\bibnamefont {Gandolfi}}, \bibinfo {author}
			{\bibfnamefont {K.}~\bibnamefont {Hebeler}}, \bibinfo {author} {\bibfnamefont
				{C.~J.}\ \bibnamefont {Horowitz}}, \bibinfo {author} {\bibfnamefont
				{J.}~\bibnamefont {Lee}}, \bibinfo {author} {\bibfnamefont {W.~G.}\
				\bibnamefont {Lynch}}, \bibinfo {author} {\bibfnamefont {Z.}~\bibnamefont
				{Kohley}}, \bibinfo {author} {\bibfnamefont {R.}~\bibnamefont {Lemmon}},
			\bibinfo {author} {\bibfnamefont {P.}~\bibnamefont {M\"oller}}, \bibinfo
			{author} {\bibfnamefont {T.}~\bibnamefont {Murakami}}, \bibinfo {author}
			{\bibfnamefont {S.}~\bibnamefont {Riordan}}, \bibinfo {author} {\bibfnamefont
				{X.}~\bibnamefont {Roca-Maza}}, \bibinfo {author} {\bibfnamefont
				{F.}~\bibnamefont {Sammarruca}}, \bibinfo {author} {\bibfnamefont {A.~W.}\
				\bibnamefont {Steiner}}, \bibinfo {author} {\bibfnamefont {I.}~\bibnamefont
				{Vida\~na}},\ and\ \bibinfo {author} {\bibfnamefont {S.~J.}\ \bibnamefont
				{Yennello}},\ }\bibfield  {title} {\bibinfo {title} {Constraints on the
				symmetry energy and neutron skins from experiments and theory},\ }\href
		{https://doi.org/10.1103/PhysRevC.86.015803} {\bibfield  {journal} {\bibinfo
				{journal} {Phys. Rev. C}\ }\textbf {\bibinfo {volume} {86}},\ \bibinfo
			{pages} {015803} (\bibinfo {year} {2012})}\BibitemShut {NoStop}%
		\bibitem [{\citenamefont {Hagen}\ \emph {et~al.}(2016)\citenamefont {Hagen},
			\citenamefont {Ekstr{\"o}m}, \citenamefont {Forss{\'e}n}, \citenamefont
			{Jansen}, \citenamefont {Nazarewicz}, \citenamefont {Papenbrock},
			\citenamefont {Wendt}, \citenamefont {Bacca}, \citenamefont {Barnea},
			\citenamefont {Carlsson}, \citenamefont {Drischler}, \citenamefont {Hebeler},
			\citenamefont {Hjorth-Jensen}, \citenamefont {Miorelli}, \citenamefont
			{Orlandini}, \citenamefont {Schwenk},\ and\ \citenamefont
			{Simonis}}]{Hagen2016}%
		\BibitemOpen
		\bibfield  {author} {\bibinfo {author} {\bibfnamefont {G.}~\bibnamefont
				{Hagen}}, \bibinfo {author} {\bibfnamefont {A.}~\bibnamefont {Ekstr{\"o}m}},
			\bibinfo {author} {\bibfnamefont {C.}~\bibnamefont {Forss{\'e}n}}, \bibinfo
			{author} {\bibfnamefont {G.~R.}\ \bibnamefont {Jansen}}, \bibinfo {author}
			{\bibfnamefont {W.}~\bibnamefont {Nazarewicz}}, \bibinfo {author}
			{\bibfnamefont {T.}~\bibnamefont {Papenbrock}}, \bibinfo {author}
			{\bibfnamefont {K.~A.}\ \bibnamefont {Wendt}}, \bibinfo {author}
			{\bibfnamefont {S.}~\bibnamefont {Bacca}}, \bibinfo {author} {\bibfnamefont
				{N.}~\bibnamefont {Barnea}}, \bibinfo {author} {\bibfnamefont
				{B.}~\bibnamefont {Carlsson}}, \bibinfo {author} {\bibfnamefont
				{C.}~\bibnamefont {Drischler}}, \bibinfo {author} {\bibfnamefont
				{K.}~\bibnamefont {Hebeler}}, \bibinfo {author} {\bibfnamefont
				{M.}~\bibnamefont {Hjorth-Jensen}}, \bibinfo {author} {\bibfnamefont
				{M.}~\bibnamefont {Miorelli}}, \bibinfo {author} {\bibfnamefont
				{G.}~\bibnamefont {Orlandini}}, \bibinfo {author} {\bibfnamefont
				{A.}~\bibnamefont {Schwenk}},\ and\ \bibinfo {author} {\bibfnamefont
				{J.}~\bibnamefont {Simonis}},\ }\bibfield  {title} {\bibinfo {title} {Neutron
				and weak-charge distributions of the 48ca nucleus},\ }\href
		{https://doi.org/10.1038/nphys3529} {\bibfield  {journal} {\bibinfo
				{journal} {Nature Physics}\ }\textbf {\bibinfo {volume} {12}},\ \bibinfo
			{pages} {186} (\bibinfo {year} {2016})}\BibitemShut {NoStop}%
		\bibitem [{\citenamefont {Brown}(2017{\natexlab{a}})}]{brown-rch}%
		\BibitemOpen
		\bibfield  {author} {\bibinfo {author} {\bibfnamefont {B.~A.}\ \bibnamefont
				{Brown}},\ }\bibfield  {title} {\bibinfo {title} {Mirror charge radii and the
				neutron equation of state},\ }\href
		{https://doi.org/10.1103/PhysRevLett.119.122502} {\bibfield  {journal}
			{\bibinfo  {journal} {Phys. Rev. Lett.}\ }\textbf {\bibinfo {volume} {119}},\
			\bibinfo {pages} {122502} (\bibinfo {year} {2017}{\natexlab{a}})}\BibitemShut
		{NoStop}%
		\bibitem [{\citenamefont {Fattoyev}\ \emph {et~al.}(2018)\citenamefont
			{Fattoyev}, \citenamefont {Piekarewicz},\ and\ \citenamefont
			{Horowitz}}]{PhysRevLett.120.172702}%
		\BibitemOpen
		\bibfield  {author} {\bibinfo {author} {\bibfnamefont {F.~J.}\ \bibnamefont
				{Fattoyev}}, \bibinfo {author} {\bibfnamefont {J.}~\bibnamefont
				{Piekarewicz}},\ and\ \bibinfo {author} {\bibfnamefont {C.~J.}\ \bibnamefont
				{Horowitz}},\ }\bibfield  {title} {\bibinfo {title} {Neutron skins and
				neutron stars in the multimessenger era},\ }\href
		{https://doi.org/10.1103/PhysRevLett.120.172702} {\bibfield  {journal}
			{\bibinfo  {journal} {Phys. Rev. Lett.}\ }\textbf {\bibinfo {volume} {120}},\
			\bibinfo {pages} {172702} (\bibinfo {year} {2018})}\BibitemShut {NoStop}%
		\bibitem [{\citenamefont {Bertulani}\ and\ \citenamefont
			{Valencia}(2019)}]{PhysRevC.100.015802}%
		\BibitemOpen
		\bibfield  {author} {\bibinfo {author} {\bibfnamefont {C.~A.}\ \bibnamefont
				{Bertulani}}\ and\ \bibinfo {author} {\bibfnamefont {J.}~\bibnamefont
				{Valencia}},\ }\bibfield  {title} {\bibinfo {title} {Neutron skins as
				laboratory constraints on properties of neutron stars and on what we can
				learn from heavy ion fragmentation reactions},\ }\href
		{https://doi.org/10.1103/PhysRevC.100.015802} {\bibfield  {journal} {\bibinfo
				{journal} {Phys. Rev. C}\ }\textbf {\bibinfo {volume} {100}},\ \bibinfo
			{pages} {015802} (\bibinfo {year} {2019})}\BibitemShut {NoStop}%
		\bibitem [{\citenamefont {Roca-Maza}\ \emph {et~al.}(2011)\citenamefont
			{Roca-Maza}, \citenamefont {Centelles}, \citenamefont {Vi\~nas},\ and\
			\citenamefont {Warda}}]{rnp-1}%
		\BibitemOpen
		\bibfield  {author} {\bibinfo {author} {\bibfnamefont {X.}~\bibnamefont
				{Roca-Maza}}, \bibinfo {author} {\bibfnamefont {M.}~\bibnamefont
				{Centelles}}, \bibinfo {author} {\bibfnamefont {X.}~\bibnamefont {Vi\~nas}},\
			and\ \bibinfo {author} {\bibfnamefont {M.}~\bibnamefont {Warda}},\ }\bibfield
		{title} {\bibinfo {title} {Neutron skin of $^{208}\mathrm{Pb}$, nuclear
				symmetry energy, and the parity radius experiment},\ }\href
		{https://doi.org/10.1103/PhysRevLett.106.252501} {\bibfield  {journal}
			{\bibinfo  {journal} {Phys. Rev. Lett.}\ }\textbf {\bibinfo {volume} {106}},\
			\bibinfo {pages} {252501} (\bibinfo {year} {2011})}\BibitemShut {NoStop}%
		\bibitem [{\citenamefont {Reinhard}\ and\ \citenamefont
			{Nazarewicz}(2016)}]{rnp-2}%
		\BibitemOpen
		\bibfield  {author} {\bibinfo {author} {\bibfnamefont {P.-G.}\ \bibnamefont
				{Reinhard}}\ and\ \bibinfo {author} {\bibfnamefont {W.}~\bibnamefont
				{Nazarewicz}},\ }\bibfield  {title} {\bibinfo {title} {Nuclear charge and
				neutron radii and nuclear matter: Trend analysis in skyrme
				density-functional-theory approach},\ }\href
		{https://doi.org/10.1103/PhysRevC.93.051303} {\bibfield  {journal} {\bibinfo
				{journal} {Phys. Rev. C}\ }\textbf {\bibinfo {volume} {93}},\ \bibinfo
			{pages} {051303} (\bibinfo {year} {2016})}\BibitemShut {NoStop}%
		\bibitem [{\citenamefont {Steiner}\ \emph
			{et~al.}(2005{\natexlab{b}})\citenamefont {Steiner}, \citenamefont {Prakash},
			\citenamefont {Lattimer},\ and\ \citenamefont {Ellis}}]{explain_phys_rep}%
		\BibitemOpen
		\bibfield  {author} {\bibinfo {author} {\bibfnamefont {A.}~\bibnamefont
				{Steiner}}, \bibinfo {author} {\bibfnamefont {M.}~\bibnamefont {Prakash}},
			\bibinfo {author} {\bibfnamefont {J.}~\bibnamefont {Lattimer}},\ and\
			\bibinfo {author} {\bibfnamefont {P.}~\bibnamefont {Ellis}},\ }\bibfield
		{title} {\bibinfo {title} {Isospin asymmetry in nuclei and neutron stars},\
		}\href {https://doi.org/https://doi.org/10.1016/j.physrep.2005.02.004}
		{\bibfield  {journal} {\bibinfo  {journal} {Physics Reports}\ }\textbf
			{\bibinfo {volume} {411}},\ \bibinfo {pages} {325} (\bibinfo {year}
			{2005}{\natexlab{b}})}\BibitemShut {NoStop}%
		\bibitem [{\citenamefont {Alex~Brown}(2000{\natexlab{c}})}]{to-high-density}%
		\BibitemOpen
		\bibfield  {author} {\bibinfo {author} {\bibfnamefont {B.}~\bibnamefont
				{Alex~Brown}},\ }\bibfield  {title} {\bibinfo {title} {Neutron radii in
				nuclei and the neutron equation of state},\ }\href
		{https://doi.org/10.1103/PhysRevLett.85.5296} {\bibfield  {journal} {\bibinfo
				{journal} {Phys. Rev. Lett.}\ }\textbf {\bibinfo {volume} {85}},\ \bibinfo
			{pages} {5296} (\bibinfo {year} {2000}{\natexlab{c}})}\BibitemShut {NoStop}%
		\bibitem [{\citenamefont {Brown}(2017{\natexlab{b}})}]{L-pressure}%
		\BibitemOpen
		\bibfield  {author} {\bibinfo {author} {\bibfnamefont {B.~A.}\ \bibnamefont
				{Brown}},\ }\bibfield  {title} {\bibinfo {title} {Mirror charge radii and the
				neutron equation of state},\ }\href
		{https://doi.org/10.1103/PhysRevLett.119.122502} {\bibfield  {journal}
			{\bibinfo  {journal} {Phys. Rev. Lett.}\ }\textbf {\bibinfo {volume} {119}},\
			\bibinfo {pages} {122502} (\bibinfo {year} {2017}{\natexlab{b}})}\BibitemShut
		{NoStop}%
		\bibitem [{\citenamefont {Chen}\ \emph {et~al.}(2005)\citenamefont {Chen},
			\citenamefont {Ko},\ and\ \citenamefont {Li}}]{CLW}%
		\BibitemOpen
		\bibfield  {author} {\bibinfo {author} {\bibfnamefont {L.-W.}\ \bibnamefont
				{Chen}}, \bibinfo {author} {\bibfnamefont {C.~M.}\ \bibnamefont {Ko}},\ and\
			\bibinfo {author} {\bibfnamefont {B.-A.}\ \bibnamefont {Li}},\ }\bibfield
		{title} {\bibinfo {title} {Nuclear matter symmetry energy and the neutron
				skin thickness of heavy nuclei},\ }\href
		{https://doi.org/10.1103/PhysRevC.72.064309} {\bibfield  {journal} {\bibinfo
				{journal} {Phys. Rev. C}\ }\textbf {\bibinfo {volume} {72}},\ \bibinfo
			{pages} {064309} (\bibinfo {year} {2005})}\BibitemShut {NoStop}%
		\bibitem [{\citenamefont {Li}\ and\ \citenamefont {Han}(2013)}]{Li-L}%
		\BibitemOpen
		\bibfield  {author} {\bibinfo {author} {\bibfnamefont {B.-A.}\ \bibnamefont
				{Li}}\ and\ \bibinfo {author} {\bibfnamefont {X.}~\bibnamefont {Han}},\
		}\bibfield  {title} {\bibinfo {title} {Constraining the neutron-proton
				effective mass splitting using empirical constraints on the density
				dependence of nuclear symmetry energy around normal density},\ }\href
		{https://doi.org/https://doi.org/10.1016/j.physletb.2013.10.006} {\bibfield
			{journal} {\bibinfo  {journal} {Physics Letters B}\ }\textbf {\bibinfo
				{volume} {727}},\ \bibinfo {pages} {276} (\bibinfo {year}
			{2013})}\BibitemShut {NoStop}%
		\bibitem [{\citenamefont {Oertel}\ \emph
			{et~al.}(2017{\natexlab{b}})\citenamefont {Oertel}, \citenamefont {Hempel},
			\citenamefont {Kl\"ahn},\ and\ \citenamefont {Typel}}]{rmp-con-L}%
		\BibitemOpen
		\bibfield  {author} {\bibinfo {author} {\bibfnamefont {M.}~\bibnamefont
				{Oertel}}, \bibinfo {author} {\bibfnamefont {M.}~\bibnamefont {Hempel}},
			\bibinfo {author} {\bibfnamefont {T.}~\bibnamefont {Kl\"ahn}},\ and\ \bibinfo
			{author} {\bibfnamefont {S.}~\bibnamefont {Typel}},\ }\bibfield  {title}
		{\bibinfo {title} {Equations of state for supernovae and compact stars},\
		}\href {https://doi.org/10.1103/RevModPhys.89.015007} {\bibfield  {journal}
			{\bibinfo  {journal} {Rev. Mod. Phys.}\ }\textbf {\bibinfo {volume} {89}},\
			\bibinfo {pages} {015007} (\bibinfo {year} {2017}{\natexlab{b}})}\BibitemShut
		{NoStop}%
		\bibitem [{\citenamefont {Drischler}\ \emph {et~al.}(2020)\citenamefont
			{Drischler}, \citenamefont {Furnstahl}, \citenamefont {Melendez},\ and\
			\citenamefont {Phillips}}]{Bay-L}%
		\BibitemOpen
		\bibfield  {author} {\bibinfo {author} {\bibfnamefont {C.}~\bibnamefont
				{Drischler}}, \bibinfo {author} {\bibfnamefont {R.~J.}\ \bibnamefont
				{Furnstahl}}, \bibinfo {author} {\bibfnamefont {J.~A.}\ \bibnamefont
				{Melendez}},\ and\ \bibinfo {author} {\bibfnamefont {D.~R.}\ \bibnamefont
				{Phillips}},\ }\bibfield  {title} {\bibinfo {title} {How well do we know the
				neutron-matter equation of state at the densities inside neutron stars? a
				bayesian approach with correlated uncertainties},\ }\href
		{https://doi.org/10.1103/PhysRevLett.125.202702} {\bibfield  {journal}
			{\bibinfo  {journal} {Phys. Rev. Lett.}\ }\textbf {\bibinfo {volume} {125}},\
			\bibinfo {pages} {202702} (\bibinfo {year} {2020})}\BibitemShut {NoStop}%
		\bibitem [{\citenamefont {Li}\ \emph {et~al.}(2021{\natexlab{a}})\citenamefont
			{Li}, \citenamefont {Cai}, \citenamefont {Xie},\ and\ \citenamefont
			{Zhang}}]{NS-L}%
		\BibitemOpen
		\bibfield  {author} {\bibinfo {author} {\bibfnamefont {B.-A.}\ \bibnamefont
				{Li}}, \bibinfo {author} {\bibfnamefont {B.-J.}\ \bibnamefont {Cai}},
			\bibinfo {author} {\bibfnamefont {W.-J.}\ \bibnamefont {Xie}},\ and\ \bibinfo
			{author} {\bibfnamefont {N.-B.}\ \bibnamefont {Zhang}},\ }\bibfield  {title}
		{\bibinfo {title} {Progress in constraining nuclear symmetry energy using
				neutron star observables since gw170817},\ }\bibfield  {journal} {\bibinfo
			{journal} {Universe}\ }\textbf {\bibinfo {volume} {7}},\ \href
		{https://doi.org/10.3390/universe7060182} {10.3390/universe7060182} (\bibinfo
		{year} {2021}{\natexlab{a}})\BibitemShut {NoStop}%
		\bibitem [{\citenamefont {Carbone}\ \emph
			{et~al.}(2010{\natexlab{b}})\citenamefont {Carbone}, \citenamefont {Col\`o},
			\citenamefont {Bracco}, \citenamefont {Cao}, \citenamefont {Bortignon},
			\citenamefont {Camera},\ and\ \citenamefont {Wieland}}]{PhysRevC.81.041301}%
		\BibitemOpen
		\bibfield  {author} {\bibinfo {author} {\bibfnamefont {A.}~\bibnamefont
				{Carbone}}, \bibinfo {author} {\bibfnamefont {G.}~\bibnamefont {Col\`o}},
			\bibinfo {author} {\bibfnamefont {A.}~\bibnamefont {Bracco}}, \bibinfo
			{author} {\bibfnamefont {L.-G.}\ \bibnamefont {Cao}}, \bibinfo {author}
			{\bibfnamefont {P.~F.}\ \bibnamefont {Bortignon}}, \bibinfo {author}
			{\bibfnamefont {F.}~\bibnamefont {Camera}},\ and\ \bibinfo {author}
			{\bibfnamefont {O.}~\bibnamefont {Wieland}},\ }\bibfield  {title} {\bibinfo
			{title} {Constraints on the symmetry energy and neutron skins from pygmy
				resonances in $^{68}\mathrm{Ni}$ and $^{132}\mathrm{Sn}$},\ }\href
		{https://doi.org/10.1103/PhysRevC.81.041301} {\bibfield  {journal} {\bibinfo
				{journal} {Phys. Rev. C}\ }\textbf {\bibinfo {volume} {81}},\ \bibinfo
			{pages} {041301} (\bibinfo {year} {2010}{\natexlab{b}})}\BibitemShut
		{NoStop}%
		\bibitem [{\citenamefont {Roca-Maza}\ \emph {et~al.}(2013)\citenamefont
			{Roca-Maza}, \citenamefont {Brenna}, \citenamefont {Agrawal}, \citenamefont
			{Bortignon}, \citenamefont {Col\`o}, \citenamefont {Cao}, \citenamefont
			{Paar},\ and\ \citenamefont {Vretenar}}]{PhysRevC.87.034301}%
		\BibitemOpen
		\bibfield  {author} {\bibinfo {author} {\bibfnamefont {X.}~\bibnamefont
				{Roca-Maza}}, \bibinfo {author} {\bibfnamefont {M.}~\bibnamefont {Brenna}},
			\bibinfo {author} {\bibfnamefont {B.~K.}\ \bibnamefont {Agrawal}}, \bibinfo
			{author} {\bibfnamefont {P.~F.}\ \bibnamefont {Bortignon}}, \bibinfo {author}
			{\bibfnamefont {G.}~\bibnamefont {Col\`o}}, \bibinfo {author} {\bibfnamefont
				{L.-G.}\ \bibnamefont {Cao}}, \bibinfo {author} {\bibfnamefont
				{N.}~\bibnamefont {Paar}},\ and\ \bibinfo {author} {\bibfnamefont
				{D.}~\bibnamefont {Vretenar}},\ }\bibfield  {title} {\bibinfo {title} {Giant
				quadrupole resonances in ${}^{208}$pb, the nuclear symmetry energy, and the
				neutron skin thickness},\ }\href {https://doi.org/10.1103/PhysRevC.87.034301}
		{\bibfield  {journal} {\bibinfo  {journal} {Phys. Rev. C}\ }\textbf {\bibinfo
				{volume} {87}},\ \bibinfo {pages} {034301} (\bibinfo {year}
			{2013})}\BibitemShut {NoStop}%
		\bibitem [{\citenamefont {Zhang}\ and\ \citenamefont {Chen}(2013)}]{CLW-np}%
		\BibitemOpen
		\bibfield  {author} {\bibinfo {author} {\bibfnamefont {Z.}~\bibnamefont
				{Zhang}}\ and\ \bibinfo {author} {\bibfnamefont {L.-W.}\ \bibnamefont
				{Chen}},\ }\bibfield  {title} {\bibinfo {title} {Constraining the symmetry
				energy at subsaturation densities using isotope binding energy difference and
				neutron skin thickness},\ }\href
		{https://doi.org/https://doi.org/10.1016/j.physletb.2013.08.002} {\bibfield
			{journal} {\bibinfo  {journal} {Physics Letters B}\ }\textbf {\bibinfo
				{volume} {726}},\ \bibinfo {pages} {234} (\bibinfo {year}
			{2013})}\BibitemShut {NoStop}%
		\bibitem [{\citenamefont {Mondal}\ \emph {et~al.}(2016)\citenamefont {Mondal},
			\citenamefont {Agrawal}, \citenamefont {Centelles}, \citenamefont {Col\`o},
			\citenamefont {Roca-Maza}, \citenamefont {Paar}, \citenamefont {Vi\~nas},
			\citenamefont {Singh},\ and\ \citenamefont {Patra}}]{PhysRevC.93.064303}%
		\BibitemOpen
		\bibfield  {author} {\bibinfo {author} {\bibfnamefont {C.}~\bibnamefont
				{Mondal}}, \bibinfo {author} {\bibfnamefont {B.~K.}\ \bibnamefont {Agrawal}},
			\bibinfo {author} {\bibfnamefont {M.}~\bibnamefont {Centelles}}, \bibinfo
			{author} {\bibfnamefont {G.}~\bibnamefont {Col\`o}}, \bibinfo {author}
			{\bibfnamefont {X.}~\bibnamefont {Roca-Maza}}, \bibinfo {author}
			{\bibfnamefont {N.}~\bibnamefont {Paar}}, \bibinfo {author} {\bibfnamefont
				{X.}~\bibnamefont {Vi\~nas}}, \bibinfo {author} {\bibfnamefont {S.~K.}\
				\bibnamefont {Singh}},\ and\ \bibinfo {author} {\bibfnamefont {S.~K.}\
				\bibnamefont {Patra}},\ }\bibfield  {title} {\bibinfo {title} {Model
				dependence of the neutron-skin thickness on the symmetry energy},\ }\href
		{https://doi.org/10.1103/PhysRevC.93.064303} {\bibfield  {journal} {\bibinfo
				{journal} {Phys. Rev. C}\ }\textbf {\bibinfo {volume} {93}},\ \bibinfo
			{pages} {064303} (\bibinfo {year} {2016})}\BibitemShut {NoStop}%
		\bibitem [{\citenamefont {Min}\ \emph {et~al.}(2011)\citenamefont {Min},
			\citenamefont {Zhu-Xia}, \citenamefont {Ning},\ and\ \citenamefont
			{Feng-Shou}}]{WL-np}%
		\BibitemOpen
		\bibfield  {author} {\bibinfo {author} {\bibfnamefont {L.}~\bibnamefont
				{Min}}, \bibinfo {author} {\bibfnamefont {L.}~\bibnamefont {Zhu-Xia}},
			\bibinfo {author} {\bibfnamefont {W.}~\bibnamefont {Ning}},\ and\ \bibinfo
			{author} {\bibfnamefont {Z.}~\bibnamefont {Feng-Shou}},\ }\bibfield  {title}
		{\bibinfo {title} {Exploring nuclear symmetry energy with isospin dependence
				in neutron skin thickness of nuclei},\ }\href
		{https://doi.org/10.1088/1674-1137/35/7/006} {\bibfield  {journal} {\bibinfo
				{journal} {Chinese Physics C}\ }\textbf {\bibinfo {volume} {35}},\ \bibinfo
			{pages} {629} (\bibinfo {year} {2011})}\BibitemShut {NoStop}%
		\bibitem [{\citenamefont {Fan}\ \emph {et~al.}(2015)\citenamefont {Fan},
			\citenamefont {Dong},\ and\ \citenamefont {Zuo}}]{Fan2015}%
		\BibitemOpen
		\bibfield  {author} {\bibinfo {author} {\bibfnamefont {X.}~\bibnamefont
				{Fan}}, \bibinfo {author} {\bibfnamefont {J.}~\bibnamefont {Dong}},\ and\
			\bibinfo {author} {\bibfnamefont {W.}~\bibnamefont {Zuo}},\ }\bibfield
		{title} {\bibinfo {title} {Symmetry energy at subsaturation densities and the
				neutron skin thickness of 208pb},\ }\href
		{https://doi.org/10.1007/s11433-015-5673-8} {\bibfield  {journal} {\bibinfo
				{journal} {Science China Physics, Mechanics {\&} Astronomy}\ }\textbf
			{\bibinfo {volume} {58}},\ \bibinfo {pages} {1} (\bibinfo {year}
			{2015})}\BibitemShut {NoStop}%
		\bibitem [{\citenamefont {Xu}\ \emph {et~al.}(2014)\citenamefont {Xu},
			\citenamefont {Ren},\ and\ \citenamefont {Liu}}]{RZZ-np}%
		\BibitemOpen
		\bibfield  {author} {\bibinfo {author} {\bibfnamefont {C.}~\bibnamefont
				{Xu}}, \bibinfo {author} {\bibfnamefont {Z.}~\bibnamefont {Ren}},\ and\
			\bibinfo {author} {\bibfnamefont {J.}~\bibnamefont {Liu}},\ }\bibfield
		{title} {\bibinfo {title} {Attempt to link the neutron skin thickness of
				$^{208}\mathrm{Pb}$ with the symmetry energy through cluster radioactivity},\
		}\href {https://doi.org/10.1103/PhysRevC.90.064310} {\bibfield  {journal}
			{\bibinfo  {journal} {Phys. Rev. C}\ }\textbf {\bibinfo {volume} {90}},\
			\bibinfo {pages} {064310} (\bibinfo {year} {2014})}\BibitemShut {NoStop}%
		\bibitem [{\citenamefont {Xu}\ \emph {et~al.}(2020)\citenamefont {Xu},
			\citenamefont {Xie},\ and\ \citenamefont {Li}}]{LBA-np}%
		\BibitemOpen
		\bibfield  {author} {\bibinfo {author} {\bibfnamefont {J.}~\bibnamefont
				{Xu}}, \bibinfo {author} {\bibfnamefont {W.-J.}\ \bibnamefont {Xie}},\ and\
			\bibinfo {author} {\bibfnamefont {B.-A.}\ \bibnamefont {Li}},\ }\bibfield
		{title} {\bibinfo {title} {Bayesian inference of nuclear symmetry energy from
				measured and imagined neutron skin thickness in
				$^{116,118,120,122,124,130,132}\mathrm{Sn}, ^{208}\mathrm{Pb}$, and
				$^{48}\mathrm{Ca}$},\ }\href {https://doi.org/10.1103/PhysRevC.102.044316}
		{\bibfield  {journal} {\bibinfo  {journal} {Phys. Rev. C}\ }\textbf {\bibinfo
				{volume} {102}},\ \bibinfo {pages} {044316} (\bibinfo {year}
			{2020})}\BibitemShut {NoStop}%
		\bibitem [{\citenamefont {Wang}\ and\ \citenamefont {Li}(2013)}]{WL-rch}%
		\BibitemOpen
		\bibfield  {author} {\bibinfo {author} {\bibfnamefont {N.}~\bibnamefont
				{Wang}}\ and\ \bibinfo {author} {\bibfnamefont {T.}~\bibnamefont {Li}},\
		}\bibfield  {title} {\bibinfo {title} {Shell and isospin effects in nuclear
				charge radii},\ }\href {https://doi.org/10.1103/PhysRevC.88.011301}
		{\bibfield  {journal} {\bibinfo  {journal} {Phys. Rev. C}\ }\textbf {\bibinfo
				{volume} {88}},\ \bibinfo {pages} {011301} (\bibinfo {year}
			{2013})}\BibitemShut {NoStop}%
		\bibitem [{\citenamefont {Brown}\ \emph {et~al.}(2020)\citenamefont {Brown},
			\citenamefont {Minamisono}, \citenamefont {Piekarewicz}, \citenamefont
			{Hergert}, \citenamefont {Garand}, \citenamefont {Klose}, \citenamefont
			{K\"onig}, \citenamefont {Lantis}, \citenamefont {Liu}, \citenamefont
			{Maa\ss{}}, \citenamefont {Miller}, \citenamefont {N\"ortersh\"auser},
			\citenamefont {Pineda}, \citenamefont {Powel}, \citenamefont {Rossi},
			\citenamefont {Sommer}, \citenamefont {Sumithrarachchi}, \citenamefont
			{Teigelh\"ofer}, \citenamefont {Watkins},\ and\ \citenamefont
			{Wirth}}]{brown-rch-2}%
		\BibitemOpen
		\bibfield  {author} {\bibinfo {author} {\bibfnamefont {B.~A.}\ \bibnamefont
				{Brown}}, \bibinfo {author} {\bibfnamefont {K.}~\bibnamefont {Minamisono}},
			\bibinfo {author} {\bibfnamefont {J.}~\bibnamefont {Piekarewicz}}, \bibinfo
			{author} {\bibfnamefont {H.}~\bibnamefont {Hergert}}, \bibinfo {author}
			{\bibfnamefont {D.}~\bibnamefont {Garand}}, \bibinfo {author} {\bibfnamefont
				{A.}~\bibnamefont {Klose}}, \bibinfo {author} {\bibfnamefont
				{K.}~\bibnamefont {K\"onig}}, \bibinfo {author} {\bibfnamefont {J.~D.}\
				\bibnamefont {Lantis}}, \bibinfo {author} {\bibfnamefont {Y.}~\bibnamefont
				{Liu}}, \bibinfo {author} {\bibfnamefont {B.}~\bibnamefont {Maa\ss{}}},
			\bibinfo {author} {\bibfnamefont {A.~J.}\ \bibnamefont {Miller}}, \bibinfo
			{author} {\bibfnamefont {W.}~\bibnamefont {N\"ortersh\"auser}}, \bibinfo
			{author} {\bibfnamefont {S.~V.}\ \bibnamefont {Pineda}}, \bibinfo {author}
			{\bibfnamefont {R.~C.}\ \bibnamefont {Powel}}, \bibinfo {author}
			{\bibfnamefont {D.~M.}\ \bibnamefont {Rossi}}, \bibinfo {author}
			{\bibfnamefont {F.}~\bibnamefont {Sommer}}, \bibinfo {author} {\bibfnamefont
				{C.}~\bibnamefont {Sumithrarachchi}}, \bibinfo {author} {\bibfnamefont
				{A.}~\bibnamefont {Teigelh\"ofer}}, \bibinfo {author} {\bibfnamefont
				{J.}~\bibnamefont {Watkins}},\ and\ \bibinfo {author} {\bibfnamefont
				{R.}~\bibnamefont {Wirth}},\ }\bibfield  {title} {\bibinfo {title}
			{Implications of the $^{36}\mathrm{Ca}\ensuremath{-}^{36}\mathrm{S}$ and
				$^{38}\mathrm{Ca}\ensuremath{-}^{38}\mathrm{Ar}$ difference in mirror charge
				radii on the neutron matter equation of state},\ }\href
		{https://doi.org/10.1103/PhysRevResearch.2.022035} {\bibfield  {journal}
			{\bibinfo  {journal} {Phys. Rev. Res.}\ }\textbf {\bibinfo {volume} {2}},\
			\bibinfo {pages} {022035} (\bibinfo {year} {2020})}\BibitemShut {NoStop}%
		\bibitem [{\citenamefont {Pineda}\ \emph {et~al.}(2021)\citenamefont {Pineda},
			\citenamefont {K\"onig}, \citenamefont {Rossi}, \citenamefont {Brown},
			\citenamefont {Incorvati}, \citenamefont {Lantis}, \citenamefont
			{Minamisono}, \citenamefont {N\"ortersh\"auser}, \citenamefont {Piekarewicz},
			\citenamefont {Powel},\ and\ \citenamefont
			{Sommer}}]{PhysRevLett.127.182503}%
		\BibitemOpen
		\bibfield  {author} {\bibinfo {author} {\bibfnamefont {S.~V.}\ \bibnamefont
				{Pineda}}, \bibinfo {author} {\bibfnamefont {K.}~\bibnamefont {K\"onig}},
			\bibinfo {author} {\bibfnamefont {D.~M.}\ \bibnamefont {Rossi}}, \bibinfo
			{author} {\bibfnamefont {B.~A.}\ \bibnamefont {Brown}}, \bibinfo {author}
			{\bibfnamefont {A.}~\bibnamefont {Incorvati}}, \bibinfo {author}
			{\bibfnamefont {J.}~\bibnamefont {Lantis}}, \bibinfo {author} {\bibfnamefont
				{K.}~\bibnamefont {Minamisono}}, \bibinfo {author} {\bibfnamefont
				{W.}~\bibnamefont {N\"ortersh\"auser}}, \bibinfo {author} {\bibfnamefont
				{J.}~\bibnamefont {Piekarewicz}}, \bibinfo {author} {\bibfnamefont
				{R.}~\bibnamefont {Powel}},\ and\ \bibinfo {author} {\bibfnamefont
				{F.}~\bibnamefont {Sommer}},\ }\bibfield  {title} {\bibinfo {title} {Charge
				radius of neutron-deficient $^{54}\mathrm{Ni}$ and symmetry energy
				constraints using the difference in mirror pair charge radii},\ }\href
		{https://doi.org/10.1103/PhysRevLett.127.182503} {\bibfield  {journal}
			{\bibinfo  {journal} {Phys. Rev. Lett.}\ }\textbf {\bibinfo {volume} {127}},\
			\bibinfo {pages} {182503} (\bibinfo {year} {2021})}\BibitemShut {NoStop}%
		\bibitem [{\citenamefont {An}\ \emph {et~al.}(2023)\citenamefont {An},
			\citenamefont {Sun}, \citenamefont {Cao},\ and\ \citenamefont
			{Zhang}}]{An2023}%
		\BibitemOpen
		\bibfield  {author} {\bibinfo {author} {\bibfnamefont {R.}~\bibnamefont
				{An}}, \bibinfo {author} {\bibfnamefont {S.}~\bibnamefont {Sun}}, \bibinfo
			{author} {\bibfnamefont {L.-G.}\ \bibnamefont {Cao}},\ and\ \bibinfo {author}
			{\bibfnamefont {F.-S.}\ \bibnamefont {Zhang}},\ }\bibfield  {title} {\bibinfo
			{title} {Constraining nuclear symmetry energy with the charge radii of
				mirror-pair nuclei},\ }\href {https://doi.org/10.1007/s41365-023-01269-1}
		{\bibfield  {journal} {\bibinfo  {journal} {Nuclear Science and Techniques}\
			}\textbf {\bibinfo {volume} {34}},\ \bibinfo {pages} {119} (\bibinfo {year}
			{2023})}\BibitemShut {NoStop}%
		\bibitem [{\citenamefont {Huang}\ \emph {et~al.}(2023)\citenamefont {Huang},
			\citenamefont {Li},\ and\ \citenamefont {Niu}}]{LYF-rch}%
		\BibitemOpen
		\bibfield  {author} {\bibinfo {author} {\bibfnamefont {Y.~N.}\ \bibnamefont
				{Huang}}, \bibinfo {author} {\bibfnamefont {Z.~Z.}\ \bibnamefont {Li}},\ and\
			\bibinfo {author} {\bibfnamefont {Y.~F.}\ \bibnamefont {Niu}},\ }\bibfield
		{title} {\bibinfo {title} {Correlation between the difference of charge radii
				in mirror nuclei and the slope parameter of the symmetry energy},\ }\href
		{https://doi.org/10.1103/PhysRevC.107.034319} {\bibfield  {journal} {\bibinfo
				{journal} {Phys. Rev. C}\ }\textbf {\bibinfo {volume} {107}},\ \bibinfo
			{pages} {034319} (\bibinfo {year} {2023})}\BibitemShut {NoStop}%
		\bibitem [{\citenamefont {Li}\ \emph {et~al.}(2021{\natexlab{b}})\citenamefont
			{Li}, \citenamefont {Luo},\ and\ \citenamefont {Wang}}]{rc21}%
		\BibitemOpen
		\bibfield  {author} {\bibinfo {author} {\bibfnamefont {T.}~\bibnamefont
				{Li}}, \bibinfo {author} {\bibfnamefont {Y.}~\bibnamefont {Luo}},\ and\
			\bibinfo {author} {\bibfnamefont {N.}~\bibnamefont {Wang}},\ }\bibfield
		{title} {\bibinfo {title} {Compilation of recent nuclear ground state charge
				radius measurements and tests for models},\ }\href
		{https://doi.org/https://doi.org/10.1016/j.adt.2021.101440} {\bibfield
			{journal} {\bibinfo  {journal} {Atomic Data and Nuclear Data Tables}\
			}\textbf {\bibinfo {volume} {140}},\ \bibinfo {pages} {101440} (\bibinfo
			{year} {2021}{\natexlab{b}})}\BibitemShut {NoStop}%
		\bibitem [{\citenamefont {Trzci\ifmmode~\acute{n}\else \'{n}\fi{}ska}\ \emph
			{et~al.}(2001)\citenamefont {Trzci\ifmmode~\acute{n}\else \'{n}\fi{}ska},
			\citenamefont {Jastrz\ifmmode~\mbox{\c{e}}\else \c{e}\fi{}bski},
			\citenamefont {Lubi\ifmmode~\acute{n}\else \'{n}\fi{}ski}, \citenamefont
			{Hartmann}, \citenamefont {Schmidt}, \citenamefont {von Egidy},\ and\
			\citenamefont {K\l{}os}}]{PhysRevLett.87.082501}%
		\BibitemOpen
		\bibfield  {author} {\bibinfo {author} {\bibfnamefont {A.}~\bibnamefont
				{Trzci\ifmmode~\acute{n}\else \'{n}\fi{}ska}}, \bibinfo {author}
			{\bibfnamefont {J.}~\bibnamefont {Jastrz\ifmmode~\mbox{\c{e}}\else
					\c{e}\fi{}bski}}, \bibinfo {author} {\bibfnamefont {P.}~\bibnamefont
				{Lubi\ifmmode~\acute{n}\else \'{n}\fi{}ski}}, \bibinfo {author}
			{\bibfnamefont {F.~J.}\ \bibnamefont {Hartmann}}, \bibinfo {author}
			{\bibfnamefont {R.}~\bibnamefont {Schmidt}}, \bibinfo {author} {\bibfnamefont
				{T.}~\bibnamefont {von Egidy}},\ and\ \bibinfo {author} {\bibfnamefont
				{B.}~\bibnamefont {K\l{}os}},\ }\bibfield  {title} {\bibinfo {title} {Neutron
				density distributions deduced from antiprotonic atoms},\ }\href
		{https://doi.org/10.1103/PhysRevLett.87.082501} {\bibfield  {journal}
			{\bibinfo  {journal} {Phys. Rev. Lett.}\ }\textbf {\bibinfo {volume} {87}},\
			\bibinfo {pages} {082501} (\bibinfo {year} {2001})}\BibitemShut {NoStop}%
		\bibitem [{\citenamefont {Novario}\ \emph {et~al.}(2023)\citenamefont
			{Novario}, \citenamefont {Lonardoni}, \citenamefont {Gandolfi},\ and\
			\citenamefont {Hagen}}]{rch-I}%
		\BibitemOpen
		\bibfield  {author} {\bibinfo {author} {\bibfnamefont {S.~J.}\ \bibnamefont
				{Novario}}, \bibinfo {author} {\bibfnamefont {D.}~\bibnamefont {Lonardoni}},
			\bibinfo {author} {\bibfnamefont {S.}~\bibnamefont {Gandolfi}},\ and\
			\bibinfo {author} {\bibfnamefont {G.}~\bibnamefont {Hagen}},\ }\bibfield
		{title} {\bibinfo {title} {Trends of neutron skins and radii of mirror nuclei
				from first principles},\ }\href
		{https://doi.org/10.1103/PhysRevLett.130.032501} {\bibfield  {journal}
			{\bibinfo  {journal} {Phys. Rev. Lett.}\ }\textbf {\bibinfo {volume} {130}},\
			\bibinfo {pages} {032501} (\bibinfo {year} {2023})}\BibitemShut {NoStop}%
		\bibitem [{\citenamefont {Myers}\ and\ \citenamefont
			{Swiatecki}(1969)}]{MYERS1969395}%
		\BibitemOpen
		\bibfield  {author} {\bibinfo {author} {\bibfnamefont {W.~D.}\ \bibnamefont
				{Myers}}\ and\ \bibinfo {author} {\bibfnamefont {W.}~\bibnamefont
				{Swiatecki}},\ }\bibfield  {title} {\bibinfo {title} {Average nuclear
				properties},\ }\href
		{https://doi.org/https://doi.org/10.1016/0003-4916(69)90202-4} {\bibfield
			{journal} {\bibinfo  {journal} {Annals of Physics}\ }\textbf {\bibinfo
				{volume} {55}},\ \bibinfo {pages} {395} (\bibinfo {year} {1969})}\BibitemShut
		{NoStop}%
		\bibitem [{\citenamefont {Myers}\ and\ \citenamefont
			{Swiatecki}(1974)}]{MYERS1974186}%
		\BibitemOpen
		\bibfield  {author} {\bibinfo {author} {\bibfnamefont {W.}~\bibnamefont
				{Myers}}\ and\ \bibinfo {author} {\bibfnamefont {W.}~\bibnamefont
				{Swiatecki}},\ }\bibfield  {title} {\bibinfo {title} {The nuclear droplet
				model for arbitrary shapes},\ }\href
		{https://doi.org/https://doi.org/10.1016/0003-4916(74)90299-1} {\bibfield
			{journal} {\bibinfo  {journal} {Annals of Physics}\ }\textbf {\bibinfo
				{volume} {84}},\ \bibinfo {pages} {186} (\bibinfo {year} {1974})}\BibitemShut
		{NoStop}%
		\bibitem [{\citenamefont {Myers}\ and\ \citenamefont
			{Swiatecki}(1980)}]{MYERS1980267}%
		\BibitemOpen
		\bibfield  {author} {\bibinfo {author} {\bibfnamefont {W.}~\bibnamefont
				{Myers}}\ and\ \bibinfo {author} {\bibfnamefont {W.}~\bibnamefont
				{Swiatecki}},\ }\bibfield  {title} {\bibinfo {title} {Droplet-model theory of
				the neutron skin},\ }\href
		{https://doi.org/https://doi.org/10.1016/0375-9474(80)90623-5} {\bibfield
			{journal} {\bibinfo  {journal} {Nuclear Physics A}\ }\textbf {\bibinfo
				{volume} {336}},\ \bibinfo {pages} {267} (\bibinfo {year}
			{1980})}\BibitemShut {NoStop}%
		\bibitem [{\citenamefont {Pethick}\ and\ \citenamefont
			{Ravenhall}(1996)}]{PETHICK1996173}%
		\BibitemOpen
		\bibfield  {author} {\bibinfo {author} {\bibfnamefont {C.}~\bibnamefont
				{Pethick}}\ and\ \bibinfo {author} {\bibfnamefont {D.}~\bibnamefont
				{Ravenhall}},\ }\bibfield  {title} {\bibinfo {title} {The dependence of
				neutron skin thickness and surface tension on neutron excess},\ }\href
		{https://doi.org/https://doi.org/10.1016/0375-9474(96)00216-3} {\bibfield
			{journal} {\bibinfo  {journal} {Nuclear Physics A}\ }\textbf {\bibinfo
				{volume} {606}},\ \bibinfo {pages} {173} (\bibinfo {year}
			{1996})}\BibitemShut {NoStop}%
		\bibitem [{\citenamefont {Zenihiro}\ \emph {et~al.}(2018)\citenamefont
			{Zenihiro}, \citenamefont {Sakaguchi}, \citenamefont {Terashima},
			\citenamefont {Uesaka}, \citenamefont {Hagen}, \citenamefont {Itoh},
			\citenamefont {Murakami}, \citenamefont {Nakatsugawa}, \citenamefont
			{Ohnishi}, \citenamefont {Sagawa}, \citenamefont {Takeda}, \citenamefont
			{Uchida}, \citenamefont {Yoshida}, \citenamefont {Yoshida},\ and\
			\citenamefont {Yosoi}}]{zenihiro2018directdeterminationneutronskin}%
		\BibitemOpen
		\bibfield  {author} {\bibinfo {author} {\bibfnamefont {J.}~\bibnamefont
				{Zenihiro}}, \bibinfo {author} {\bibfnamefont {H.}~\bibnamefont {Sakaguchi}},
			\bibinfo {author} {\bibfnamefont {S.}~\bibnamefont {Terashima}}, \bibinfo
			{author} {\bibfnamefont {T.}~\bibnamefont {Uesaka}}, \bibinfo {author}
			{\bibfnamefont {G.}~\bibnamefont {Hagen}}, \bibinfo {author} {\bibfnamefont
				{M.}~\bibnamefont {Itoh}}, \bibinfo {author} {\bibfnamefont {T.}~\bibnamefont
				{Murakami}}, \bibinfo {author} {\bibfnamefont {Y.}~\bibnamefont
				{Nakatsugawa}}, \bibinfo {author} {\bibfnamefont {T.}~\bibnamefont
				{Ohnishi}}, \bibinfo {author} {\bibfnamefont {H.}~\bibnamefont {Sagawa}},
			\bibinfo {author} {\bibfnamefont {H.}~\bibnamefont {Takeda}}, \bibinfo
			{author} {\bibfnamefont {M.}~\bibnamefont {Uchida}}, \bibinfo {author}
			{\bibfnamefont {H.~P.}\ \bibnamefont {Yoshida}}, \bibinfo {author}
			{\bibfnamefont {S.}~\bibnamefont {Yoshida}},\ and\ \bibinfo {author}
			{\bibfnamefont {M.}~\bibnamefont {Yosoi}},\ }\href
		{https://arxiv.org/abs/1810.11796} {\bibinfo {title} {Direct determination of
				the neutron skin thicknesses in $^{40,48}$ca from proton elastic scattering
				at $e_p = 295$ mev}} (\bibinfo {year} {2018}),\ \Eprint
		{https://arxiv.org/abs/1810.11796} {arXiv:1810.11796 [nucl-ex]} \BibitemShut
		{NoStop}%
		\bibitem [{\citenamefont {JASTRZEBSKI}\ \emph {et~al.}(2004)\citenamefont
			{JASTRZEBSKI}, \citenamefont {TRZCI\'{N}SKA}, \citenamefont {LUBI\'{N}SKI},
			\citenamefont {K\L{}OS}, \citenamefont {HARTMANN}, \citenamefont {von
				EGIDY},\ and\ \citenamefont {WYCECH}}]{doi:10.1142/S0218301304002168}%
		\BibitemOpen
		\bibfield  {author} {\bibinfo {author} {\bibfnamefont {J.}~\bibnamefont
				{JASTRZEBSKI}}, \bibinfo {author} {\bibfnamefont {A.}~\bibnamefont
				{TRZCI\'{N}SKA}}, \bibinfo {author} {\bibfnamefont {P.}~\bibnamefont
				{LUBI\'{N}SKI}}, \bibinfo {author} {\bibfnamefont {B.}~\bibnamefont
				{K\L{}OS}}, \bibinfo {author} {\bibfnamefont {F.~J.}\ \bibnamefont
				{HARTMANN}}, \bibinfo {author} {\bibfnamefont {T.}~\bibnamefont {von
					EGIDY}},\ and\ \bibinfo {author} {\bibfnamefont {S.}~\bibnamefont {WYCECH}},\
		}\bibfield  {title} {\bibinfo {title} {Neutron density distributions from
				antiprotonic atoms compared with hadron scattering data},\ }\href
		{https://doi.org/10.1142/S0218301304002168} {\bibfield  {journal} {\bibinfo
				{journal} {International Journal of Modern Physics E}\ }\textbf {\bibinfo
				{volume} {13}},\ \bibinfo {pages} {343} (\bibinfo {year} {2004})},\ \Eprint
		{https://arxiv.org/abs/https://doi.org/10.1142/S0218301304002168}
		{https://doi.org/10.1142/S0218301304002168} \BibitemShut {NoStop}%
		\bibitem [{\citenamefont {Gibbs}\ and\ \citenamefont
			{Dedonder}(1992)}]{PhysRevC.46.1825}%
		\BibitemOpen
		\bibfield  {author} {\bibinfo {author} {\bibfnamefont {W.~R.}\ \bibnamefont
				{Gibbs}}\ and\ \bibinfo {author} {\bibfnamefont {J.-P.}\ \bibnamefont
				{Dedonder}},\ }\bibfield  {title} {\bibinfo {title} {Neutron radii of the
				calcium isotopes},\ }\href {https://doi.org/10.1103/PhysRevC.46.1825}
		{\bibfield  {journal} {\bibinfo  {journal} {Phys. Rev. C}\ }\textbf {\bibinfo
				{volume} {46}},\ \bibinfo {pages} {1825} (\bibinfo {year}
			{1992})}\BibitemShut {NoStop}%
		\bibitem [{\citenamefont {Adhikari}\ \emph {et~al.}(2022)\citenamefont
			{Adhikari}, \citenamefont {Albataineh}, \citenamefont {Androic},
			\citenamefont {Aniol}, \citenamefont {Armstrong}, \citenamefont {Averett},
			\citenamefont {Ayerbe~Gayoso}, \citenamefont {Barcus}, \citenamefont
			{Bellini}, \citenamefont {Beminiwattha}, \citenamefont {Benesch},
			\citenamefont {Bhatt}, \citenamefont {Bhatta~Pathak}, \citenamefont
			{Bhetuwal}, \citenamefont {Blaikie}, \citenamefont {Boyd}, \citenamefont
			{Campagna}, \citenamefont {Camsonne}, \citenamefont {Cates}, \citenamefont
			{Chen}, \citenamefont {Clarke}, \citenamefont {Cornejo}, \citenamefont
			{Covrig~Dusa}, \citenamefont {Dalton}, \citenamefont {Datta}, \citenamefont
			{Deshpande}, \citenamefont {Dutta}, \citenamefont {Feldman}, \citenamefont
			{Fuchey}, \citenamefont {Gal}, \citenamefont {Gaskell}, \citenamefont
			{Gautam}, \citenamefont {Gericke}, \citenamefont {Ghosh}, \citenamefont
			{Halilovic}, \citenamefont {Hansen}, \citenamefont {Hassan}, \citenamefont
			{Hauenstein}, \citenamefont {Henry}, \citenamefont {Horowitz}, \citenamefont
			{Jantzi}, \citenamefont {Jian}, \citenamefont {Johnston}, \citenamefont
			{Jones}, \citenamefont {Kakkar}, \citenamefont {Katugampola}, \citenamefont
			{Keppel}, \citenamefont {King}, \citenamefont {King}, \citenamefont {Kumar},
			\citenamefont {Kutz}, \citenamefont {Lashley-Colthirst}, \citenamefont
			{Leverick}, \citenamefont {Liu}, \citenamefont {Liyanage}, \citenamefont
			{Mammei}, \citenamefont {Mammei}, \citenamefont {McCaughan}, \citenamefont
			{McNulty}, \citenamefont {Meekins}, \citenamefont {Metts}, \citenamefont
			{Michaels}, \citenamefont {Mihovilovic}, \citenamefont {Mondal},
			\citenamefont {Napolitano}, \citenamefont {Narayan}, \citenamefont
			{Nikolaev}, \citenamefont {Owen}, \citenamefont {Palatchi}, \citenamefont
			{Pan}, \citenamefont {Pandey}, \citenamefont {Park}, \citenamefont {Paschke},
			\citenamefont {Petrusky}, \citenamefont {Pitt}, \citenamefont {Premathilake},
			\citenamefont {Quinn}, \citenamefont {Radloff}, \citenamefont {Rahman},
			\citenamefont {Rashad}, \citenamefont {Rathnayake}, \citenamefont {Reed},
			\citenamefont {Reimer}, \citenamefont {Richards}, \citenamefont {Riordan},
			\citenamefont {Roblin}, \citenamefont {Seeds}, \citenamefont {Shahinyan},
			\citenamefont {Souder}, \citenamefont {Thiel}, \citenamefont {Tian},
			\citenamefont {Urciuoli}, \citenamefont {Wertz}, \citenamefont
			{Wojtsekhowski}, \citenamefont {Yale}, \citenamefont {Ye}, \citenamefont
			{Yoon}, \citenamefont {Xiong}, \citenamefont {Zec}, \citenamefont {Zhang},
			\citenamefont {Zhang},\ and\ \citenamefont {Zheng}}]{PhysRevLett.129.042501}%
		\BibitemOpen
		\bibfield  {author} {\bibinfo {author} {\bibfnamefont {D.}~\bibnamefont
				{Adhikari}}, \bibinfo {author} {\bibfnamefont {H.}~\bibnamefont
				{Albataineh}}, \bibinfo {author} {\bibfnamefont {D.}~\bibnamefont {Androic}},
			\bibinfo {author} {\bibfnamefont {K.~A.}\ \bibnamefont {Aniol}}, \bibinfo
			{author} {\bibfnamefont {D.~S.}\ \bibnamefont {Armstrong}}, \bibinfo {author}
			{\bibfnamefont {T.}~\bibnamefont {Averett}}, \bibinfo {author} {\bibfnamefont
				{C.}~\bibnamefont {Ayerbe~Gayoso}}, \bibinfo {author} {\bibfnamefont {S.~K.}\
				\bibnamefont {Barcus}}, \bibinfo {author} {\bibfnamefont {V.}~\bibnamefont
				{Bellini}}, \bibinfo {author} {\bibfnamefont {R.~S.}\ \bibnamefont
				{Beminiwattha}}, \bibinfo {author} {\bibfnamefont {J.~F.}\ \bibnamefont
				{Benesch}}, \bibinfo {author} {\bibfnamefont {H.}~\bibnamefont {Bhatt}},
			\bibinfo {author} {\bibfnamefont {D.}~\bibnamefont {Bhatta~Pathak}}, \bibinfo
			{author} {\bibfnamefont {D.}~\bibnamefont {Bhetuwal}}, \bibinfo {author}
			{\bibfnamefont {B.}~\bibnamefont {Blaikie}}, \bibinfo {author} {\bibfnamefont
				{J.}~\bibnamefont {Boyd}}, \bibinfo {author} {\bibfnamefont {Q.}~\bibnamefont
				{Campagna}}, \bibinfo {author} {\bibfnamefont {A.}~\bibnamefont {Camsonne}},
			\bibinfo {author} {\bibfnamefont {G.~D.}\ \bibnamefont {Cates}}, \bibinfo
			{author} {\bibfnamefont {Y.}~\bibnamefont {Chen}}, \bibinfo {author}
			{\bibfnamefont {C.}~\bibnamefont {Clarke}}, \bibinfo {author} {\bibfnamefont
				{J.~C.}\ \bibnamefont {Cornejo}}, \bibinfo {author} {\bibfnamefont
				{S.}~\bibnamefont {Covrig~Dusa}}, \bibinfo {author} {\bibfnamefont {M.~M.}\
				\bibnamefont {Dalton}}, \bibinfo {author} {\bibfnamefont {P.}~\bibnamefont
				{Datta}}, \bibinfo {author} {\bibfnamefont {A.}~\bibnamefont {Deshpande}},
			\bibinfo {author} {\bibfnamefont {D.}~\bibnamefont {Dutta}}, \bibinfo
			{author} {\bibfnamefont {C.}~\bibnamefont {Feldman}}, \bibinfo {author}
			{\bibfnamefont {E.}~\bibnamefont {Fuchey}}, \bibinfo {author} {\bibfnamefont
				{C.}~\bibnamefont {Gal}}, \bibinfo {author} {\bibfnamefont {D.}~\bibnamefont
				{Gaskell}}, \bibinfo {author} {\bibfnamefont {T.}~\bibnamefont {Gautam}},
			\bibinfo {author} {\bibfnamefont {M.}~\bibnamefont {Gericke}}, \bibinfo
			{author} {\bibfnamefont {C.}~\bibnamefont {Ghosh}}, \bibinfo {author}
			{\bibfnamefont {I.}~\bibnamefont {Halilovic}}, \bibinfo {author}
			{\bibfnamefont {J.-O.}\ \bibnamefont {Hansen}}, \bibinfo {author}
			{\bibfnamefont {O.}~\bibnamefont {Hassan}}, \bibinfo {author} {\bibfnamefont
				{F.}~\bibnamefont {Hauenstein}}, \bibinfo {author} {\bibfnamefont
				{W.}~\bibnamefont {Henry}}, \bibinfo {author} {\bibfnamefont {C.~J.}\
				\bibnamefont {Horowitz}}, \bibinfo {author} {\bibfnamefont {C.}~\bibnamefont
				{Jantzi}}, \bibinfo {author} {\bibfnamefont {S.}~\bibnamefont {Jian}},
			\bibinfo {author} {\bibfnamefont {S.}~\bibnamefont {Johnston}}, \bibinfo
			{author} {\bibfnamefont {D.~C.}\ \bibnamefont {Jones}}, \bibinfo {author}
			{\bibfnamefont {S.}~\bibnamefont {Kakkar}}, \bibinfo {author} {\bibfnamefont
				{S.}~\bibnamefont {Katugampola}}, \bibinfo {author} {\bibfnamefont
				{C.}~\bibnamefont {Keppel}}, \bibinfo {author} {\bibfnamefont {P.~M.}\
				\bibnamefont {King}}, \bibinfo {author} {\bibfnamefont {D.~E.}\ \bibnamefont
				{King}}, \bibinfo {author} {\bibfnamefont {K.~S.}\ \bibnamefont {Kumar}},
			\bibinfo {author} {\bibfnamefont {T.}~\bibnamefont {Kutz}}, \bibinfo {author}
			{\bibfnamefont {N.}~\bibnamefont {Lashley-Colthirst}}, \bibinfo {author}
			{\bibfnamefont {G.}~\bibnamefont {Leverick}}, \bibinfo {author}
			{\bibfnamefont {H.}~\bibnamefont {Liu}}, \bibinfo {author} {\bibfnamefont
				{N.}~\bibnamefont {Liyanage}}, \bibinfo {author} {\bibfnamefont
				{J.}~\bibnamefont {Mammei}}, \bibinfo {author} {\bibfnamefont
				{R.}~\bibnamefont {Mammei}}, \bibinfo {author} {\bibfnamefont
				{M.}~\bibnamefont {McCaughan}}, \bibinfo {author} {\bibfnamefont
				{D.}~\bibnamefont {McNulty}}, \bibinfo {author} {\bibfnamefont
				{D.}~\bibnamefont {Meekins}}, \bibinfo {author} {\bibfnamefont
				{C.}~\bibnamefont {Metts}}, \bibinfo {author} {\bibfnamefont
				{R.}~\bibnamefont {Michaels}}, \bibinfo {author} {\bibfnamefont
				{M.}~\bibnamefont {Mihovilovic}}, \bibinfo {author} {\bibfnamefont {M.~M.}\
				\bibnamefont {Mondal}}, \bibinfo {author} {\bibfnamefont {J.}~\bibnamefont
				{Napolitano}}, \bibinfo {author} {\bibfnamefont {A.}~\bibnamefont {Narayan}},
			\bibinfo {author} {\bibfnamefont {D.}~\bibnamefont {Nikolaev}}, \bibinfo
			{author} {\bibfnamefont {V.}~\bibnamefont {Owen}}, \bibinfo {author}
			{\bibfnamefont {C.}~\bibnamefont {Palatchi}}, \bibinfo {author}
			{\bibfnamefont {J.}~\bibnamefont {Pan}}, \bibinfo {author} {\bibfnamefont
				{B.}~\bibnamefont {Pandey}}, \bibinfo {author} {\bibfnamefont
				{S.}~\bibnamefont {Park}}, \bibinfo {author} {\bibfnamefont {K.~D.}\
				\bibnamefont {Paschke}}, \bibinfo {author} {\bibfnamefont {M.}~\bibnamefont
				{Petrusky}}, \bibinfo {author} {\bibfnamefont {M.~L.}\ \bibnamefont {Pitt}},
			\bibinfo {author} {\bibfnamefont {S.}~\bibnamefont {Premathilake}}, \bibinfo
			{author} {\bibfnamefont {B.}~\bibnamefont {Quinn}}, \bibinfo {author}
			{\bibfnamefont {R.}~\bibnamefont {Radloff}}, \bibinfo {author} {\bibfnamefont
				{S.}~\bibnamefont {Rahman}}, \bibinfo {author} {\bibfnamefont {M.~N.~H.}\
				\bibnamefont {Rashad}}, \bibinfo {author} {\bibfnamefont {A.}~\bibnamefont
				{Rathnayake}}, \bibinfo {author} {\bibfnamefont {B.~T.}\ \bibnamefont
				{Reed}}, \bibinfo {author} {\bibfnamefont {P.~E.}\ \bibnamefont {Reimer}},
			\bibinfo {author} {\bibfnamefont {R.}~\bibnamefont {Richards}}, \bibinfo
			{author} {\bibfnamefont {S.}~\bibnamefont {Riordan}}, \bibinfo {author}
			{\bibfnamefont {Y.~R.}\ \bibnamefont {Roblin}}, \bibinfo {author}
			{\bibfnamefont {S.}~\bibnamefont {Seeds}}, \bibinfo {author} {\bibfnamefont
				{A.}~\bibnamefont {Shahinyan}}, \bibinfo {author} {\bibfnamefont
				{P.}~\bibnamefont {Souder}}, \bibinfo {author} {\bibfnamefont
				{M.}~\bibnamefont {Thiel}}, \bibinfo {author} {\bibfnamefont
				{Y.}~\bibnamefont {Tian}}, \bibinfo {author} {\bibfnamefont {G.~M.}\
				\bibnamefont {Urciuoli}}, \bibinfo {author} {\bibfnamefont {E.~W.}\
				\bibnamefont {Wertz}}, \bibinfo {author} {\bibfnamefont {B.}~\bibnamefont
				{Wojtsekhowski}}, \bibinfo {author} {\bibfnamefont {B.}~\bibnamefont {Yale}},
			\bibinfo {author} {\bibfnamefont {T.}~\bibnamefont {Ye}}, \bibinfo {author}
			{\bibfnamefont {A.}~\bibnamefont {Yoon}}, \bibinfo {author} {\bibfnamefont
				{W.}~\bibnamefont {Xiong}}, \bibinfo {author} {\bibfnamefont
				{A.}~\bibnamefont {Zec}}, \bibinfo {author} {\bibfnamefont {W.}~\bibnamefont
				{Zhang}}, \bibinfo {author} {\bibfnamefont {J.}~\bibnamefont {Zhang}},\ and\
			\bibinfo {author} {\bibfnamefont {X.}~\bibnamefont {Zheng}} (\bibinfo
			{collaboration} {CREX Collaboration}),\ }\bibfield  {title} {\bibinfo {title}
			{Precision determination of the neutral weak form factor of
				$^{48}\mathrm{Ca}$},\ }\href {https://doi.org/10.1103/PhysRevLett.129.042501}
		{\bibfield  {journal} {\bibinfo  {journal} {Phys. Rev. Lett.}\ }\textbf
			{\bibinfo {volume} {129}},\ \bibinfo {pages} {042501} (\bibinfo {year}
			{2022})}\BibitemShut {NoStop}%
		\bibitem [{\citenamefont {Giacalone}\ \emph {et~al.}(2023)\citenamefont
			{Giacalone}, \citenamefont {Nijs},\ and\ \citenamefont {van~der
				Schee}}]{PhysRevLett.131.202302}%
		\BibitemOpen
		\bibfield  {author} {\bibinfo {author} {\bibfnamefont {G.}~\bibnamefont
				{Giacalone}}, \bibinfo {author} {\bibfnamefont {G.}~\bibnamefont {Nijs}},\
			and\ \bibinfo {author} {\bibfnamefont {W.}~\bibnamefont {van~der Schee}},\
		}\bibfield  {title} {\bibinfo {title} {Determination of the neutron skin of
				$^{208}\mathrm{Pb}$ from ultrarelativistic nuclear collisions},\ }\href
		{https://doi.org/10.1103/PhysRevLett.131.202302} {\bibfield  {journal}
			{\bibinfo  {journal} {Phys. Rev. Lett.}\ }\textbf {\bibinfo {volume} {131}},\
			\bibinfo {pages} {202302} (\bibinfo {year} {2023})}\BibitemShut {NoStop}%
		\bibitem [{\citenamefont {Adhikari}\ \emph {et~al.}(2021)\citenamefont
			{Adhikari}, \citenamefont {Albataineh}, \citenamefont {Androic},
			\citenamefont {Aniol}, \citenamefont {Armstrong}, \citenamefont {Averett},
			\citenamefont {Ayerbe~Gayoso}, \citenamefont {Barcus}, \citenamefont
			{Bellini}, \citenamefont {Beminiwattha}, \citenamefont {Benesch},
			\citenamefont {Bhatt}, \citenamefont {Bhatta~Pathak}, \citenamefont
			{Bhetuwal}, \citenamefont {Blaikie}, \citenamefont {Campagna}, \citenamefont
			{Camsonne}, \citenamefont {Cates}, \citenamefont {Chen}, \citenamefont
			{Clarke}, \citenamefont {Cornejo}, \citenamefont {Covrig~Dusa}, \citenamefont
			{Datta}, \citenamefont {Deshpande}, \citenamefont {Dutta}, \citenamefont
			{Feldman}, \citenamefont {Fuchey}, \citenamefont {Gal}, \citenamefont
			{Gaskell}, \citenamefont {Gautam}, \citenamefont {Gericke}, \citenamefont
			{Ghosh}, \citenamefont {Halilovic}, \citenamefont {Hansen}, \citenamefont
			{Hauenstein}, \citenamefont {Henry}, \citenamefont {Horowitz}, \citenamefont
			{Jantzi}, \citenamefont {Jian}, \citenamefont {Johnston}, \citenamefont
			{Jones}, \citenamefont {Karki}, \citenamefont {Katugampola}, \citenamefont
			{Keppel}, \citenamefont {King}, \citenamefont {King}, \citenamefont {Knauss},
			\citenamefont {Kumar}, \citenamefont {Kutz}, \citenamefont
			{Lashley-Colthirst}, \citenamefont {Leverick}, \citenamefont {Liu},
			\citenamefont {Liyange}, \citenamefont {Malace}, \citenamefont {Mammei},
			\citenamefont {Mammei}, \citenamefont {McCaughan}, \citenamefont {McNulty},
			\citenamefont {Meekins}, \citenamefont {Metts}, \citenamefont {Michaels},
			\citenamefont {Mondal}, \citenamefont {Napolitano}, \citenamefont {Narayan},
			\citenamefont {Nikolaev}, \citenamefont {Rashad}, \citenamefont {Owen},
			\citenamefont {Palatchi}, \citenamefont {Pan}, \citenamefont {Pandey},
			\citenamefont {Park}, \citenamefont {Paschke}, \citenamefont {Petrusky},
			\citenamefont {Pitt}, \citenamefont {Premathilake}, \citenamefont {Puckett},
			\citenamefont {Quinn}, \citenamefont {Radloff}, \citenamefont {Rahman},
			\citenamefont {Rathnayake}, \citenamefont {Reed}, \citenamefont {Reimer},
			\citenamefont {Richards}, \citenamefont {Riordan}, \citenamefont {Roblin},
			\citenamefont {Seeds}, \citenamefont {Shahinyan}, \citenamefont {Souder},
			\citenamefont {Tang}, \citenamefont {Thiel}, \citenamefont {Tian},
			\citenamefont {Urciuoli}, \citenamefont {Wertz}, \citenamefont
			{Wojtsekhowski}, \citenamefont {Yale}, \citenamefont {Ye}, \citenamefont
			{Yoon}, \citenamefont {Zec}, \citenamefont {Zhang}, \citenamefont {Zhang},\
			and\ \citenamefont {Zheng}}]{PhysRevLett.126.172502}%
		\BibitemOpen
		\bibfield  {author} {\bibinfo {author} {\bibfnamefont {D.}~\bibnamefont
				{Adhikari}}, \bibinfo {author} {\bibfnamefont {H.}~\bibnamefont
				{Albataineh}}, \bibinfo {author} {\bibfnamefont {D.}~\bibnamefont {Androic}},
			\bibinfo {author} {\bibfnamefont {K.}~\bibnamefont {Aniol}}, \bibinfo
			{author} {\bibfnamefont {D.~S.}\ \bibnamefont {Armstrong}}, \bibinfo {author}
			{\bibfnamefont {T.}~\bibnamefont {Averett}}, \bibinfo {author} {\bibfnamefont
				{C.}~\bibnamefont {Ayerbe~Gayoso}}, \bibinfo {author} {\bibfnamefont
				{S.}~\bibnamefont {Barcus}}, \bibinfo {author} {\bibfnamefont
				{V.}~\bibnamefont {Bellini}}, \bibinfo {author} {\bibfnamefont {R.~S.}\
				\bibnamefont {Beminiwattha}}, \bibinfo {author} {\bibfnamefont {J.~F.}\
				\bibnamefont {Benesch}}, \bibinfo {author} {\bibfnamefont {H.}~\bibnamefont
				{Bhatt}}, \bibinfo {author} {\bibfnamefont {D.}~\bibnamefont
				{Bhatta~Pathak}}, \bibinfo {author} {\bibfnamefont {D.}~\bibnamefont
				{Bhetuwal}}, \bibinfo {author} {\bibfnamefont {B.}~\bibnamefont {Blaikie}},
			\bibinfo {author} {\bibfnamefont {Q.}~\bibnamefont {Campagna}}, \bibinfo
			{author} {\bibfnamefont {A.}~\bibnamefont {Camsonne}}, \bibinfo {author}
			{\bibfnamefont {G.~D.}\ \bibnamefont {Cates}}, \bibinfo {author}
			{\bibfnamefont {Y.}~\bibnamefont {Chen}}, \bibinfo {author} {\bibfnamefont
				{C.}~\bibnamefont {Clarke}}, \bibinfo {author} {\bibfnamefont {J.~C.}\
				\bibnamefont {Cornejo}}, \bibinfo {author} {\bibfnamefont {S.}~\bibnamefont
				{Covrig~Dusa}}, \bibinfo {author} {\bibfnamefont {P.}~\bibnamefont {Datta}},
			\bibinfo {author} {\bibfnamefont {A.}~\bibnamefont {Deshpande}}, \bibinfo
			{author} {\bibfnamefont {D.}~\bibnamefont {Dutta}}, \bibinfo {author}
			{\bibfnamefont {C.}~\bibnamefont {Feldman}}, \bibinfo {author} {\bibfnamefont
				{E.}~\bibnamefont {Fuchey}}, \bibinfo {author} {\bibfnamefont
				{C.}~\bibnamefont {Gal}}, \bibinfo {author} {\bibfnamefont {D.}~\bibnamefont
				{Gaskell}}, \bibinfo {author} {\bibfnamefont {T.}~\bibnamefont {Gautam}},
			\bibinfo {author} {\bibfnamefont {M.}~\bibnamefont {Gericke}}, \bibinfo
			{author} {\bibfnamefont {C.}~\bibnamefont {Ghosh}}, \bibinfo {author}
			{\bibfnamefont {I.}~\bibnamefont {Halilovic}}, \bibinfo {author}
			{\bibfnamefont {J.-O.}\ \bibnamefont {Hansen}}, \bibinfo {author}
			{\bibfnamefont {F.}~\bibnamefont {Hauenstein}}, \bibinfo {author}
			{\bibfnamefont {W.}~\bibnamefont {Henry}}, \bibinfo {author} {\bibfnamefont
				{C.~J.}\ \bibnamefont {Horowitz}}, \bibinfo {author} {\bibfnamefont
				{C.}~\bibnamefont {Jantzi}}, \bibinfo {author} {\bibfnamefont
				{S.}~\bibnamefont {Jian}}, \bibinfo {author} {\bibfnamefont {S.}~\bibnamefont
				{Johnston}}, \bibinfo {author} {\bibfnamefont {D.~C.}\ \bibnamefont {Jones}},
			\bibinfo {author} {\bibfnamefont {B.}~\bibnamefont {Karki}}, \bibinfo
			{author} {\bibfnamefont {S.}~\bibnamefont {Katugampola}}, \bibinfo {author}
			{\bibfnamefont {C.}~\bibnamefont {Keppel}}, \bibinfo {author} {\bibfnamefont
				{P.~M.}\ \bibnamefont {King}}, \bibinfo {author} {\bibfnamefont {D.~E.}\
				\bibnamefont {King}}, \bibinfo {author} {\bibfnamefont {M.}~\bibnamefont
				{Knauss}}, \bibinfo {author} {\bibfnamefont {K.~S.}\ \bibnamefont {Kumar}},
			\bibinfo {author} {\bibfnamefont {T.}~\bibnamefont {Kutz}}, \bibinfo {author}
			{\bibfnamefont {N.}~\bibnamefont {Lashley-Colthirst}}, \bibinfo {author}
			{\bibfnamefont {G.}~\bibnamefont {Leverick}}, \bibinfo {author}
			{\bibfnamefont {H.}~\bibnamefont {Liu}}, \bibinfo {author} {\bibfnamefont
				{N.}~\bibnamefont {Liyange}}, \bibinfo {author} {\bibfnamefont
				{S.}~\bibnamefont {Malace}}, \bibinfo {author} {\bibfnamefont
				{R.}~\bibnamefont {Mammei}}, \bibinfo {author} {\bibfnamefont
				{J.}~\bibnamefont {Mammei}}, \bibinfo {author} {\bibfnamefont
				{M.}~\bibnamefont {McCaughan}}, \bibinfo {author} {\bibfnamefont
				{D.}~\bibnamefont {McNulty}}, \bibinfo {author} {\bibfnamefont
				{D.}~\bibnamefont {Meekins}}, \bibinfo {author} {\bibfnamefont
				{C.}~\bibnamefont {Metts}}, \bibinfo {author} {\bibfnamefont
				{R.}~\bibnamefont {Michaels}}, \bibinfo {author} {\bibfnamefont {M.~M.}\
				\bibnamefont {Mondal}}, \bibinfo {author} {\bibfnamefont {J.}~\bibnamefont
				{Napolitano}}, \bibinfo {author} {\bibfnamefont {A.}~\bibnamefont {Narayan}},
			\bibinfo {author} {\bibfnamefont {D.}~\bibnamefont {Nikolaev}}, \bibinfo
			{author} {\bibfnamefont {M.~N.~H.}\ \bibnamefont {Rashad}}, \bibinfo {author}
			{\bibfnamefont {V.}~\bibnamefont {Owen}}, \bibinfo {author} {\bibfnamefont
				{C.}~\bibnamefont {Palatchi}}, \bibinfo {author} {\bibfnamefont
				{J.}~\bibnamefont {Pan}}, \bibinfo {author} {\bibfnamefont {B.}~\bibnamefont
				{Pandey}}, \bibinfo {author} {\bibfnamefont {S.}~\bibnamefont {Park}},
			\bibinfo {author} {\bibfnamefont {K.~D.}\ \bibnamefont {Paschke}}, \bibinfo
			{author} {\bibfnamefont {M.}~\bibnamefont {Petrusky}}, \bibinfo {author}
			{\bibfnamefont {M.~L.}\ \bibnamefont {Pitt}}, \bibinfo {author}
			{\bibfnamefont {S.}~\bibnamefont {Premathilake}}, \bibinfo {author}
			{\bibfnamefont {A.~J.~R.}\ \bibnamefont {Puckett}}, \bibinfo {author}
			{\bibfnamefont {B.}~\bibnamefont {Quinn}}, \bibinfo {author} {\bibfnamefont
				{R.}~\bibnamefont {Radloff}}, \bibinfo {author} {\bibfnamefont
				{S.}~\bibnamefont {Rahman}}, \bibinfo {author} {\bibfnamefont
				{A.}~\bibnamefont {Rathnayake}}, \bibinfo {author} {\bibfnamefont {B.~T.}\
				\bibnamefont {Reed}}, \bibinfo {author} {\bibfnamefont {P.~E.}\ \bibnamefont
				{Reimer}}, \bibinfo {author} {\bibfnamefont {R.}~\bibnamefont {Richards}},
			\bibinfo {author} {\bibfnamefont {S.}~\bibnamefont {Riordan}}, \bibinfo
			{author} {\bibfnamefont {Y.}~\bibnamefont {Roblin}}, \bibinfo {author}
			{\bibfnamefont {S.}~\bibnamefont {Seeds}}, \bibinfo {author} {\bibfnamefont
				{A.}~\bibnamefont {Shahinyan}}, \bibinfo {author} {\bibfnamefont
				{P.}~\bibnamefont {Souder}}, \bibinfo {author} {\bibfnamefont
				{L.}~\bibnamefont {Tang}}, \bibinfo {author} {\bibfnamefont {M.}~\bibnamefont
				{Thiel}}, \bibinfo {author} {\bibfnamefont {Y.}~\bibnamefont {Tian}},
			\bibinfo {author} {\bibfnamefont {G.~M.}\ \bibnamefont {Urciuoli}}, \bibinfo
			{author} {\bibfnamefont {E.~W.}\ \bibnamefont {Wertz}}, \bibinfo {author}
			{\bibfnamefont {B.}~\bibnamefont {Wojtsekhowski}}, \bibinfo {author}
			{\bibfnamefont {B.}~\bibnamefont {Yale}}, \bibinfo {author} {\bibfnamefont
				{T.}~\bibnamefont {Ye}}, \bibinfo {author} {\bibfnamefont {A.}~\bibnamefont
				{Yoon}}, \bibinfo {author} {\bibfnamefont {A.}~\bibnamefont {Zec}}, \bibinfo
			{author} {\bibfnamefont {W.}~\bibnamefont {Zhang}}, \bibinfo {author}
			{\bibfnamefont {J.}~\bibnamefont {Zhang}},\ and\ \bibinfo {author}
			{\bibfnamefont {X.}~\bibnamefont {Zheng}} (\bibinfo {collaboration} {PREX
				Collaboration}),\ }\bibfield  {title} {\bibinfo {title} {Accurate
				determination of the neutron skin thickness of $^{208}\mathrm{Pb}$ through
				parity-violation in electron scattering},\ }\href
		{https://doi.org/10.1103/PhysRevLett.126.172502} {\bibfield  {journal}
			{\bibinfo  {journal} {Phys. Rev. Lett.}\ }\textbf {\bibinfo {volume} {126}},\
			\bibinfo {pages} {172502} (\bibinfo {year} {2021})}\BibitemShut {NoStop}%
		\bibitem [{\citenamefont {Alexandrovich}\ \emph {et~al.}(1994)\citenamefont
			{Alexandrovich}, \citenamefont {Gagarski}, \citenamefont {Krasnoschekova},
			\citenamefont {Petrov}, \citenamefont {Petrova}, \citenamefont {Petukhov},
			\citenamefont {Pleva}, \citenamefont {Geltenbort}, \citenamefont {Last},\
			and\ \citenamefont {Schreckenbach}}]{npa567.521}%
		\BibitemOpen
		\bibfield  {author} {\bibinfo {author} {\bibfnamefont {A.}~\bibnamefont
				{Alexandrovich}}, \bibinfo {author} {\bibfnamefont {A.}~\bibnamefont
				{Gagarski}}, \bibinfo {author} {\bibfnamefont {I.}~\bibnamefont
				{Krasnoschekova}}, \bibinfo {author} {\bibfnamefont {G.}~\bibnamefont
				{Petrov}}, \bibinfo {author} {\bibfnamefont {V.}~\bibnamefont {Petrova}},
			\bibinfo {author} {\bibfnamefont {A.}~\bibnamefont {Petukhov}}, \bibinfo
			{author} {\bibfnamefont {Y.}~\bibnamefont {Pleva}}, \bibinfo {author}
			{\bibfnamefont {P.}~\bibnamefont {Geltenbort}}, \bibinfo {author}
			{\bibfnamefont {J.}~\bibnamefont {Last}},\ and\ \bibinfo {author}
			{\bibfnamefont {K.}~\bibnamefont {Schreckenbach}},\ }\bibfield  {title}
		{\bibinfo {title} {New observation of space-parity violation in
				neutron-induced fission of 229th, 241pu and 241am},\ }\href
		{https://doi.org/https://doi.org/10.1016/0375-9474(94)90023-X} {\bibfield
			{journal} {\bibinfo  {journal} {Nuclear Physics A}\ }\textbf {\bibinfo
				{volume} {567}},\ \bibinfo {pages} {541} (\bibinfo {year}
			{1994})}\BibitemShut {NoStop}%
		\bibitem [{\citenamefont {Ray}(1979)}]{PhysRevC.19.1855}%
		\BibitemOpen
		\bibfield  {author} {\bibinfo {author} {\bibfnamefont {L.}~\bibnamefont
				{Ray}},\ }\bibfield  {title} {\bibinfo {title} {Neutron isotopic density
				differences deduced from 0.8 gev polarized proton elastic scattering},\
		}\href {https://doi.org/10.1103/PhysRevC.19.1855} {\bibfield  {journal}
			{\bibinfo  {journal} {Phys. Rev. C}\ }\textbf {\bibinfo {volume} {19}},\
			\bibinfo {pages} {1855} (\bibinfo {year} {1979})}\BibitemShut {NoStop}%
		\bibitem [{\citenamefont {Hoffmann}\ \emph {et~al.}(1980)\citenamefont
			{Hoffmann}, \citenamefont {Ray}, \citenamefont {Barlett}, \citenamefont
			{McGill}, \citenamefont {Adams}, \citenamefont {Igo}, \citenamefont {Irom},
			\citenamefont {Wang}, \citenamefont {Whitten}, \citenamefont {Boudrie},
			\citenamefont {Amann}, \citenamefont {Glashausser}, \citenamefont {Hintz},
			\citenamefont {Kyle},\ and\ \citenamefont {Blanpied}}]{PhysRevC.21.1488}%
		\BibitemOpen
		\bibfield  {author} {\bibinfo {author} {\bibfnamefont {G.~W.}\ \bibnamefont
				{Hoffmann}}, \bibinfo {author} {\bibfnamefont {L.}~\bibnamefont {Ray}},
			\bibinfo {author} {\bibfnamefont {M.}~\bibnamefont {Barlett}}, \bibinfo
			{author} {\bibfnamefont {J.}~\bibnamefont {McGill}}, \bibinfo {author}
			{\bibfnamefont {G.~S.}\ \bibnamefont {Adams}}, \bibinfo {author}
			{\bibfnamefont {G.~J.}\ \bibnamefont {Igo}}, \bibinfo {author} {\bibfnamefont
				{F.}~\bibnamefont {Irom}}, \bibinfo {author} {\bibfnamefont {A.~T.~M.}\
				\bibnamefont {Wang}}, \bibinfo {author} {\bibfnamefont {C.~A.}\ \bibnamefont
				{Whitten}}, \bibinfo {author} {\bibfnamefont {R.~L.}\ \bibnamefont
				{Boudrie}}, \bibinfo {author} {\bibfnamefont {J.~F.}\ \bibnamefont {Amann}},
			\bibinfo {author} {\bibfnamefont {C.}~\bibnamefont {Glashausser}}, \bibinfo
			{author} {\bibfnamefont {N.~M.}\ \bibnamefont {Hintz}}, \bibinfo {author}
			{\bibfnamefont {G.~S.}\ \bibnamefont {Kyle}},\ and\ \bibinfo {author}
			{\bibfnamefont {G.~S.}\ \bibnamefont {Blanpied}},\ }\bibfield  {title}
		{\bibinfo {title} {0.8 gev $p$+$^{208}\mathrm{Pb}$ elastic scattering and the
				quantity $\ensuremath{\Delta}{r}_{\mathrm{np}}$},\ }\href
		{https://doi.org/10.1103/PhysRevC.21.1488} {\bibfield  {journal} {\bibinfo
				{journal} {Phys. Rev. C}\ }\textbf {\bibinfo {volume} {21}},\ \bibinfo
			{pages} {1488} (\bibinfo {year} {1980})}\BibitemShut {NoStop}%
		\bibitem [{\citenamefont {Abrahamyan}\ \emph {et~al.}(2012)\citenamefont
			{Abrahamyan}, \citenamefont {Ahmed}, \citenamefont {Albataineh},
			\citenamefont {Aniol}, \citenamefont {Armstrong}, \citenamefont {Armstrong},
			\citenamefont {Averett}, \citenamefont {Babineau}, \citenamefont {Barbieri},
			\citenamefont {Bellini}, \citenamefont {Beminiwattha}, \citenamefont
			{Benesch}, \citenamefont {Benmokhtar}, \citenamefont {Bielarski},
			\citenamefont {Boeglin}, \citenamefont {Camsonne}, \citenamefont {Canan},
			\citenamefont {Carter}, \citenamefont {Cates}, \citenamefont {Chen},
			\citenamefont {Chen}, \citenamefont {Hen}, \citenamefont {Cusanno},
			\citenamefont {Dalton}, \citenamefont {De~Leo}, \citenamefont {de~Jager},
			\citenamefont {Deconinck}, \citenamefont {Decowski}, \citenamefont {Deng},
			\citenamefont {Deur}, \citenamefont {Dutta}, \citenamefont {Etile},
			\citenamefont {Flay}, \citenamefont {Franklin}, \citenamefont {Friend},
			\citenamefont {Frullani}, \citenamefont {Fuchey}, \citenamefont {Garibaldi},
			\citenamefont {Gasser}, \citenamefont {Gilman}, \citenamefont {Giusa},
			\citenamefont {Glamazdin}, \citenamefont {Gomez}, \citenamefont {Grames},
			\citenamefont {Gu}, \citenamefont {Hansen}, \citenamefont {Hansknecht},
			\citenamefont {Higinbotham}, \citenamefont {Holmes}, \citenamefont
			{Holmstrom}, \citenamefont {Horowitz}, \citenamefont {Hoskins}, \citenamefont
			{Huang}, \citenamefont {Hyde}, \citenamefont {Itard}, \citenamefont {Jen},
			\citenamefont {Jensen}, \citenamefont {Jin}, \citenamefont {Johnston},
			\citenamefont {Kelleher}, \citenamefont {Kliakhandler}, \citenamefont {King},
			\citenamefont {Kowalski}, \citenamefont {Kumar}, \citenamefont {Leacock},
			\citenamefont {Leckey}, \citenamefont {Lee}, \citenamefont {LeRose},
			\citenamefont {Lindgren}, \citenamefont {Liyanage}, \citenamefont {Lubinsky},
			\citenamefont {Mammei}, \citenamefont {Mammoliti}, \citenamefont
			{Margaziotis}, \citenamefont {Markowitz}, \citenamefont {McCreary},
			\citenamefont {McNulty}, \citenamefont {Mercado}, \citenamefont {Meziani},
			\citenamefont {Michaels}, \citenamefont {Mihovilovic}, \citenamefont
			{Muangma}, \citenamefont {Mu\~noz Camacho}, \citenamefont {Nanda},
			\citenamefont {Nelyubin}, \citenamefont {Nuruzzaman}, \citenamefont {Oh},
			\citenamefont {Palmer}, \citenamefont {Parno}, \citenamefont {Paschke},
			\citenamefont {Phillips}, \citenamefont {Poelker}, \citenamefont
			{Pomatsalyuk}, \citenamefont {Posik}, \citenamefont {Puckett}, \citenamefont
			{Quinn}, \citenamefont {Rakhman}, \citenamefont {Reimer}, \citenamefont
			{Riordan}, \citenamefont {Rogan}, \citenamefont {Ron}, \citenamefont {Russo},
			\citenamefont {Saenboonruang}, \citenamefont {Saha}, \citenamefont
			{Sawatzky}, \citenamefont {Shahinyan}, \citenamefont {Silwal}, \citenamefont
			{Sirca}, \citenamefont {Slifer}, \citenamefont {Solvignon}, \citenamefont
			{Souder}, \citenamefont {Sperduto}, \citenamefont {Subedi}, \citenamefont
			{Suleiman}, \citenamefont {Sulkosky}, \citenamefont {Sutera}, \citenamefont
			{Tobias}, \citenamefont {Troth}, \citenamefont {Urciuoli}, \citenamefont
			{Waidyawansa}, \citenamefont {Wang}, \citenamefont {Wexler}, \citenamefont
			{Wilson}, \citenamefont {Wojtsekhowski}, \citenamefont {Yan}, \citenamefont
			{Yao}, \citenamefont {Ye}, \citenamefont {Ye}, \citenamefont {Yim},
			\citenamefont {Zana}, \citenamefont {Zhan}, \citenamefont {Zhang},
			\citenamefont {Zhang}, \citenamefont {Zheng},\ and\ \citenamefont
			{Zhu}}]{PhysRevLett.108.112502}%
		\BibitemOpen
		\bibfield  {author} {\bibinfo {author} {\bibfnamefont {S.}~\bibnamefont
				{Abrahamyan}}, \bibinfo {author} {\bibfnamefont {Z.}~\bibnamefont {Ahmed}},
			\bibinfo {author} {\bibfnamefont {H.}~\bibnamefont {Albataineh}}, \bibinfo
			{author} {\bibfnamefont {K.}~\bibnamefont {Aniol}}, \bibinfo {author}
			{\bibfnamefont {D.~S.}\ \bibnamefont {Armstrong}}, \bibinfo {author}
			{\bibfnamefont {W.}~\bibnamefont {Armstrong}}, \bibinfo {author}
			{\bibfnamefont {T.}~\bibnamefont {Averett}}, \bibinfo {author} {\bibfnamefont
				{B.}~\bibnamefont {Babineau}}, \bibinfo {author} {\bibfnamefont
				{A.}~\bibnamefont {Barbieri}}, \bibinfo {author} {\bibfnamefont
				{V.}~\bibnamefont {Bellini}}, \bibinfo {author} {\bibfnamefont
				{R.}~\bibnamefont {Beminiwattha}}, \bibinfo {author} {\bibfnamefont
				{J.}~\bibnamefont {Benesch}}, \bibinfo {author} {\bibfnamefont
				{F.}~\bibnamefont {Benmokhtar}}, \bibinfo {author} {\bibfnamefont
				{T.}~\bibnamefont {Bielarski}}, \bibinfo {author} {\bibfnamefont
				{W.}~\bibnamefont {Boeglin}}, \bibinfo {author} {\bibfnamefont
				{A.}~\bibnamefont {Camsonne}}, \bibinfo {author} {\bibfnamefont
				{M.}~\bibnamefont {Canan}}, \bibinfo {author} {\bibfnamefont
				{P.}~\bibnamefont {Carter}}, \bibinfo {author} {\bibfnamefont {G.~D.}\
				\bibnamefont {Cates}}, \bibinfo {author} {\bibfnamefont {C.}~\bibnamefont
				{Chen}}, \bibinfo {author} {\bibfnamefont {J.-P.}\ \bibnamefont {Chen}},
			\bibinfo {author} {\bibfnamefont {O.}~\bibnamefont {Hen}}, \bibinfo {author}
			{\bibfnamefont {F.}~\bibnamefont {Cusanno}}, \bibinfo {author} {\bibfnamefont
				{M.~M.}\ \bibnamefont {Dalton}}, \bibinfo {author} {\bibfnamefont
				{R.}~\bibnamefont {De~Leo}}, \bibinfo {author} {\bibfnamefont
				{K.}~\bibnamefont {de~Jager}}, \bibinfo {author} {\bibfnamefont
				{W.}~\bibnamefont {Deconinck}}, \bibinfo {author} {\bibfnamefont
				{P.}~\bibnamefont {Decowski}}, \bibinfo {author} {\bibfnamefont
				{X.}~\bibnamefont {Deng}}, \bibinfo {author} {\bibfnamefont {A.}~\bibnamefont
				{Deur}}, \bibinfo {author} {\bibfnamefont {D.}~\bibnamefont {Dutta}},
			\bibinfo {author} {\bibfnamefont {A.}~\bibnamefont {Etile}}, \bibinfo
			{author} {\bibfnamefont {D.}~\bibnamefont {Flay}}, \bibinfo {author}
			{\bibfnamefont {G.~B.}\ \bibnamefont {Franklin}}, \bibinfo {author}
			{\bibfnamefont {M.}~\bibnamefont {Friend}}, \bibinfo {author} {\bibfnamefont
				{S.}~\bibnamefont {Frullani}}, \bibinfo {author} {\bibfnamefont
				{E.}~\bibnamefont {Fuchey}}, \bibinfo {author} {\bibfnamefont
				{F.}~\bibnamefont {Garibaldi}}, \bibinfo {author} {\bibfnamefont
				{E.}~\bibnamefont {Gasser}}, \bibinfo {author} {\bibfnamefont
				{R.}~\bibnamefont {Gilman}}, \bibinfo {author} {\bibfnamefont
				{A.}~\bibnamefont {Giusa}}, \bibinfo {author} {\bibfnamefont
				{A.}~\bibnamefont {Glamazdin}}, \bibinfo {author} {\bibfnamefont
				{J.}~\bibnamefont {Gomez}}, \bibinfo {author} {\bibfnamefont
				{J.}~\bibnamefont {Grames}}, \bibinfo {author} {\bibfnamefont
				{C.}~\bibnamefont {Gu}}, \bibinfo {author} {\bibfnamefont {O.}~\bibnamefont
				{Hansen}}, \bibinfo {author} {\bibfnamefont {J.}~\bibnamefont {Hansknecht}},
			\bibinfo {author} {\bibfnamefont {D.~W.}\ \bibnamefont {Higinbotham}},
			\bibinfo {author} {\bibfnamefont {R.~S.}\ \bibnamefont {Holmes}}, \bibinfo
			{author} {\bibfnamefont {T.}~\bibnamefont {Holmstrom}}, \bibinfo {author}
			{\bibfnamefont {C.~J.}\ \bibnamefont {Horowitz}}, \bibinfo {author}
			{\bibfnamefont {J.}~\bibnamefont {Hoskins}}, \bibinfo {author} {\bibfnamefont
				{J.}~\bibnamefont {Huang}}, \bibinfo {author} {\bibfnamefont {C.~E.}\
				\bibnamefont {Hyde}}, \bibinfo {author} {\bibfnamefont {F.}~\bibnamefont
				{Itard}}, \bibinfo {author} {\bibfnamefont {C.-M.}\ \bibnamefont {Jen}},
			\bibinfo {author} {\bibfnamefont {E.}~\bibnamefont {Jensen}}, \bibinfo
			{author} {\bibfnamefont {G.}~\bibnamefont {Jin}}, \bibinfo {author}
			{\bibfnamefont {S.}~\bibnamefont {Johnston}}, \bibinfo {author}
			{\bibfnamefont {A.}~\bibnamefont {Kelleher}}, \bibinfo {author}
			{\bibfnamefont {K.}~\bibnamefont {Kliakhandler}}, \bibinfo {author}
			{\bibfnamefont {P.~M.}\ \bibnamefont {King}}, \bibinfo {author}
			{\bibfnamefont {S.}~\bibnamefont {Kowalski}}, \bibinfo {author}
			{\bibfnamefont {K.~S.}\ \bibnamefont {Kumar}}, \bibinfo {author}
			{\bibfnamefont {J.}~\bibnamefont {Leacock}}, \bibinfo {author} {\bibfnamefont
				{J.}~\bibnamefont {Leckey}}, \bibinfo {author} {\bibfnamefont {J.~H.}\
				\bibnamefont {Lee}}, \bibinfo {author} {\bibfnamefont {J.~J.}\ \bibnamefont
				{LeRose}}, \bibinfo {author} {\bibfnamefont {R.}~\bibnamefont {Lindgren}},
			\bibinfo {author} {\bibfnamefont {N.}~\bibnamefont {Liyanage}}, \bibinfo
			{author} {\bibfnamefont {N.}~\bibnamefont {Lubinsky}}, \bibinfo {author}
			{\bibfnamefont {J.}~\bibnamefont {Mammei}}, \bibinfo {author} {\bibfnamefont
				{F.}~\bibnamefont {Mammoliti}}, \bibinfo {author} {\bibfnamefont {D.~J.}\
				\bibnamefont {Margaziotis}}, \bibinfo {author} {\bibfnamefont
				{P.}~\bibnamefont {Markowitz}}, \bibinfo {author} {\bibfnamefont
				{A.}~\bibnamefont {McCreary}}, \bibinfo {author} {\bibfnamefont
				{D.}~\bibnamefont {McNulty}}, \bibinfo {author} {\bibfnamefont
				{L.}~\bibnamefont {Mercado}}, \bibinfo {author} {\bibfnamefont {Z.-E.}\
				\bibnamefont {Meziani}}, \bibinfo {author} {\bibfnamefont {R.~W.}\
				\bibnamefont {Michaels}}, \bibinfo {author} {\bibfnamefont {M.}~\bibnamefont
				{Mihovilovic}}, \bibinfo {author} {\bibfnamefont {N.}~\bibnamefont
				{Muangma}}, \bibinfo {author} {\bibfnamefont {C.}~\bibnamefont {Mu\~noz
					Camacho}}, \bibinfo {author} {\bibfnamefont {S.}~\bibnamefont {Nanda}},
			\bibinfo {author} {\bibfnamefont {V.}~\bibnamefont {Nelyubin}}, \bibinfo
			{author} {\bibfnamefont {N.}~\bibnamefont {Nuruzzaman}}, \bibinfo {author}
			{\bibfnamefont {Y.}~\bibnamefont {Oh}}, \bibinfo {author} {\bibfnamefont
				{A.}~\bibnamefont {Palmer}}, \bibinfo {author} {\bibfnamefont
				{D.}~\bibnamefont {Parno}}, \bibinfo {author} {\bibfnamefont {K.~D.}\
				\bibnamefont {Paschke}}, \bibinfo {author} {\bibfnamefont {S.~K.}\
				\bibnamefont {Phillips}}, \bibinfo {author} {\bibfnamefont {B.}~\bibnamefont
				{Poelker}}, \bibinfo {author} {\bibfnamefont {R.}~\bibnamefont
				{Pomatsalyuk}}, \bibinfo {author} {\bibfnamefont {M.}~\bibnamefont {Posik}},
			\bibinfo {author} {\bibfnamefont {A.~J.~R.}\ \bibnamefont {Puckett}},
			\bibinfo {author} {\bibfnamefont {B.}~\bibnamefont {Quinn}}, \bibinfo
			{author} {\bibfnamefont {A.}~\bibnamefont {Rakhman}}, \bibinfo {author}
			{\bibfnamefont {P.~E.}\ \bibnamefont {Reimer}}, \bibinfo {author}
			{\bibfnamefont {S.}~\bibnamefont {Riordan}}, \bibinfo {author} {\bibfnamefont
				{P.}~\bibnamefont {Rogan}}, \bibinfo {author} {\bibfnamefont
				{G.}~\bibnamefont {Ron}}, \bibinfo {author} {\bibfnamefont {G.}~\bibnamefont
				{Russo}}, \bibinfo {author} {\bibfnamefont {K.}~\bibnamefont
				{Saenboonruang}}, \bibinfo {author} {\bibfnamefont {A.}~\bibnamefont {Saha}},
			\bibinfo {author} {\bibfnamefont {B.}~\bibnamefont {Sawatzky}}, \bibinfo
			{author} {\bibfnamefont {A.}~\bibnamefont {Shahinyan}}, \bibinfo {author}
			{\bibfnamefont {R.}~\bibnamefont {Silwal}}, \bibinfo {author} {\bibfnamefont
				{S.}~\bibnamefont {Sirca}}, \bibinfo {author} {\bibfnamefont
				{K.}~\bibnamefont {Slifer}}, \bibinfo {author} {\bibfnamefont
				{P.}~\bibnamefont {Solvignon}}, \bibinfo {author} {\bibfnamefont {P.~A.}\
				\bibnamefont {Souder}}, \bibinfo {author} {\bibfnamefont {M.~L.}\
				\bibnamefont {Sperduto}}, \bibinfo {author} {\bibfnamefont {R.}~\bibnamefont
				{Subedi}}, \bibinfo {author} {\bibfnamefont {R.}~\bibnamefont {Suleiman}},
			\bibinfo {author} {\bibfnamefont {V.}~\bibnamefont {Sulkosky}}, \bibinfo
			{author} {\bibfnamefont {C.~M.}\ \bibnamefont {Sutera}}, \bibinfo {author}
			{\bibfnamefont {W.~A.}\ \bibnamefont {Tobias}}, \bibinfo {author}
			{\bibfnamefont {W.}~\bibnamefont {Troth}}, \bibinfo {author} {\bibfnamefont
				{G.~M.}\ \bibnamefont {Urciuoli}}, \bibinfo {author} {\bibfnamefont
				{B.}~\bibnamefont {Waidyawansa}}, \bibinfo {author} {\bibfnamefont
				{D.}~\bibnamefont {Wang}}, \bibinfo {author} {\bibfnamefont {J.}~\bibnamefont
				{Wexler}}, \bibinfo {author} {\bibfnamefont {R.}~\bibnamefont {Wilson}},
			\bibinfo {author} {\bibfnamefont {B.}~\bibnamefont {Wojtsekhowski}}, \bibinfo
			{author} {\bibfnamefont {X.}~\bibnamefont {Yan}}, \bibinfo {author}
			{\bibfnamefont {H.}~\bibnamefont {Yao}}, \bibinfo {author} {\bibfnamefont
				{Y.}~\bibnamefont {Ye}}, \bibinfo {author} {\bibfnamefont {Z.}~\bibnamefont
				{Ye}}, \bibinfo {author} {\bibfnamefont {V.}~\bibnamefont {Yim}}, \bibinfo
			{author} {\bibfnamefont {L.}~\bibnamefont {Zana}}, \bibinfo {author}
			{\bibfnamefont {X.}~\bibnamefont {Zhan}}, \bibinfo {author} {\bibfnamefont
				{J.}~\bibnamefont {Zhang}}, \bibinfo {author} {\bibfnamefont
				{Y.}~\bibnamefont {Zhang}}, \bibinfo {author} {\bibfnamefont
				{X.}~\bibnamefont {Zheng}},\ and\ \bibinfo {author} {\bibfnamefont
				{P.}~\bibnamefont {Zhu}} (\bibinfo {collaboration} {PREX Collaboration}),\
		}\bibfield  {title} {\bibinfo {title} {Measurement of the neutron radius of
				$^{208}\mathrm{Pb}$ through parity violation in electron scattering},\ }\href
		{https://doi.org/10.1103/PhysRevLett.108.112502} {\bibfield  {journal}
			{\bibinfo  {journal} {Phys. Rev. Lett.}\ }\textbf {\bibinfo {volume} {108}},\
			\bibinfo {pages} {112502} (\bibinfo {year} {2012})}\BibitemShut {NoStop}%
		\bibitem [{\citenamefont {Angeli}\ and\ \citenamefont {Marinova}(2013)}]{rc13}%
		\BibitemOpen
		\bibfield  {author} {\bibinfo {author} {\bibfnamefont {I.}~\bibnamefont
				{Angeli}}\ and\ \bibinfo {author} {\bibfnamefont {K.}~\bibnamefont
				{Marinova}},\ }\bibfield  {title} {\bibinfo {title} {Table of experimental
				nuclear ground state charge radii: An update},\ }\href
		{https://doi.org/https://doi.org/10.1016/j.adt.2011.12.006} {\bibfield
			{journal} {\bibinfo  {journal} {Atomic Data and Nuclear Data Tables}\
			}\textbf {\bibinfo {volume} {99}},\ \bibinfo {pages} {69} (\bibinfo {year}
			{2013})}\BibitemShut {NoStop}%
		\bibitem [{\citenamefont {K\"onig}\ \emph {et~al.}(2024)\citenamefont
			{K\"onig}, \citenamefont {Berengut}, \citenamefont {Borschevsky},
			\citenamefont {Brinson}, \citenamefont {Brown}, \citenamefont {Dockery},
			\citenamefont {Elhatisari}, \citenamefont {Eliav}, \citenamefont {Ruiz},
			\citenamefont {Holt}, \citenamefont {Hu}, \citenamefont {Karthein},
			\citenamefont {Lee}, \citenamefont {Ma}, \citenamefont {Mei\ss{}ner},
			\citenamefont {Minamisono}, \citenamefont {Oleynichenko}, \citenamefont
			{Pineda}, \citenamefont {Prosnyak}, \citenamefont {Reitsma}, \citenamefont
			{Skripnikov}, \citenamefont {Vernon},\ and\ \citenamefont
			{Zaitsevskii}}]{PhysRevLett.132.162502}%
		\BibitemOpen
		\bibfield  {author} {\bibinfo {author} {\bibfnamefont {K.}~\bibnamefont
				{K\"onig}}, \bibinfo {author} {\bibfnamefont {J.~C.}\ \bibnamefont
				{Berengut}}, \bibinfo {author} {\bibfnamefont {A.}~\bibnamefont
				{Borschevsky}}, \bibinfo {author} {\bibfnamefont {A.}~\bibnamefont
				{Brinson}}, \bibinfo {author} {\bibfnamefont {B.~A.}\ \bibnamefont {Brown}},
			\bibinfo {author} {\bibfnamefont {A.}~\bibnamefont {Dockery}}, \bibinfo
			{author} {\bibfnamefont {S.}~\bibnamefont {Elhatisari}}, \bibinfo {author}
			{\bibfnamefont {E.}~\bibnamefont {Eliav}}, \bibinfo {author} {\bibfnamefont
				{R.~F.~G.}\ \bibnamefont {Ruiz}}, \bibinfo {author} {\bibfnamefont {J.~D.}\
				\bibnamefont {Holt}}, \bibinfo {author} {\bibfnamefont {B.-S.}\ \bibnamefont
				{Hu}}, \bibinfo {author} {\bibfnamefont {J.}~\bibnamefont {Karthein}},
			\bibinfo {author} {\bibfnamefont {D.}~\bibnamefont {Lee}}, \bibinfo {author}
			{\bibfnamefont {Y.-Z.}\ \bibnamefont {Ma}}, \bibinfo {author} {\bibfnamefont
				{U.-G.}\ \bibnamefont {Mei\ss{}ner}}, \bibinfo {author} {\bibfnamefont
				{K.}~\bibnamefont {Minamisono}}, \bibinfo {author} {\bibfnamefont {A.~V.}\
				\bibnamefont {Oleynichenko}}, \bibinfo {author} {\bibfnamefont {S.~V.}\
				\bibnamefont {Pineda}}, \bibinfo {author} {\bibfnamefont {S.~D.}\
				\bibnamefont {Prosnyak}}, \bibinfo {author} {\bibfnamefont {M.~L.}\
				\bibnamefont {Reitsma}}, \bibinfo {author} {\bibfnamefont {L.~V.}\
				\bibnamefont {Skripnikov}}, \bibinfo {author} {\bibfnamefont
				{A.}~\bibnamefont {Vernon}},\ and\ \bibinfo {author} {\bibfnamefont
				{A.}~\bibnamefont {Zaitsevskii}},\ }\bibfield  {title} {\bibinfo {title}
			{Nuclear charge radii of silicon isotopes},\ }\href
		{https://doi.org/10.1103/PhysRevLett.132.162502} {\bibfield  {journal}
			{\bibinfo  {journal} {Phys. Rev. Lett.}\ }\textbf {\bibinfo {volume} {132}},\
			\bibinfo {pages} {162502} (\bibinfo {year} {2024})}\BibitemShut {NoStop}%
		\bibitem [{\citenamefont {Zhao}\ \emph {et~al.}(2024)\citenamefont {Zhao},
			\citenamefont {Sun}, \citenamefont {Tanihata}, \citenamefont {Xu},
			\citenamefont {Zhang}, \citenamefont {Prochazka}, \citenamefont {Zhu},
			\citenamefont {Terashima}, \citenamefont {Meng}, \citenamefont {He},
			\citenamefont {Liu}, \citenamefont {Li}, \citenamefont {Lu}, \citenamefont
			{Lin}, \citenamefont {Lin}, \citenamefont {Liu}, \citenamefont {Ren},
			\citenamefont {Sun}, \citenamefont {Wang}, \citenamefont {Wang},
			\citenamefont {Wang}, \citenamefont {Wang}, \citenamefont {Wei},
			\citenamefont {Xu}, \citenamefont {Zhang}, \citenamefont {Zhang},\ and\
			\citenamefont {Zhang}}]{ZHAO2024139082}%
		\BibitemOpen
		\bibfield  {author} {\bibinfo {author} {\bibfnamefont {J.}~\bibnamefont
				{Zhao}}, \bibinfo {author} {\bibfnamefont {B.-H.}\ \bibnamefont {Sun}},
			\bibinfo {author} {\bibfnamefont {I.}~\bibnamefont {Tanihata}}, \bibinfo
			{author} {\bibfnamefont {J.}~\bibnamefont {Xu}}, \bibinfo {author}
			{\bibfnamefont {K.}~\bibnamefont {Zhang}}, \bibinfo {author} {\bibfnamefont
				{A.}~\bibnamefont {Prochazka}}, \bibinfo {author} {\bibfnamefont
				{L.}~\bibnamefont {Zhu}}, \bibinfo {author} {\bibfnamefont {S.}~\bibnamefont
				{Terashima}}, \bibinfo {author} {\bibfnamefont {J.}~\bibnamefont {Meng}},
			\bibinfo {author} {\bibfnamefont {L.}~\bibnamefont {He}}, \bibinfo {author}
			{\bibfnamefont {C.}~\bibnamefont {Liu}}, \bibinfo {author} {\bibfnamefont
				{G.}~\bibnamefont {Li}}, \bibinfo {author} {\bibfnamefont {C.}~\bibnamefont
				{Lu}}, \bibinfo {author} {\bibfnamefont {W.}~\bibnamefont {Lin}}, \bibinfo
			{author} {\bibfnamefont {W.}~\bibnamefont {Lin}}, \bibinfo {author}
			{\bibfnamefont {Z.}~\bibnamefont {Liu}}, \bibinfo {author} {\bibfnamefont
				{P.}~\bibnamefont {Ren}}, \bibinfo {author} {\bibfnamefont {Z.}~\bibnamefont
				{Sun}}, \bibinfo {author} {\bibfnamefont {F.}~\bibnamefont {Wang}}, \bibinfo
			{author} {\bibfnamefont {J.}~\bibnamefont {Wang}}, \bibinfo {author}
			{\bibfnamefont {M.}~\bibnamefont {Wang}}, \bibinfo {author} {\bibfnamefont
				{S.}~\bibnamefont {Wang}}, \bibinfo {author} {\bibfnamefont {X.}~\bibnamefont
				{Wei}}, \bibinfo {author} {\bibfnamefont {X.}~\bibnamefont {Xu}}, \bibinfo
			{author} {\bibfnamefont {J.}~\bibnamefont {Zhang}}, \bibinfo {author}
			{\bibfnamefont {M.}~\bibnamefont {Zhang}},\ and\ \bibinfo {author}
			{\bibfnamefont {X.}~\bibnamefont {Zhang}},\ }\bibfield  {title} {\bibinfo
			{title} {Charge radii of $^{11-16}${C}, $^{13-17}${N} and $^{15-18}${O}
				determined from their charge-changing cross-sections and the
				mirror-difference charge radii},\ }\href
		{https://doi.org/https://doi.org/10.1016/j.physletb.2024.139082} {\bibfield
			{journal} {\bibinfo  {journal} {Physics Letters B}\ }\textbf {\bibinfo
				{volume} {858}},\ \bibinfo {pages} {139082} (\bibinfo {year}
			{2024})}\BibitemShut {NoStop}%
		\bibitem [{\citenamefont {Kota}(2001)}]{KOTA2001223}%
		\BibitemOpen
		\bibfield  {author} {\bibinfo {author} {\bibfnamefont {V.}~\bibnamefont
				{Kota}},\ }\bibfield  {title} {\bibinfo {title} {Embedded random matrix
				ensembles for complexity and chaos in finite interacting particle systems},\
		}\href {https://doi.org/https://doi.org/10.1016/S0370-1573(00)00113-7}
		{\bibfield  {journal} {\bibinfo  {journal} {Physics Reports}\ }\textbf
			{\bibinfo {volume} {347}},\ \bibinfo {pages} {223} (\bibinfo {year}
			{2001})}\BibitemShut {NoStop}%
		\bibitem [{\citenamefont {Zelevinsky}\ and\ \citenamefont
			{Volya}(2004)}]{ZELEVINSKY2004311}%
		\BibitemOpen
		\bibfield  {author} {\bibinfo {author} {\bibfnamefont {V.}~\bibnamefont
				{Zelevinsky}}\ and\ \bibinfo {author} {\bibfnamefont {A.}~\bibnamefont
				{Volya}},\ }\bibfield  {title} {\bibinfo {title} {Nuclear structure, random
				interactions and mesoscopic physics},\ }\href
		{https://doi.org/https://doi.org/10.1016/j.physrep.2003.10.008} {\bibfield
			{journal} {\bibinfo  {journal} {Physics Reports}\ }\textbf {\bibinfo {volume}
				{391}},\ \bibinfo {pages} {311} (\bibinfo {year} {2004})},\ \bibinfo {note}
		{from atoms to nuclei to quarks and gluons: the omnipresent manybody
			theory}\BibitemShut {NoStop}%
		\bibitem [{\citenamefont {Zhao}\ \emph
			{et~al.}(2004{\natexlab{a}})\citenamefont {Zhao}, \citenamefont {Arima},\
			and\ \citenamefont {Yoshinaga}}]{ZHAO20041}%
		\BibitemOpen
		\bibfield  {author} {\bibinfo {author} {\bibfnamefont {Y.}~\bibnamefont
				{Zhao}}, \bibinfo {author} {\bibfnamefont {A.}~\bibnamefont {Arima}},\ and\
			\bibinfo {author} {\bibfnamefont {N.}~\bibnamefont {Yoshinaga}},\ }\bibfield
		{title} {\bibinfo {title} {Regularities of many-body systems interacting by a
				two-body random ensemble},\ }\href
		{https://doi.org/https://doi.org/10.1016/j.physrep.2004.07.004} {\bibfield
			{journal} {\bibinfo  {journal} {Physics Reports}\ }\textbf {\bibinfo {volume}
				{400}},\ \bibinfo {pages} {1} (\bibinfo {year}
			{2004}{\natexlab{a}})}\BibitemShut {NoStop}%
		\bibitem [{\citenamefont {Weidenm\"uller}\ and\ \citenamefont
			{Mitchell}(2009)}]{RevModPhys.81.539}%
		\BibitemOpen
		\bibfield  {author} {\bibinfo {author} {\bibfnamefont {H.~A.}\ \bibnamefont
				{Weidenm\"uller}}\ and\ \bibinfo {author} {\bibfnamefont {G.~E.}\
				\bibnamefont {Mitchell}},\ }\bibfield  {title} {\bibinfo {title} {Random
				matrices and chaos in nuclear physics: Nuclear structure},\ }\href
		{https://doi.org/10.1103/RevModPhys.81.539} {\bibfield  {journal} {\bibinfo
				{journal} {Rev. Mod. Phys.}\ }\textbf {\bibinfo {volume} {81}},\ \bibinfo
			{pages} {539} (\bibinfo {year} {2009})}\BibitemShut {NoStop}%
		\bibitem [{\citenamefont {Sherrill}\ and\ \citenamefont
			{Casten}(2005)}]{Sherrill01042005}%
		\BibitemOpen
		\bibfield  {author} {\bibinfo {author} {\bibfnamefont {B.}~\bibnamefont
				{Sherrill}}\ and\ \bibinfo {author} {\bibfnamefont {R.~F.}\ \bibnamefont
				{Casten}},\ }\bibfield  {title} {\bibinfo {title} {Future articles: Frontiers
				of nuclear structure: Exotic nuclei},\ }\href
		{https://doi.org/10.1080/10506890500454675} {\bibfield  {journal} {\bibinfo
				{journal} {Nuclear Physics News}\ }\textbf {\bibinfo {volume} {15}},\
			\bibinfo {pages} {13} (\bibinfo {year} {2005})},\ \Eprint
		{https://arxiv.org/abs/https://doi.org/10.1080/10506890500454675}
		{https://doi.org/10.1080/10506890500454675} \BibitemShut {NoStop}%
		\bibitem [{\citenamefont {Johnson}\ \emph {et~al.}(1998)\citenamefont
			{Johnson}, \citenamefont {Bertsch},\ and\ \citenamefont
			{Dean}}]{PhysRevLett.80.2749}%
		\BibitemOpen
		\bibfield  {author} {\bibinfo {author} {\bibfnamefont {C.~W.}\ \bibnamefont
				{Johnson}}, \bibinfo {author} {\bibfnamefont {G.~F.}\ \bibnamefont
				{Bertsch}},\ and\ \bibinfo {author} {\bibfnamefont {D.~J.}\ \bibnamefont
				{Dean}},\ }\bibfield  {title} {\bibinfo {title} {Orderly spectra from random
				interactions},\ }\href {https://doi.org/10.1103/PhysRevLett.80.2749}
		{\bibfield  {journal} {\bibinfo  {journal} {Phys. Rev. Lett.}\ }\textbf
			{\bibinfo {volume} {80}},\ \bibinfo {pages} {2749} (\bibinfo {year}
			{1998})}\BibitemShut {NoStop}%
		\bibitem [{\citenamefont {Zhao}\ \emph
			{et~al.}(2004{\natexlab{b}})\citenamefont {Zhao}, \citenamefont {Arima},
			\citenamefont {Shimizu}, \citenamefont {Ogawa}, \citenamefont {Yoshinaga},\
			and\ \citenamefont {Scholten}}]{PhysRevC.70.054322}%
		\BibitemOpen
		\bibfield  {author} {\bibinfo {author} {\bibfnamefont {Y.~M.}\ \bibnamefont
				{Zhao}}, \bibinfo {author} {\bibfnamefont {A.}~\bibnamefont {Arima}},
			\bibinfo {author} {\bibfnamefont {N.}~\bibnamefont {Shimizu}}, \bibinfo
			{author} {\bibfnamefont {K.}~\bibnamefont {Ogawa}}, \bibinfo {author}
			{\bibfnamefont {N.}~\bibnamefont {Yoshinaga}},\ and\ \bibinfo {author}
			{\bibfnamefont {O.}~\bibnamefont {Scholten}},\ }\bibfield  {title} {\bibinfo
			{title} {Patterns of the ground states in the presence of random
				interactions: Nucleon systems},\ }\href
		{https://doi.org/10.1103/PhysRevC.70.054322} {\bibfield  {journal} {\bibinfo
				{journal} {Phys. Rev. C}\ }\textbf {\bibinfo {volume} {70}},\ \bibinfo
			{pages} {054322} (\bibinfo {year} {2004}{\natexlab{b}})}\BibitemShut
		{NoStop}%
		\bibitem [{\citenamefont {Bijker}\ and\ \citenamefont
			{Frank}(2000)}]{PhysRevLett.84.420}%
		\BibitemOpen
		\bibfield  {author} {\bibinfo {author} {\bibfnamefont {R.}~\bibnamefont
				{Bijker}}\ and\ \bibinfo {author} {\bibfnamefont {A.}~\bibnamefont {Frank}},\
		}\bibfield  {title} {\bibinfo {title} {Band structure from random
				interactions},\ }\href {https://doi.org/10.1103/PhysRevLett.84.420}
		{\bibfield  {journal} {\bibinfo  {journal} {Phys. Rev. Lett.}\ }\textbf
			{\bibinfo {volume} {84}},\ \bibinfo {pages} {420} (\bibinfo {year}
			{2000})}\BibitemShut {NoStop}%
		\bibitem [{\citenamefont {Shen}\ \emph {et~al.}(2021)\citenamefont {Shen},
			\citenamefont {Jiang},\ and\ \citenamefont {Fu}}]{PhysRevC.104.054319}%
		\BibitemOpen
		\bibfield  {author} {\bibinfo {author} {\bibfnamefont {J.~J.}\ \bibnamefont
				{Shen}}, \bibinfo {author} {\bibfnamefont {H.}~\bibnamefont {Jiang}},\ and\
			\bibinfo {author} {\bibfnamefont {G.~J.}\ \bibnamefont {Fu}},\ }\bibfield
		{title} {\bibinfo {title} {Robustness of ``noncollective'' rotational
				behavior for nuclei in the presence of random interactions},\ }\href
		{https://doi.org/10.1103/PhysRevC.104.054319} {\bibfield  {journal} {\bibinfo
				{journal} {Phys. Rev. C}\ }\textbf {\bibinfo {volume} {104}},\ \bibinfo
			{pages} {054319} (\bibinfo {year} {2021})}\BibitemShut {NoStop}%
		\bibitem [{\citenamefont {Fu}\ \emph {et~al.}(2015)\citenamefont {Fu},
			\citenamefont {Shen}, \citenamefont {Zhao},\ and\ \citenamefont
			{Arima}}]{PhysRevC.91.054319}%
		\BibitemOpen
		\bibfield  {author} {\bibinfo {author} {\bibfnamefont {G.~J.}\ \bibnamefont
				{Fu}}, \bibinfo {author} {\bibfnamefont {J.~J.}\ \bibnamefont {Shen}},
			\bibinfo {author} {\bibfnamefont {Y.~M.}\ \bibnamefont {Zhao}},\ and\
			\bibinfo {author} {\bibfnamefont {A.}~\bibnamefont {Arima}},\ }\bibfield
		{title} {\bibinfo {title} {Regularities in low-lying states of atomic nuclei
				with random interactions},\ }\href
		{https://doi.org/10.1103/PhysRevC.91.054319} {\bibfield  {journal} {\bibinfo
				{journal} {Phys. Rev. C}\ }\textbf {\bibinfo {volume} {91}},\ \bibinfo
			{pages} {054319} (\bibinfo {year} {2015})}\BibitemShut {NoStop}%
		\bibitem [{\citenamefont {Lei}(2016)}]{PhysRevC.93.024319}%
		\BibitemOpen
		\bibfield  {author} {\bibinfo {author} {\bibfnamefont {Y.}~\bibnamefont
				{Lei}},\ }\bibfield  {title} {\bibinfo {title} {Robust correlations between
				quadrupole moments of low-lying ${2}^{+}$ states within random-interaction
				ensembles},\ }\href {https://doi.org/10.1103/PhysRevC.93.024319} {\bibfield
			{journal} {\bibinfo  {journal} {Phys. Rev. C}\ }\textbf {\bibinfo {volume}
				{93}},\ \bibinfo {pages} {024319} (\bibinfo {year} {2016})}\BibitemShut
		{NoStop}%
		\bibitem [{\citenamefont {Qin}\ and\ \citenamefont {Lei}(2018)}]{Qin2018}%
		\BibitemOpen
		\bibfield  {author} {\bibinfo {author} {\bibfnamefont {Z.-Z.}\ \bibnamefont
				{Qin}}\ and\ \bibinfo {author} {\bibfnamefont {Y.}~\bibnamefont {Lei}},\
		}\bibfield  {title} {\bibinfo {title} {Predominance of linear q and $\mu$
				systematics in random-interaction ensembles},\ }\href
		{https://doi.org/10.1007/s41365-018-0503-0} {\bibfield  {journal} {\bibinfo
				{journal} {Nuclear Science and Techniques}\ }\textbf {\bibinfo {volume}
				{29}},\ \bibinfo {pages} {163} (\bibinfo {year} {2018})}\BibitemShut
		{NoStop}%
		\bibitem [{\citenamefont {Johnson}\ \emph {et~al.}(2018)\citenamefont
			{Johnson}, \citenamefont {Ormand}, \citenamefont {McElvain},\ and\
			\citenamefont
			{Shan}}]{johnson2018bigstickflexibleconfigurationinteractionshellmodel}%
		\BibitemOpen
		\bibfield  {author} {\bibinfo {author} {\bibfnamefont {C.~W.}\ \bibnamefont
				{Johnson}}, \bibinfo {author} {\bibfnamefont {W.~E.}\ \bibnamefont {Ormand}},
			\bibinfo {author} {\bibfnamefont {K.~S.}\ \bibnamefont {McElvain}},\ and\
			\bibinfo {author} {\bibfnamefont {H.}~\bibnamefont {Shan}},\ }\href
		{https://arxiv.org/abs/1801.08432} {\bibinfo {title} {Bigstick: A flexible
				configuration-interaction shell-model code}} (\bibinfo {year} {2018}),\
		\Eprint {https://arxiv.org/abs/1801.08432} {arXiv:1801.08432
			[physics.comp-ph]} \BibitemShut {NoStop}%
		\bibitem [{\citenamefont {Karl}(1895)}]{pearson1895}%
		\BibitemOpen
		\bibfield  {author} {\bibinfo {author} {\bibfnamefont {P.}~\bibnamefont
				{Karl}},\ }\bibfield  {title} {\bibinfo {title} {Vii. note on regression and
				inheritance in the case of two parents},\ }\href
		{https://doi.org/10.1098/rspl.1895.0041} {\bibfield  {journal} {\bibinfo
				{journal} {Proc. R. Soc. Lond.}\ }\textbf {\bibinfo {volume} {58}},\ \bibinfo
			{pages} {240} (\bibinfo {year} {1895})}\BibitemShut {NoStop}%
		\bibitem [{\citenamefont {Marevic}\ \emph {et~al.}(2022)\citenamefont
			{Marevic}, \citenamefont {Schunck}, \citenamefont {Ney}, \citenamefont
			{{Navarro Perez}}, \citenamefont {Verriere},\ and\ \citenamefont
			{O'Neal}}]{hfbtho}%
		\BibitemOpen
		\bibfield  {author} {\bibinfo {author} {\bibfnamefont {P.}~\bibnamefont
				{Marevic}}, \bibinfo {author} {\bibfnamefont {N.}~\bibnamefont {Schunck}},
			\bibinfo {author} {\bibfnamefont {E.}~\bibnamefont {Ney}}, \bibinfo {author}
			{\bibfnamefont {R.}~\bibnamefont {{Navarro Perez}}}, \bibinfo {author}
			{\bibfnamefont {M.}~\bibnamefont {Verriere}},\ and\ \bibinfo {author}
			{\bibfnamefont {J.}~\bibnamefont {O'Neal}},\ }\bibfield  {title} {\bibinfo
			{title} {Axially-deformed solution of the skyrme-hartree-fock-bogoliubov
				equations using the transformed harmonic oscillator basis (iv) hfbtho (v4.0):
				A new version of the program},\ }\href
		{https://doi.org/https://doi.org/10.1016/j.cpc.2022.108367} {\bibfield
			{journal} {\bibinfo  {journal} {Computer Physics Communications}\ }\textbf
			{\bibinfo {volume} {276}},\ \bibinfo {pages} {108367} (\bibinfo {year}
			{2022})}\BibitemShut {NoStop}%
		\bibitem [{\citenamefont {Dutra}\ \emph {et~al.}(2012)\citenamefont {Dutra},
			\citenamefont {Louren\ifmmode~\mbox{\c{c}}\else \c{c}\fi{}o}, \citenamefont
			{S\'a~Martins}, \citenamefont {Delfino}, \citenamefont {Stone},\ and\
			\citenamefont {Stevenson}}]{PhysRevC.85.035201}%
		\BibitemOpen
		\bibfield  {author} {\bibinfo {author} {\bibfnamefont {M.}~\bibnamefont
				{Dutra}}, \bibinfo {author} {\bibfnamefont {O.}~\bibnamefont
				{Louren\ifmmode~\mbox{\c{c}}\else \c{c}\fi{}o}}, \bibinfo {author}
			{\bibfnamefont {J.~S.}\ \bibnamefont {S\'a~Martins}}, \bibinfo {author}
			{\bibfnamefont {A.}~\bibnamefont {Delfino}}, \bibinfo {author} {\bibfnamefont
				{J.~R.}\ \bibnamefont {Stone}},\ and\ \bibinfo {author} {\bibfnamefont
				{P.~D.}\ \bibnamefont {Stevenson}},\ }\bibfield  {title} {\bibinfo {title}
			{Skyrme interaction and nuclear matter constraints},\ }\href
		{https://doi.org/10.1103/PhysRevC.85.035201} {\bibfield  {journal} {\bibinfo
				{journal} {Phys. Rev. C}\ }\textbf {\bibinfo {volume} {85}},\ \bibinfo
			{pages} {035201} (\bibinfo {year} {2012})}\BibitemShut {NoStop}%
		\bibitem [{\citenamefont {Devroye}(1986)}]{mul-gau-rand}%
		\BibitemOpen
		\bibfield  {author} {\bibinfo {author} {\bibfnamefont {L.}~\bibnamefont
				{Devroye}},\ }\bibinfo {title} {Multivariate distributions},\ in\ \href
		{https://doi.org/10.1007/978-1-4613-8643-8_11} {\emph {\bibinfo {booktitle}
				{Non-Uniform Random Variate Generation}}}\ (\bibinfo  {publisher} {Springer
			New York},\ \bibinfo {address} {New York, NY},\ \bibinfo {year} {1986})\ pp.\
		\bibinfo {pages} {554--610}\BibitemShut {NoStop}%
		\bibitem [{\citenamefont {Kohler}(1976)}]{KOHLER1976301}%
		\BibitemOpen
		\bibfield  {author} {\bibinfo {author} {\bibfnamefont {H.}~\bibnamefont
				{Kohler}},\ }\bibfield  {title} {\bibinfo {title} {Skyrme force and the mass
				formula},\ }\href
		{https://doi.org/https://doi.org/10.1016/0375-9474(76)90008-7} {\bibfield
			{journal} {\bibinfo  {journal} {Nuclear Physics A}\ }\textbf {\bibinfo
				{volume} {258}},\ \bibinfo {pages} {301} (\bibinfo {year}
			{1976})}\BibitemShut {NoStop}%
		\bibitem [{\citenamefont {Reed}\ \emph {et~al.}(2021)\citenamefont {Reed},
			\citenamefont {Fattoyev}, \citenamefont {Horowitz},\ and\ \citenamefont
			{Piekarewicz}}]{Pb-L}%
		\BibitemOpen
		\bibfield  {author} {\bibinfo {author} {\bibfnamefont {B.~T.}\ \bibnamefont
				{Reed}}, \bibinfo {author} {\bibfnamefont {F.~J.}\ \bibnamefont {Fattoyev}},
			\bibinfo {author} {\bibfnamefont {C.~J.}\ \bibnamefont {Horowitz}},\ and\
			\bibinfo {author} {\bibfnamefont {J.}~\bibnamefont {Piekarewicz}},\
		}\bibfield  {title} {\bibinfo {title} {Implications of prex-2 on the equation
				of state of neutron-rich matter},\ }\href
		{https://doi.org/10.1103/PhysRevLett.126.172503} {\bibfield  {journal}
			{\bibinfo  {journal} {Phys. Rev. Lett.}\ }\textbf {\bibinfo {volume} {126}},\
			\bibinfo {pages} {172503} (\bibinfo {year} {2021})}\BibitemShut {NoStop}%
		\bibitem [{\citenamefont {Heyde}\ and\ \citenamefont
			{Wood}(2011)}]{coexistence_rmp}%
		\BibitemOpen
		\bibfield  {author} {\bibinfo {author} {\bibfnamefont {K.}~\bibnamefont
				{Heyde}}\ and\ \bibinfo {author} {\bibfnamefont {J.~L.}\ \bibnamefont
				{Wood}},\ }\bibfield  {title} {\bibinfo {title} {Shape coexistence in atomic
				nuclei},\ }\href {https://doi.org/10.1103/RevModPhys.83.1467} {\bibfield
			{journal} {\bibinfo  {journal} {Rev. Mod. Phys.}\ }\textbf {\bibinfo {volume}
				{83}},\ \bibinfo {pages} {1467} (\bibinfo {year} {2011})}\BibitemShut
		{NoStop}%
		\bibitem [{\citenamefont {Lei}\ \emph {et~al.}(2024)\citenamefont {Lei},
			\citenamefont {Qi}, \citenamefont {Lu}, \citenamefont {Jiang}, \citenamefont
			{Qin}, \citenamefont {Liu},\ and\ \citenamefont
			{Johnson}}]{PhysRevC.110.054318}%
		\BibitemOpen
		\bibfield  {author} {\bibinfo {author} {\bibfnamefont {Y.}~\bibnamefont
				{Lei}}, \bibinfo {author} {\bibfnamefont {J.}~\bibnamefont {Qi}}, \bibinfo
			{author} {\bibfnamefont {Y.}~\bibnamefont {Lu}}, \bibinfo {author}
			{\bibfnamefont {H.}~\bibnamefont {Jiang}}, \bibinfo {author} {\bibfnamefont
				{Z.~Z.}\ \bibnamefont {Qin}}, \bibinfo {author} {\bibfnamefont
				{D.}~\bibnamefont {Liu}},\ and\ \bibinfo {author} {\bibfnamefont {C.~W.}\
				\bibnamefont {Johnson}},\ }\bibfield  {title} {\bibinfo {title}
			{Pervasiveness of shape coexistence in nuclear pair condensates},\ }\href
		{https://doi.org/10.1103/PhysRevC.110.054318} {\bibfield  {journal} {\bibinfo
				{journal} {Phys. Rev. C}\ }\textbf {\bibinfo {volume} {110}},\ \bibinfo
			{pages} {054318} (\bibinfo {year} {2024})}\BibitemShut {NoStop}%
		\bibitem [{\citenamefont {Marsh}\ \emph {et~al.}(2018)\citenamefont {Marsh},
			\citenamefont {Day~Goodacre}, \citenamefont {Sels}, \citenamefont {Tsunoda},
			\citenamefont {Andel}, \citenamefont {Andreyev}, \citenamefont {Althubiti},
			\citenamefont {Atanasov}, \citenamefont {Barzakh}, \citenamefont {Billowes},
			\citenamefont {Blaum}, \citenamefont {Cocolios}, \citenamefont {Cubiss},
			\citenamefont {Dobaczewski}, \citenamefont {Farooq-Smith}, \citenamefont
			{Fedorov}, \citenamefont {Fedosseev}, \citenamefont {Flanagan}, \citenamefont
			{Gaffney}, \citenamefont {Ghys}, \citenamefont {Huyse}, \citenamefont
			{Kreim}, \citenamefont {Lunney}, \citenamefont {Lynch}, \citenamefont
			{Manea}, \citenamefont {Martinez~Palenzuela}, \citenamefont {Molkanov},
			\citenamefont {Otsuka}, \citenamefont {Pastore}, \citenamefont {Rosenbusch},
			\citenamefont {Rossel}, \citenamefont {Rothe}, \citenamefont {Schweikhard},
			\citenamefont {Seliverstov}, \citenamefont {Spagnoletti}, \citenamefont
			{Van~Beveren}, \citenamefont {Van~Duppen}, \citenamefont {Veinhard},
			\citenamefont {Verstraelen}, \citenamefont {Welker}, \citenamefont {Wendt},
			\citenamefont {Wienholtz}, \citenamefont {Wolf}, \citenamefont {Zadvornaya},\
			and\ \citenamefont {Zuber}}]{Marsh2018}%
		\BibitemOpen
		\bibfield  {author} {\bibinfo {author} {\bibfnamefont {B.~A.}\ \bibnamefont
				{Marsh}}, \bibinfo {author} {\bibfnamefont {T.}~\bibnamefont {Day~Goodacre}},
			\bibinfo {author} {\bibfnamefont {S.}~\bibnamefont {Sels}}, \bibinfo {author}
			{\bibfnamefont {Y.}~\bibnamefont {Tsunoda}}, \bibinfo {author} {\bibfnamefont
				{B.}~\bibnamefont {Andel}}, \bibinfo {author} {\bibfnamefont {A.~N.}\
				\bibnamefont {Andreyev}}, \bibinfo {author} {\bibfnamefont {N.~A.}\
				\bibnamefont {Althubiti}}, \bibinfo {author} {\bibfnamefont {D.}~\bibnamefont
				{Atanasov}}, \bibinfo {author} {\bibfnamefont {A.~E.}\ \bibnamefont
				{Barzakh}}, \bibinfo {author} {\bibfnamefont {J.}~\bibnamefont {Billowes}},
			\bibinfo {author} {\bibfnamefont {K.}~\bibnamefont {Blaum}}, \bibinfo
			{author} {\bibfnamefont {T.~E.}\ \bibnamefont {Cocolios}}, \bibinfo {author}
			{\bibfnamefont {J.~G.}\ \bibnamefont {Cubiss}}, \bibinfo {author}
			{\bibfnamefont {J.}~\bibnamefont {Dobaczewski}}, \bibinfo {author}
			{\bibfnamefont {G.~J.}\ \bibnamefont {Farooq-Smith}}, \bibinfo {author}
			{\bibfnamefont {D.~V.}\ \bibnamefont {Fedorov}}, \bibinfo {author}
			{\bibfnamefont {V.~N.}\ \bibnamefont {Fedosseev}}, \bibinfo {author}
			{\bibfnamefont {K.~T.}\ \bibnamefont {Flanagan}}, \bibinfo {author}
			{\bibfnamefont {L.~P.}\ \bibnamefont {Gaffney}}, \bibinfo {author}
			{\bibfnamefont {L.}~\bibnamefont {Ghys}}, \bibinfo {author} {\bibfnamefont
				{M.}~\bibnamefont {Huyse}}, \bibinfo {author} {\bibfnamefont
				{S.}~\bibnamefont {Kreim}}, \bibinfo {author} {\bibfnamefont
				{D.}~\bibnamefont {Lunney}}, \bibinfo {author} {\bibfnamefont {K.~M.}\
				\bibnamefont {Lynch}}, \bibinfo {author} {\bibfnamefont {V.}~\bibnamefont
				{Manea}}, \bibinfo {author} {\bibfnamefont {Y.}~\bibnamefont
				{Martinez~Palenzuela}}, \bibinfo {author} {\bibfnamefont {P.~L.}\
				\bibnamefont {Molkanov}}, \bibinfo {author} {\bibfnamefont {T.}~\bibnamefont
				{Otsuka}}, \bibinfo {author} {\bibfnamefont {A.}~\bibnamefont {Pastore}},
			\bibinfo {author} {\bibfnamefont {M.}~\bibnamefont {Rosenbusch}}, \bibinfo
			{author} {\bibfnamefont {R.~E.}\ \bibnamefont {Rossel}}, \bibinfo {author}
			{\bibfnamefont {S.}~\bibnamefont {Rothe}}, \bibinfo {author} {\bibfnamefont
				{L.}~\bibnamefont {Schweikhard}}, \bibinfo {author} {\bibfnamefont {M.~D.}\
				\bibnamefont {Seliverstov}}, \bibinfo {author} {\bibfnamefont
				{P.}~\bibnamefont {Spagnoletti}}, \bibinfo {author} {\bibfnamefont
				{C.}~\bibnamefont {Van~Beveren}}, \bibinfo {author} {\bibfnamefont
				{P.}~\bibnamefont {Van~Duppen}}, \bibinfo {author} {\bibfnamefont
				{M.}~\bibnamefont {Veinhard}}, \bibinfo {author} {\bibfnamefont
				{E.}~\bibnamefont {Verstraelen}}, \bibinfo {author} {\bibfnamefont
				{A.}~\bibnamefont {Welker}}, \bibinfo {author} {\bibfnamefont
				{K.}~\bibnamefont {Wendt}}, \bibinfo {author} {\bibfnamefont
				{F.}~\bibnamefont {Wienholtz}}, \bibinfo {author} {\bibfnamefont {R.~N.}\
				\bibnamefont {Wolf}}, \bibinfo {author} {\bibfnamefont {A.}~\bibnamefont
				{Zadvornaya}},\ and\ \bibinfo {author} {\bibfnamefont {K.}~\bibnamefont
				{Zuber}},\ }\bibfield  {title} {\bibinfo {title} {Characterization of the
				shape-staggering effect in mercury nuclei},\ }\href
		{https://doi.org/10.1038/s41567-018-0292-8} {\bibfield  {journal} {\bibinfo
				{journal} {Nature Physics}\ }\textbf {\bibinfo {volume} {14}},\ \bibinfo
			{pages} {1163} (\bibinfo {year} {2018})}\BibitemShut {NoStop}%
		\bibitem [{\citenamefont {de~Groote}\ \emph {et~al.}(2020)\citenamefont
			{de~Groote}, \citenamefont {Billowes}, \citenamefont {Binnersley},
			\citenamefont {Bissell}, \citenamefont {Cocolios}, \citenamefont
			{Day~Goodacre}, \citenamefont {Farooq-Smith}, \citenamefont {Fedorov},
			\citenamefont {Flanagan}, \citenamefont {Franchoo}, \citenamefont
			{Garcia~Ruiz}, \citenamefont {Gins}, \citenamefont {Holt}, \citenamefont
			{Koszor{\'u}s}, \citenamefont {Lynch}, \citenamefont {Miyagi}, \citenamefont
			{Nazarewicz}, \citenamefont {Neyens}, \citenamefont {Reinhard}, \citenamefont
			{Rothe}, \citenamefont {Stroke}, \citenamefont {Vernon}, \citenamefont
			{Wendt}, \citenamefont {Wilkins}, \citenamefont {Xu},\ and\ \citenamefont
			{Yang}}]{deGroote2020}%
		\BibitemOpen
		\bibfield  {author} {\bibinfo {author} {\bibfnamefont {R.~P.}\ \bibnamefont
				{de~Groote}}, \bibinfo {author} {\bibfnamefont {J.}~\bibnamefont {Billowes}},
			\bibinfo {author} {\bibfnamefont {C.~L.}\ \bibnamefont {Binnersley}},
			\bibinfo {author} {\bibfnamefont {M.~L.}\ \bibnamefont {Bissell}}, \bibinfo
			{author} {\bibfnamefont {T.~E.}\ \bibnamefont {Cocolios}}, \bibinfo {author}
			{\bibfnamefont {T.}~\bibnamefont {Day~Goodacre}}, \bibinfo {author}
			{\bibfnamefont {G.~J.}\ \bibnamefont {Farooq-Smith}}, \bibinfo {author}
			{\bibfnamefont {D.~V.}\ \bibnamefont {Fedorov}}, \bibinfo {author}
			{\bibfnamefont {K.~T.}\ \bibnamefont {Flanagan}}, \bibinfo {author}
			{\bibfnamefont {S.}~\bibnamefont {Franchoo}}, \bibinfo {author}
			{\bibfnamefont {R.~F.}\ \bibnamefont {Garcia~Ruiz}}, \bibinfo {author}
			{\bibfnamefont {W.}~\bibnamefont {Gins}}, \bibinfo {author} {\bibfnamefont
				{J.~D.}\ \bibnamefont {Holt}}, \bibinfo {author} {\bibfnamefont
				{{\'A}.}~\bibnamefont {Koszor{\'u}s}}, \bibinfo {author} {\bibfnamefont
				{K.~M.}\ \bibnamefont {Lynch}}, \bibinfo {author} {\bibfnamefont
				{T.}~\bibnamefont {Miyagi}}, \bibinfo {author} {\bibfnamefont
				{W.}~\bibnamefont {Nazarewicz}}, \bibinfo {author} {\bibfnamefont
				{G.}~\bibnamefont {Neyens}}, \bibinfo {author} {\bibfnamefont {P.-G.}\
				\bibnamefont {Reinhard}}, \bibinfo {author} {\bibfnamefont {S.}~\bibnamefont
				{Rothe}}, \bibinfo {author} {\bibfnamefont {H.~H.}\ \bibnamefont {Stroke}},
			\bibinfo {author} {\bibfnamefont {A.~R.}\ \bibnamefont {Vernon}}, \bibinfo
			{author} {\bibfnamefont {K.~D.~A.}\ \bibnamefont {Wendt}}, \bibinfo {author}
			{\bibfnamefont {S.~G.}\ \bibnamefont {Wilkins}}, \bibinfo {author}
			{\bibfnamefont {Z.~Y.}\ \bibnamefont {Xu}},\ and\ \bibinfo {author}
			{\bibfnamefont {X.~F.}\ \bibnamefont {Yang}},\ }\bibfield  {title} {\bibinfo
			{title} {Measurement and microscopic description of odd--even staggering of
				charge radii of exotic copper isotopes},\ }\href
		{https://doi.org/10.1038/s41567-020-0868-y} {\bibfield  {journal} {\bibinfo
				{journal} {Nature Physics}\ }\textbf {\bibinfo {volume} {16}},\ \bibinfo
			{pages} {620} (\bibinfo {year} {2020})}\BibitemShut {NoStop}%
	\end{thebibliography}
	%
	
\end{document}